\documentclass[graybox]{svmult}

\pdfoutput=1

\usepackage{type1cm}        
\usepackage{makeidx}         
\usepackage{graphicx}        
\usepackage{multicol}        
\usepackage[bottom]{footmisc}

\usepackage{newtxtext}       %
\usepackage{newtxmath}       
\usepackage{caption}
\usepackage{subcaption}
\usepackage{amsmath}
\usepackage{tikz}
\usepackage{enumerate}
\usepackage{url}

\newcommand\AQ[1]{{\color{black}#1}}
\newcommand\NP[1]{{\color{black}#1}}

\newcommand\LD[1]{{\color{black}#1}}

\usetikzlibrary{angles,quotes}

\everymath{\displaystyle}

\newcommand\REV[1]{{\color{black}#1}}

\usepackage{epstopdf}
\AppendGraphicsExtensions{.tiff}

\makeindex             

\begin{document}

\title*{A Mathematical Dashboard for the Analysis of Italian COVID-19 Epidemic Data}
\titlerunning{A Dashboard for the Analysis of Italian COVID-19 Epidemic Data}
\author{Nicola Parolini \and Giovanni Ardenghi \and Luca Dede' \and Alfio Quarteroni}
\institute{
Nicola Parolini \at MOX, Dipartimento di Matematica, Politecnico di Milano, Milan, Italy \email{nicola.parolini@polimi.it}
\and Giovanni Ardenghi \at MOX, Dipartimento di Matematica, 
Politecnico di Milano, Milan, Italy \email{giovanni.ardenghi@polimi.it}
\and Luca Dede' \at MOX, Dipartimento di Matematica, Politecnico di Milano, Milan, Italy \email{luca.dede@polimi.it}
\and Alfio Quarteroni \at MOX, Dipartimento di Matematica, Politecnico di Milano, Milan, Italy \\ Mathematics Institute, EPFL, Lausanne, Switzerland (professor emeritus) \email{alfio.quarteroni@epfl.ch}
}
%
%
\maketitle


\abstract{\LD{An analysis} of the COVID-19 epidemic is proposed on the basis of the \LD{{\it epiMOX}} dashboard (publicly accessible at \url{https://www.epimox.polimi.it}) that deals with data of the epidemic trends and outbreaks in Italy from late February 2020. \REV{Our analysis provides an immediate appreciation of the past epidemic development, together with its current trends by fostering a deeper interpretation of available data through several critical epidemic indicators. In addition, we complement the {\it epiMOX} dashboard with a predictive tool based on an epidemiological compartmental model, named SUIHTER, for the forecast on the near future epidemic evolution.}}

\section{Introduction}
\label{sec:1}
\REV{For} the COVID-19 pandemic, the development of reliable mathematical models, supported by the availability and analysis of complete and accurate data, {is a fundamental tool} for the interpretation and understanding of the epidemic, as well as for providing support to digital health \cite{QDP20Springer}. 
\REV{The availability of accurate and complete historical data together with a suitable epidemiological model for short-term forecasting are essential to provide institutional bodies and authorities with factual quantitative information prior to the adoption of non-pharmaceutical interventions (NPIs).} 

\REV{The first goal of this paper is to present a newly developed mathematical dashboard named {\it epiMOX} (accessible online at the address https://www.epimox.polimi.it ), which gathers the time histories of recorded epidemiological compartments and their first and second rates of variation.} \LD{We believe that} this analysis, which focuses on the situation in Italy at both national and regional scales -- for all the $20$ Italian regions -- will enhance the interpretation and transparency of available data thanks to \REV{a} thorough understanding of the past epidemic development, and of its current trends. \REV{As several other countries, Italy has been hit since February 2020 by a sequence of epidemic waves: the first one initiated on February 21, 2020 and exhausted in early June 2020, while the second wave raised in early October 2020 and peaked in late November 2020, while the outbreak of a third wave is being observed at the time of writing (March 7, 2021).

The second goal of the paper is to enable a comparison between the first and the second epidemic waves in Italy that feature indeed several analogies, as well as significant differences that we aim to highlight and analyze.} 

\REV{Finally, a preview on the expected trend of the epidemic for the near future -- late March 2021 -- is  discussed, based on a compartmental epidemiological model, named SUIHTER, which was purposely designed for the Italian COVID-19 epidemic \cite{SUIHTER20}.}


Our analysis highlights several important \LD{features of the Italian COVID-19 epidemic}. The most relevant are the following:
\begin{enumerate}[(i)]
\item In the early phase of the exponential outbreak, the timeline for the implementation of \REV{NPIs} is crucial. It is \REV{indeed far} more efficient to take more stringent restrictions for a short time span in the early phase of the outbreak rather than implementing less severe \REV{(or even the same)} restrictions for a longer interval later.
\item The second epidemic wave \LD{showed} a slower pace, but a \LD{widespread diffusion in the Country} that \LD{eventually yielded} far worse figures (in terms of fatalities, \LD{patients} hosted in intensive care units, etc.) \LD{than the first wave}.
\item \LD{The epidemiological mathematical model SUIHTER} allows the investigation of various scenarios that conform to the different restrictive \REV{NPIs} devised by the Italian government. \REV{We} identify those that are potentially more effective to contrast the near future epidemic development.
\item \AQ{A retrospective (a posteriori) analysis allows the validation of the SUIHTER. epidemiological model on different phases of the epidemic.}
\item \AQ{Thanks to the SUIHTER model we can carry out a \textit{what-if} analysis aimed at simulating different epidemic \LD{trends} that would have occurred in case different NPIs had been implemented by the Italian authorities at the outbreak of the second wave.}
\end{enumerate}

\REV{In this paper, epidemiological data available up to March 7, 2021 are used to validate the suitability of the {\it epiMOX} dashboard to provide fast and in-depth analyses of the past trends of the Italian outbreak and predict its evolution through the SUIHTER epidemiological model.} Needless to say, the {\it epiMOX} dashboard will continue to monitor the evolution of the epidemic providing updated predictions based on data made available \REV{on} a daily basis.

An outline of this paper is as follows: in Section~\ref{sec:data} a description of the COVID-19 epidemic data time series is supplied together with a description of the data processing tools (filtering, derivative, scaling, normalization) used to straightforwardly identify the main characteristics of each time series. The first and second epidemic waves are analysed and the results are compared at both the national (Section~\ref{sec:national}) and regional (Section~\ref{sec:regional}) scales. Some critical indicators for the epidemic, that can be displayed by the dashboard, are analyzed in Section~\ref{sec:indicators}. In Section~\ref{sec:suihter} we first recall the SUIHTER model and analyze its interplay with the {\it epiMOX} dashboard. In particular, through the dashboard, we validate the model in Section~\ref{sec:validation} against \REV{the} data of the second wave at the national level. Then, we propose what-if-scenarios for the analysis of alternative NPIs in Section~\ref{sec:whatif}, we analyze the impact of the UK variant in Section~\ref{sec:variants}, and we compare different short term future scenarios in Section~\ref{sec:forecast}. Conclusions follow in Section~\ref{sec:conclusions}. 

\section{Data acquisition, processing, and analysis}
\label{sec:data}
The data used in this paper are those made available by the Italian \textit{Dipartimento della Protezione Civile} \LD{(DPC)} through the open data repository \url{https://github.com/pcm-dpc/COVID-19}. 
Data are communicated on a daily basis and include the number of individuals who are currently \textit{positive}, \textit{isolated at home}, \textit{hospitalized}, \textit{hosted in ICUs} (Intensive Care Units), the \textit{\REV{daily positive}} cases, the cumulative number of \textit{deaths} and \textit{recovered} since the beginning of the pandemic, and the number of \textit{swabs} performed. All these data are supplied at \AQ{both} the regional and national levels, while the available data \REV {on} a finer scale (provinces) are unfortunately limited to the count of  \textit{\REV{positive}} cases since the beginning of the epidemic.

From now on we will refer to these data as \textit{raw} data. These raw data \LD{can then be} smoothed by resorting to a local polynomial regression based on the {Savitzky}–Golay convolution filter \cite{SG64}: at day $n$ we attribute the value of  the polynomial of degree $r$ that approximates, in the least squares sense, the $2q+1$ values of the raw data centered on day n, i.e. in the range $[n-q,n+q]$, where $r\le2q$. Here, we consider a cubic least squares polynomial and a window size of $21$ days (that is, $r=3$ and $q=10$). A standard approach based on a \textit{weekly moving average} could be obtained by taking $r=0$ and $q=3$. 

From now on, if not otherwise specified, the time series that will be presented will be smooth curves obtained using the Savitzky–Golay convolution filter \REV{described above}.

The second step is to calculate the first and second rates of variation of the different compartments. First rate of variation describes how fast a trend is either increasing or decreasing: change of sign from positive to negative for the first rate of variation indicates switching from increasing to decreasing in the corresponding curves, whereas a change of sign on the second rate of variations denotes a change of convexity (a point of inflection). 


\subsection{The epidemic at the national scale}\label{sec:national}
\begin{figure}[h]
\centering
\includegraphics[width=\textwidth]{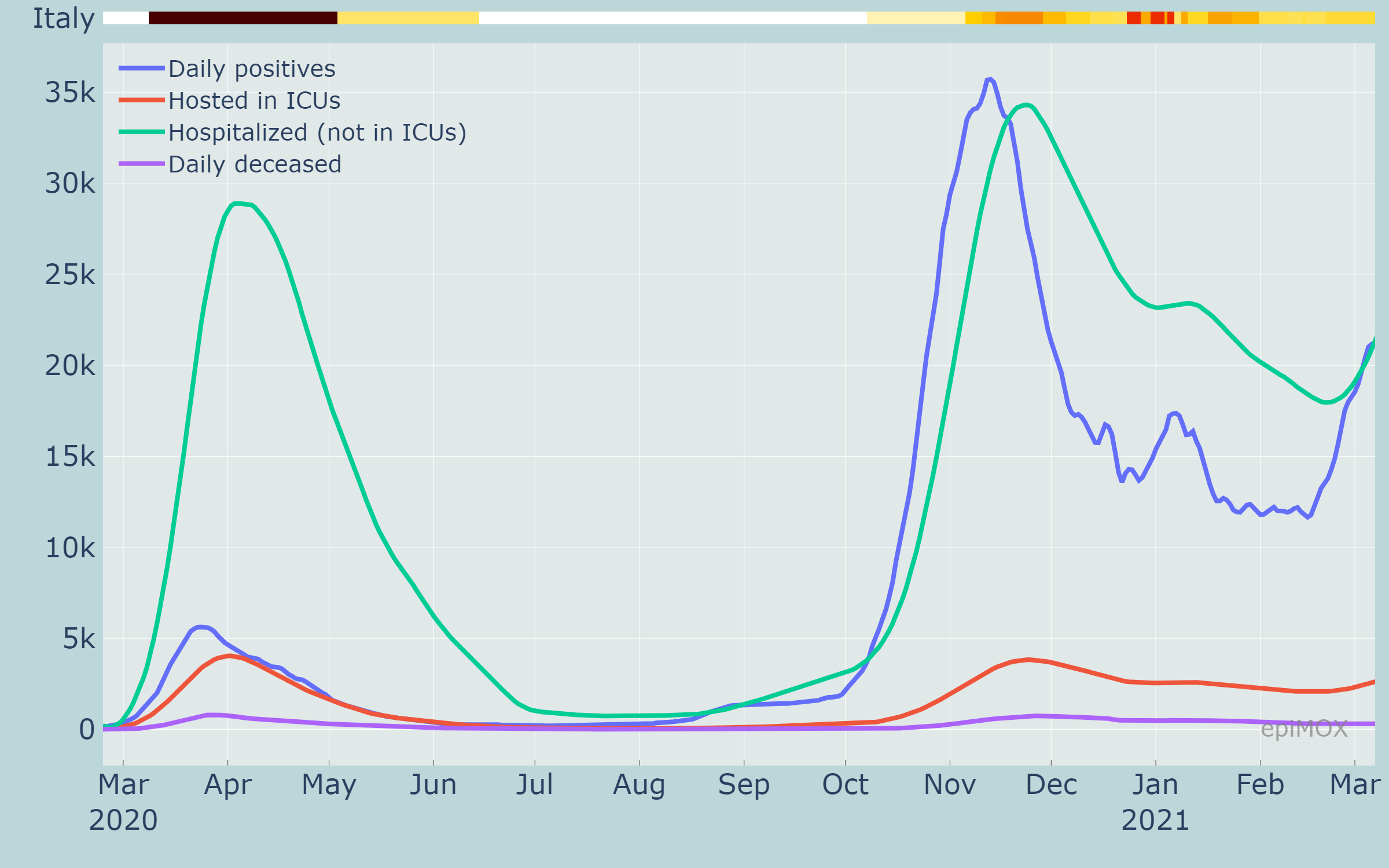}
\caption{Time series of different compartments during the whole epidemic history in Italy}
\label{fig:twowaves}
\end{figure}

The time series of the filtered data (solid lines) for some relevant \AQ{epidemiological compartments}  at the national Italian level, namely the \textit{\REV{daily positive}} cases, the \textit{daily deceased}, the number of individuals that are \textit{hospitalized} with symptoms and those \textit{hosted in ICUs}, are reported in Figure \ref{fig:twowaves}.  \REV{On top of each figure, we display a color bar. The color code used therein is corresponding to the severity of NPIs, as stated in DPCM (the decree issued by the Italian Government) on  November 4, 2020\footnote{\url{https://www.gazzettaufficiale.it/eli/gu/2020/11/04/275/so/41/sg/pdf}}.

The color codes apply at regional level and are defined as follows, according to an algorithm based on different risk indicators: 
\begin{itemize}
    \item \textit{yellow}: moderate risk zones with basic NPIs in place (mandatory use of masks, curfew from 10PM to 5AM, limitations on the activity of shops, bars and restaurants, strong limitations of sport and leisure activities, (partial) distance learning for secondary schools);
    \item \textit{orange}: elevate risk zones with stricter NPIs (municipality confinement, food service activities suspended);
    \item \textit{red}: maximum risk zones with the even stricter NPIs (home confinement, distance learning from grade 8, non-essential commercial activities suspended, all sport and leisure activities suspended).
\end{itemize}
Since November 6, 2020 the color of each region has been updated weekly. At the national level (e.g. in Figure \ref{fig:twowaves}), the color code is obtained as a suitable averaging of those in place in the regions, weighted by their population. The strict lockdown imposed at the national level in Spring 2020 is conventionally identified with the black color.
}

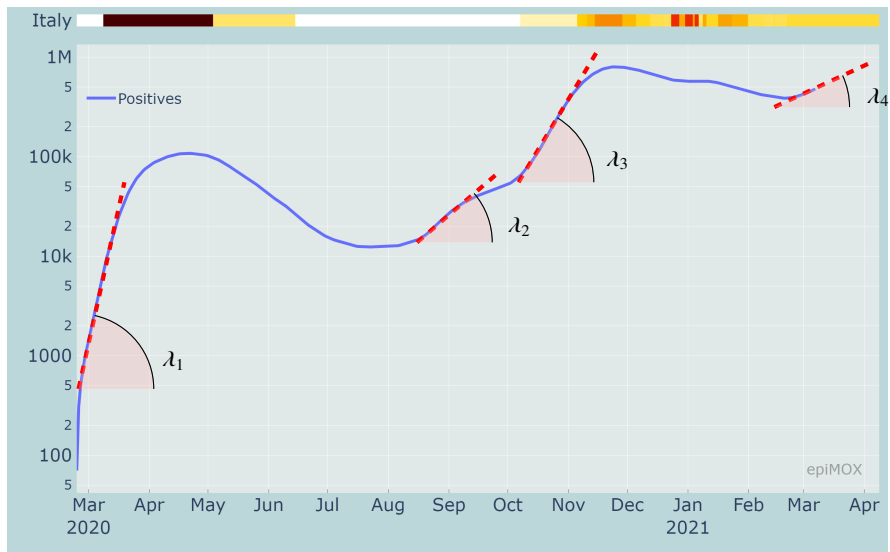
\begin{figure}[hb]
\centering
\begin{tikzpicture}
	\coordinate (a1) at (0.95,2.25);
	\coordinate (b1) at (1.72,2.25);
	\coordinate (c1) at (1.56,5);
    \node[anchor=south west,inner sep=0] at (0,0) {\includegraphics[width=\textwidth]{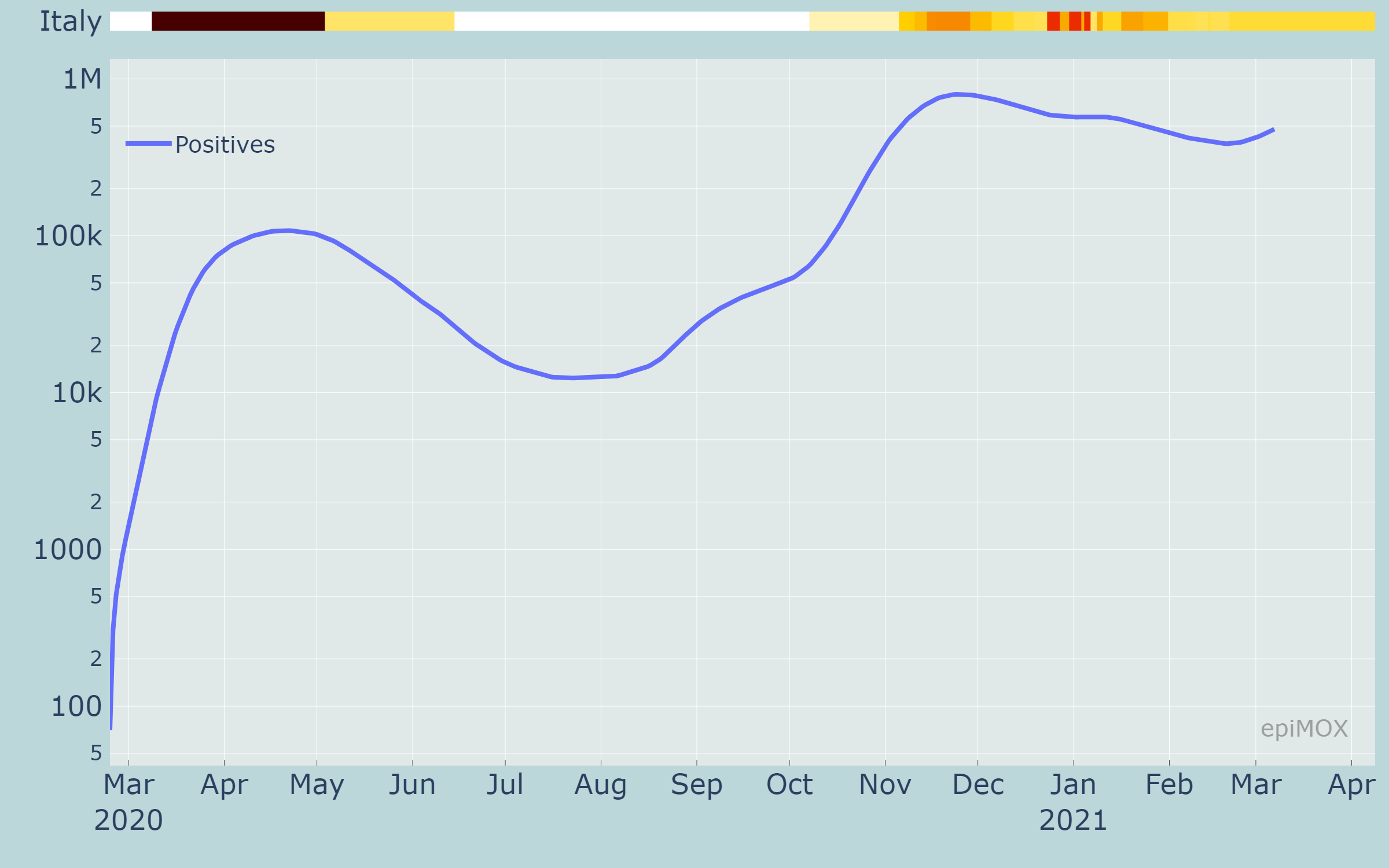}};
    \draw [dashed,red,ultra thick]  (a1) -- (c1);
	\draw pic[draw,fill=red!30,angle radius=1cm,opacity=0.3] {angle=b1--a1--c1};
	\draw pic[draw,angle radius=1cm,"$\lambda_1$" shift={(8mm,0mm)}] {angle=b1--a1--c1};
	\coordinate (a2) at (5.45,4.2);
	\coordinate (b2) at (6.4,4.2);
	\coordinate (c2) at (6.5,5.1);
    \draw [dashed,red,ultra thick]  (a2) -- (c2);
	\draw pic[draw,fill=red!30,angle radius=1cm,opacity=0.3] {angle=b2--a2--c2};
	\draw pic[draw,angle radius=1cm,"$\lambda_2$" shift={(8mm,0mm)}] {angle=b2--a2--c2};
	\coordinate (a3) at (6.8,5.);
	\coordinate (b3) at (7.85,5.);
	\coordinate (c3) at (7.85,6.75);
 	\draw [dashed,red,ultra thick]  (a3) -- (c3);
	\draw pic[draw,fill=red!30,angle radius=1cm,opacity=0.3] {angle=b3--a3--c3};
	\draw pic[draw,angle radius=1cm,"$\lambda_3$" shift={(8mm,0mm)}] {angle=b3--a3--c3};
	\coordinate (a3) at (10.2,6);
	\coordinate (b3) at (11.,6);
	\coordinate (c3) at (11.5,6.6);
 	\draw [dashed,red,ultra thick]  (a3) -- (c3);
	\draw pic[draw,fill=red!30,angle radius=1cm,opacity=0.3] {angle=b3--a3--c3};
	\draw pic[draw,angle radius=1cm,"$\lambda_4$" shift={(8mm,0mm)}] {angle=b3--a3--c3};
\end{tikzpicture}

\caption{Identification of four exponential growth phases in the time series of the \textit{\REV{positive}} cases in logarithmic scale}
\label{fig:twowaveslog}
\end{figure}

By analysing the time series of the \textit{\REV{positive}} cases using a log scale (see Figure~\ref{fig:twowaveslog}) along with the full evolution of the epidemic in Italy, we can observe, for the first epidemic wave, an exponential growth $f(t)=Ce^{\lambda_1 t}$ (which is linear in log scale with slope $\lambda_1$) \AQ{for} the first two weeks of March with a doubling time of approximately $3$ days. 

During the second half of August, very likely because of the relaxation of social distances during \AQ{the summer holidays and the }newly imported cases from abroad, a few days of exponential growth can be observed, although featuring a much lower growth factor  $\lambda_2$, yielding a doubling time of approximately $15$ days. Finally, what we will refer to as the second wave had its exponential growth last October, with a growth factor $\lambda_3$ and a doubling time of about $8$ days. This \REV{was} very likely connected to the increased number of contacts associated with the school opening in September and related commuting, the restart of recreational activities in closed ambiences, \LD{seasonality}, and the full recovery of the working activities that were dramatically reduced during the spring lockdown, as well as (although at a minor extent) during the summer period. \LD{Finally, we observe an exponential growth starting by the end of February 2021 and due to new virus variants featuring a higher transmission rate (see the discussion in Section~\ref{sec:variants}).} \NP{Although, in this case, the  growth rate $\lambda_4$ appears to be lower than previous ones with a corresponding doubling time of approximately $36$ days, the fact that this new outbreak is starting from a \LD{much larger} \AQ{number} (more than half a million) of positive \LD{individuals} makes the situation at present \LD{(early March 2021)} extremely critical.}



\subsubsection{The first epidemic wave}

\begin{figure}[h]
\centering
\includegraphics[width=\textwidth]{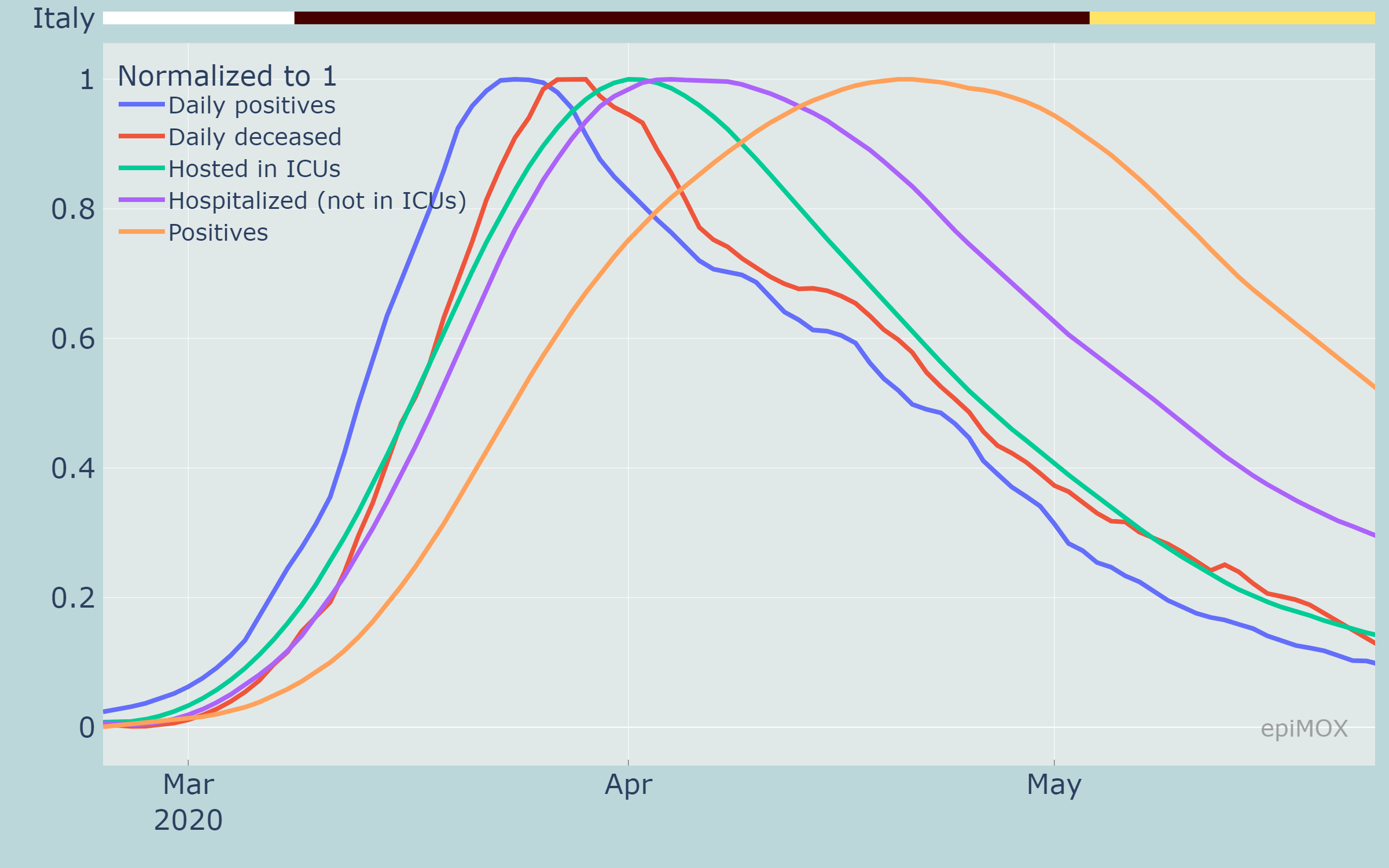}
\caption{Normalized time series of different compartments highlighting the time-shifts between the different peaks  during the first epidemic wave}
\label{fig:firstwavepeaks}
\end{figure}

We first focus on the first epidemic wave that occurred in Italy \REV{last} Spring 2020. In our dashboard, a specific time range can be selected and each time series can be normalised with respect to its maximum attained in the prescribed time range. This allows for an immediate identification of the day the different peaks have occurred and of their relative positions (relative delays). As displayed in Figure~\ref{fig:firstwavepeaks}, it can be noticed that the first compartment that reaches a peak is that of the number of \textit{\REV{daily positive}} cases (on March 24, 2020), followed $5$ days later on March 29, 2020 by the peak of \textit{\REV{daily deceased}}. The peaks for the number of individuals \textit{hosted in ICUs} and those \textit{hospitalized} occur on April 1 and 3, respectively, that is $7$ and $9$ days past the peak of \textit{\REV{daily positive}} cases. 

Even if, at a first glance, the fact that the peak of \textit{\REV{daily deceased}} occurs before those of the hospitalized and \textit{hosted in ICUs} may appear surprising, this is instead reasonable since the latter data do not refer to new daily entries but to the total number of individuals who are hosted in hospitals or ICUs at a specific date. Unfortunately, information (raw data) on the daily admissions in hospitals and ICUs are not available, \AQ{the latter being supplied starting only on December 3, 2020}. Finally, the peak of \textit{\REV{positive}} cases was reached on April 21, 2020, that is almost \LD{one} month past the peak of \textit{\REV{daily positive}} cases.

\begin{figure}[t]
\centering
    \begin{tikzpicture}
         \node[anchor=south west,inner sep=0] at (0,0) {\includegraphics[width=\textwidth]{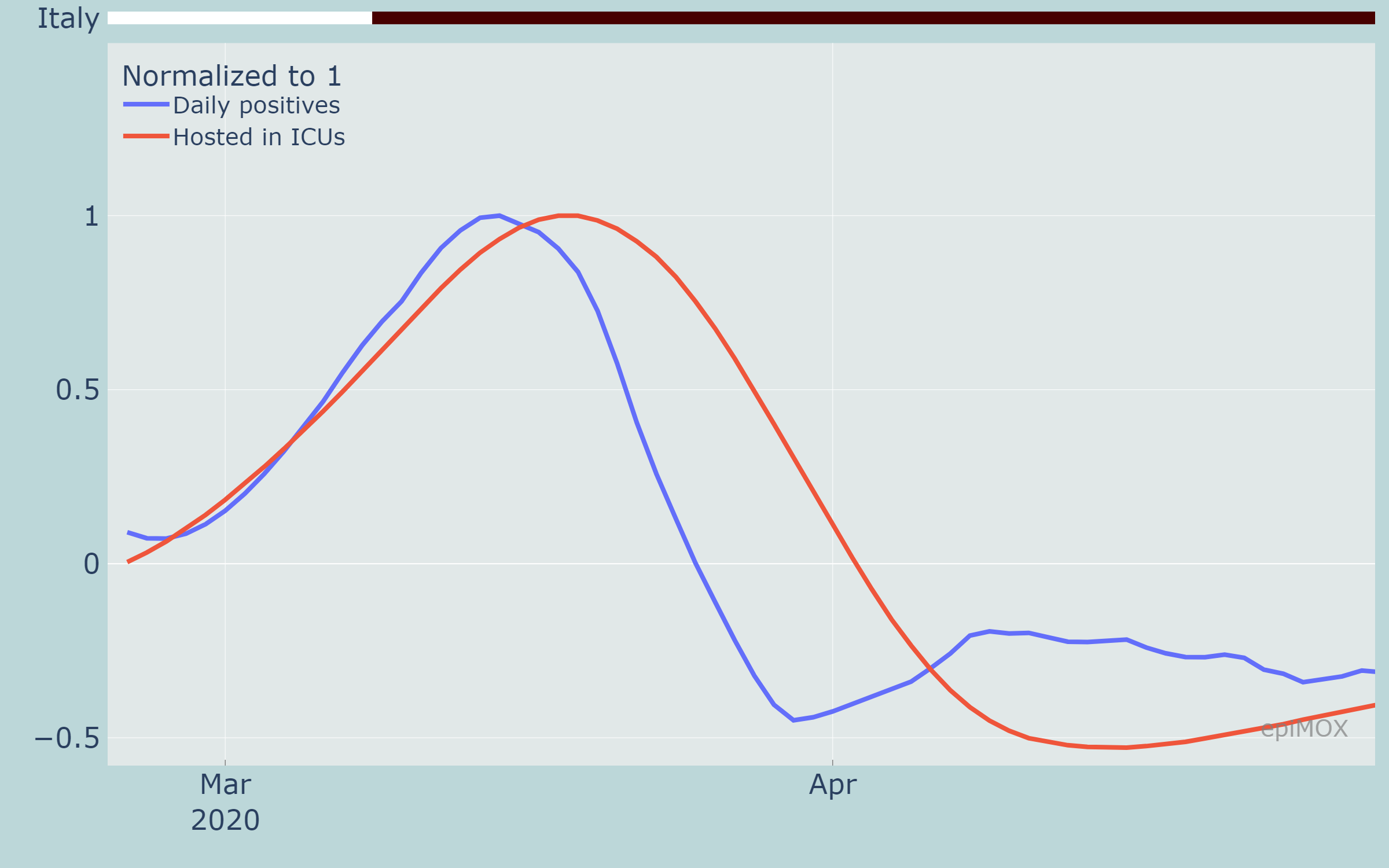}};
        \draw [<->] (4.2,5.8) -- (4.8,5.8) node[pos=0.5, above] {\scriptsize 4 days};
    \end{tikzpicture}
    \caption{Rate of change of \textit{\REV{daily positive}} cases and \textit{hosted in ICUs} highlighting the time shifts between the inflection points during the first epidemic wave}
\label{fig:firstwaveflessi}
\end{figure}

A similar analysis carried out on the rate of change (first derivative) of the different indicators may be used to determine when the initial exponential phase for each indicator is over.
In Figure \ref{fig:firstwaveflessi}, the normalized first derivatives of the same indicators discussed above are \AQ{displayed}. In particular, we notice that the growth rate of the \textit{\REV{daily positive}} cases reached its maximum on March 15, 2020, while the growth rate of the number of \textit{hosted in ICUs} is attained just $4$ days later (March 19, 2020). This corresponds to an inflection point in the time series and indicates the time at which the growth rate starts decreasing. 

\subsubsection{The second epidemic wave}

\NP{The second wave of the COVID-19 epidemic in Italy \REV{started} on the first half of October 2020. The peak of \textit{\REV{daily positive}} cases was reached on November 14, 2020 (see Figure \ref{fig:secondwave_newcases}).
\begin{figure}[t]
\centering
\includegraphics[width=\textwidth]{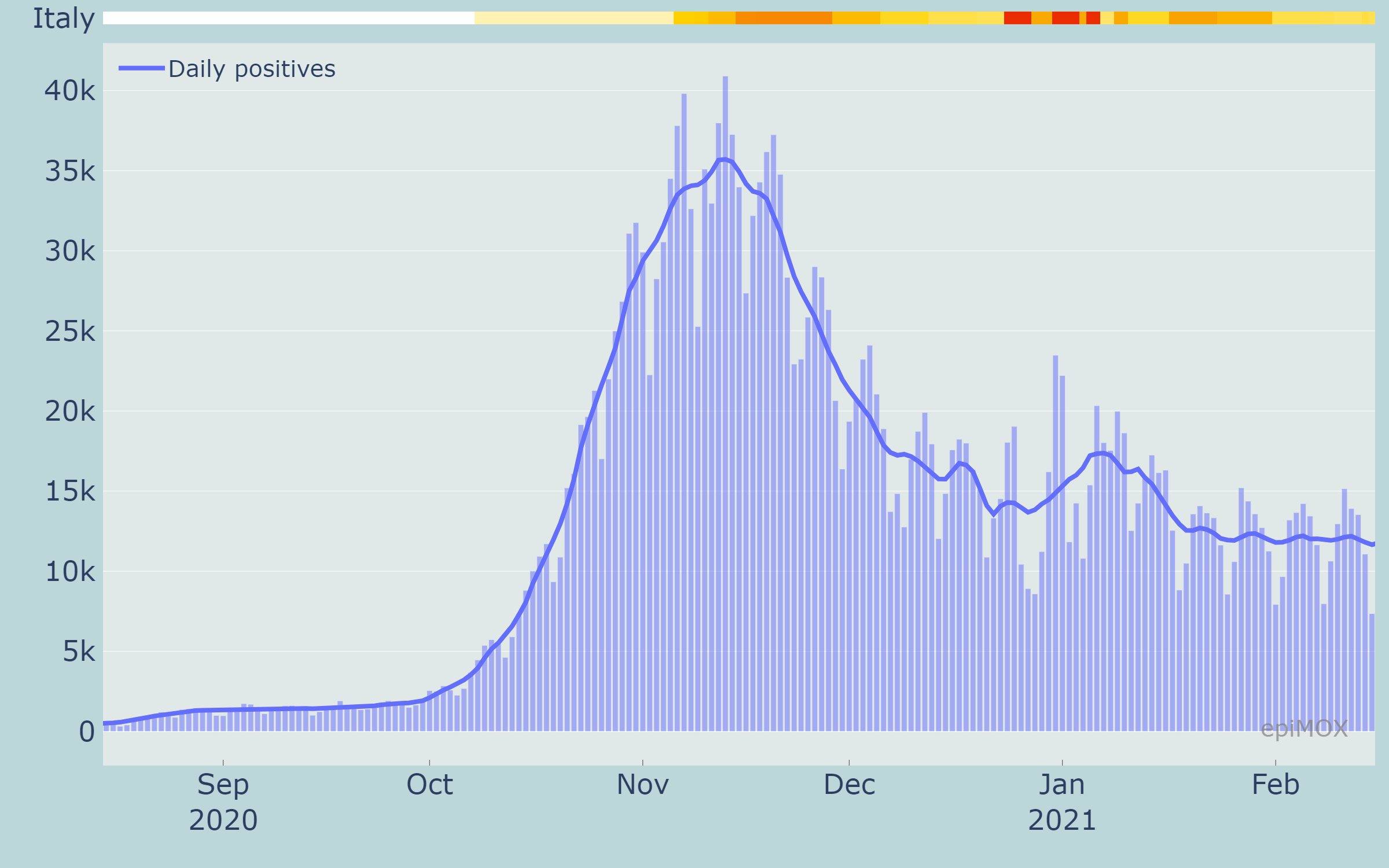}
\caption{Raw (bars) and smoothed (solid line) time series of daily new cases  during the second wave}
\label{fig:secondwave_newcases}
\end{figure}



Other relevant compartments reached their peaks with a time shift \LD{that is} consistent with the epidemic evolution. In particular, the maximum number of \textit{hospitalized},  \textit{hosted in ICUs} and \textit{\REV{daily deceased}} were reached on November 23, 24 and 29, respectively (see the normalized time series in Figure  \ref{fig:secondwavepeaksnormalized}). 

\begin{figure}
    \centering
    \includegraphics[width=\textwidth]{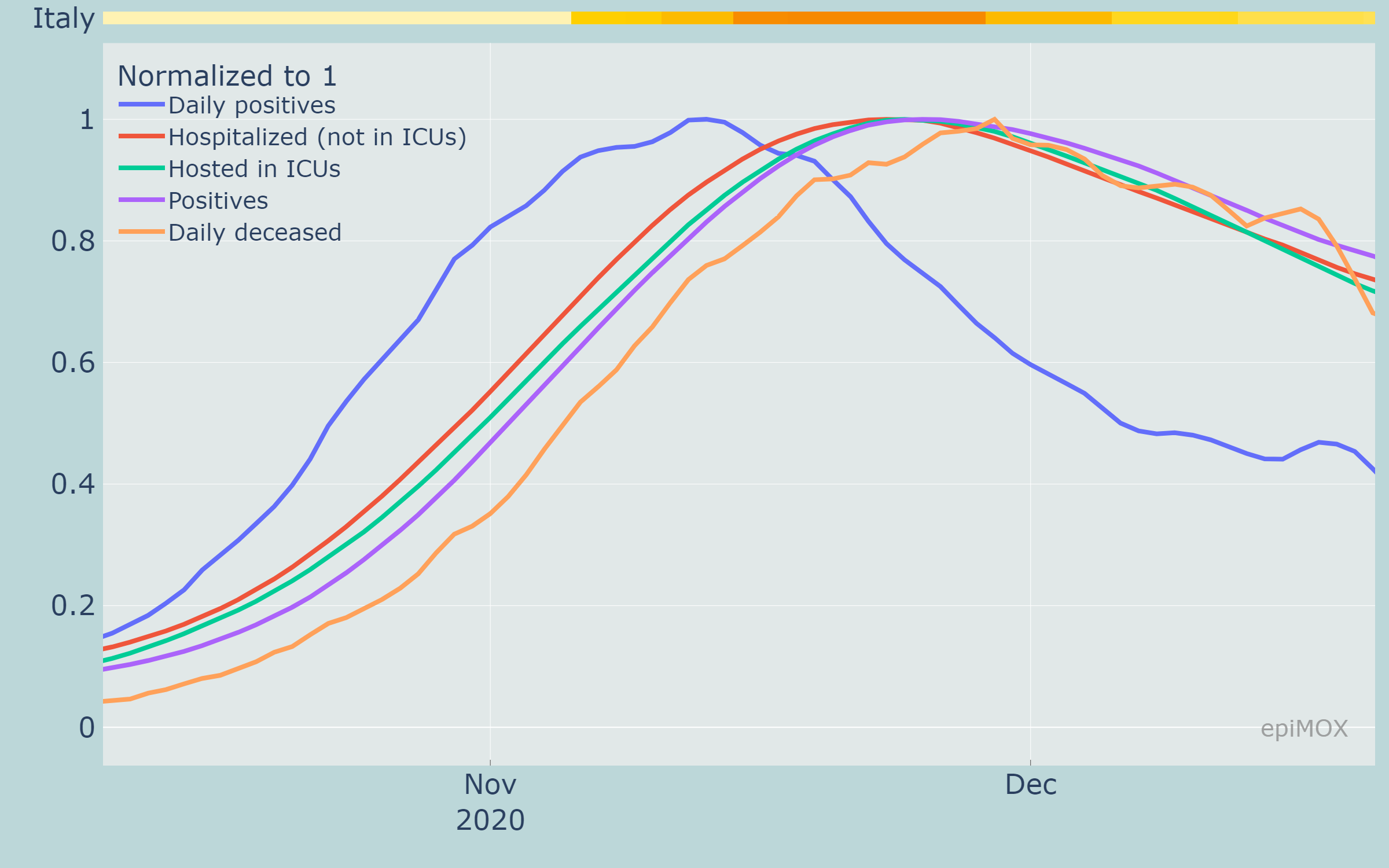}
    \caption{Normalized time series of different compartments highlighting the time-shifts between the different peaks during the second epidemic wave}
\label{fig:secondwavepeaksnormalized}
\end{figure}

As already observed for the first wave, monitoring the  growing rate (first derivative) of the time series can provide a preliminary indication on when the peaks should be expected.
For instance, as displayed in Figure \ref{fig:secondwaveflessi}, the peak on the growth rate of the \textit{\REV{daily positive}}  (on October 24, 2020) anticipates the peak of the \textit{hosted in ICUs} by $12$ days (November 5, 2020).

\begin{figure}[b]
\centering
    \begin{tikzpicture}
         \node[anchor=south west,inner sep=0] at (0,0) {\includegraphics[width=\textwidth]{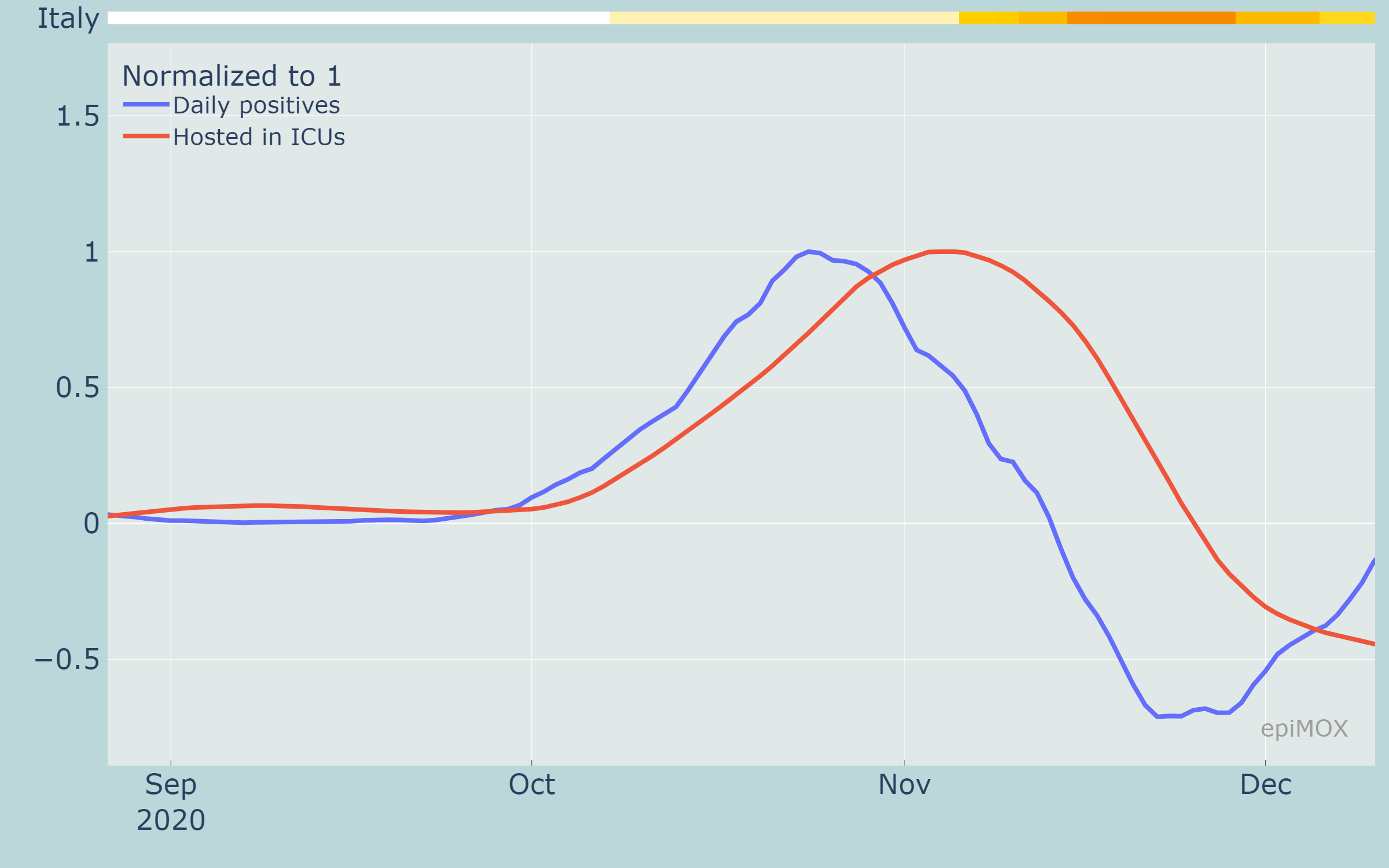}};
        \draw [<->] (6.85,5.5) -- (8,5.5) node[pos=0.5, above] {\scriptsize 12 days};
    \end{tikzpicture}
\caption{Rate of change of \textit{daily positive} cases and \textit{hosted in ICUs} highlighting the time shifts between the inflection points during the second epidemic wave}
\label{fig:secondwaveflessi}
\end{figure}

For the second wave, a consistent time lag between the peaks of the first derivative and the peak of the corresponding time series can be noticed. 
As shown in Table \ref{tab:peak_2nd_wave} for most of the compartments (\textit{Positives},  \textit{Daily positives} \textit{Hospitalized} and \textit{Hosted in ICUs}) there is a 20-days delay between inflection point and the maximum, while this delay is slightly longer (24 days) for the \textit{Daily deceased} time series. \REV{A consistent time lag between inflection point and maximum point (the one corresponding to the peak) can be used as a preliminary indicator in future epidemic waves to predict the day at which the peak will possibly occur.}

A similar correlation between the peak of the derivative and that of the corresponding time series is missing during the first wave. This is likely due to the severe under-estimation of data \LD{collected for some epidemiological compartments and then provided by the DPC.}}

\begin{table}[]
    \centering
    \begin{tabular}{|c|c|c|c|c|c|}
    \hline
     &  \multicolumn{5}{c|}{Peak date}\\ \hline  
    Trend & 
    Positives & Daily positives &
    Hospitalized & ICUs &
    Daily deceased \\
    \hline
    Time series  & 2020-11-25 & 2020-11-13 & 2020-11-23 & 2020-11-24 & 2020-11-29 \\ 
    First derivative  & 2020-11-05  & 2020-10-24 & 2020-11-04 & 2020-11-05 & 2020-11-05 \\
    \hline
    \end{tabular}
    \caption{\AQ{Peak date of the time series of several epidemiological compartments and that of the corresponding first derivative relative to the second epidemic wave in Italy}}
    \label{tab:peak_2nd_wave}
\end{table}

\subsection{The epidemic at the regional scale}\label{sec:regional}
\begin{figure}[h]
\centering
    \begin{subfigure}[b]{0.46\textwidth}
         \centering
         \includegraphics[width=\textwidth]{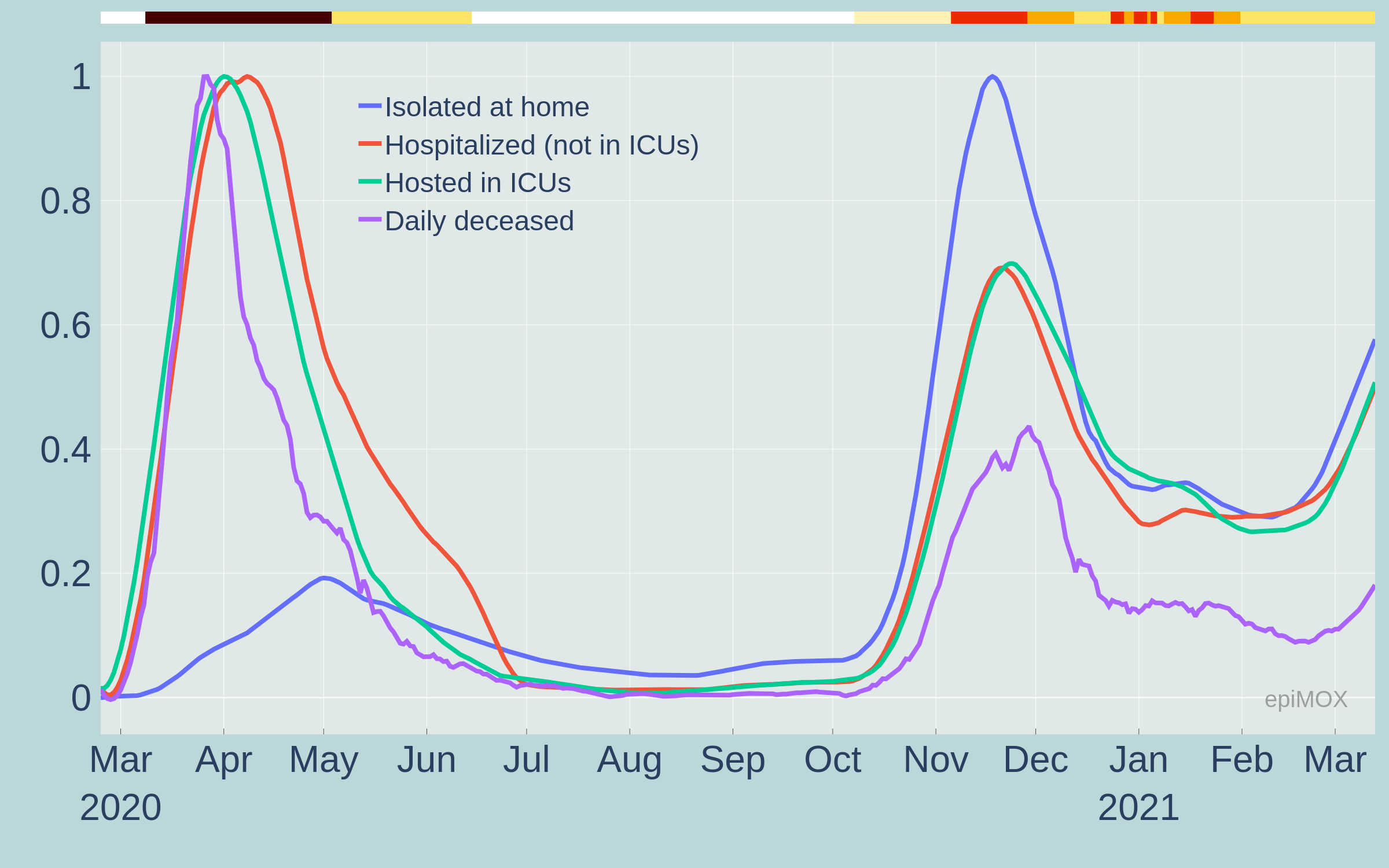}
         \caption{Lombardia}
     \end{subfigure}
     \hspace{8mm}
     \begin{subfigure}[b]{0.46\textwidth}
         \centering
         \includegraphics[width=\textwidth]{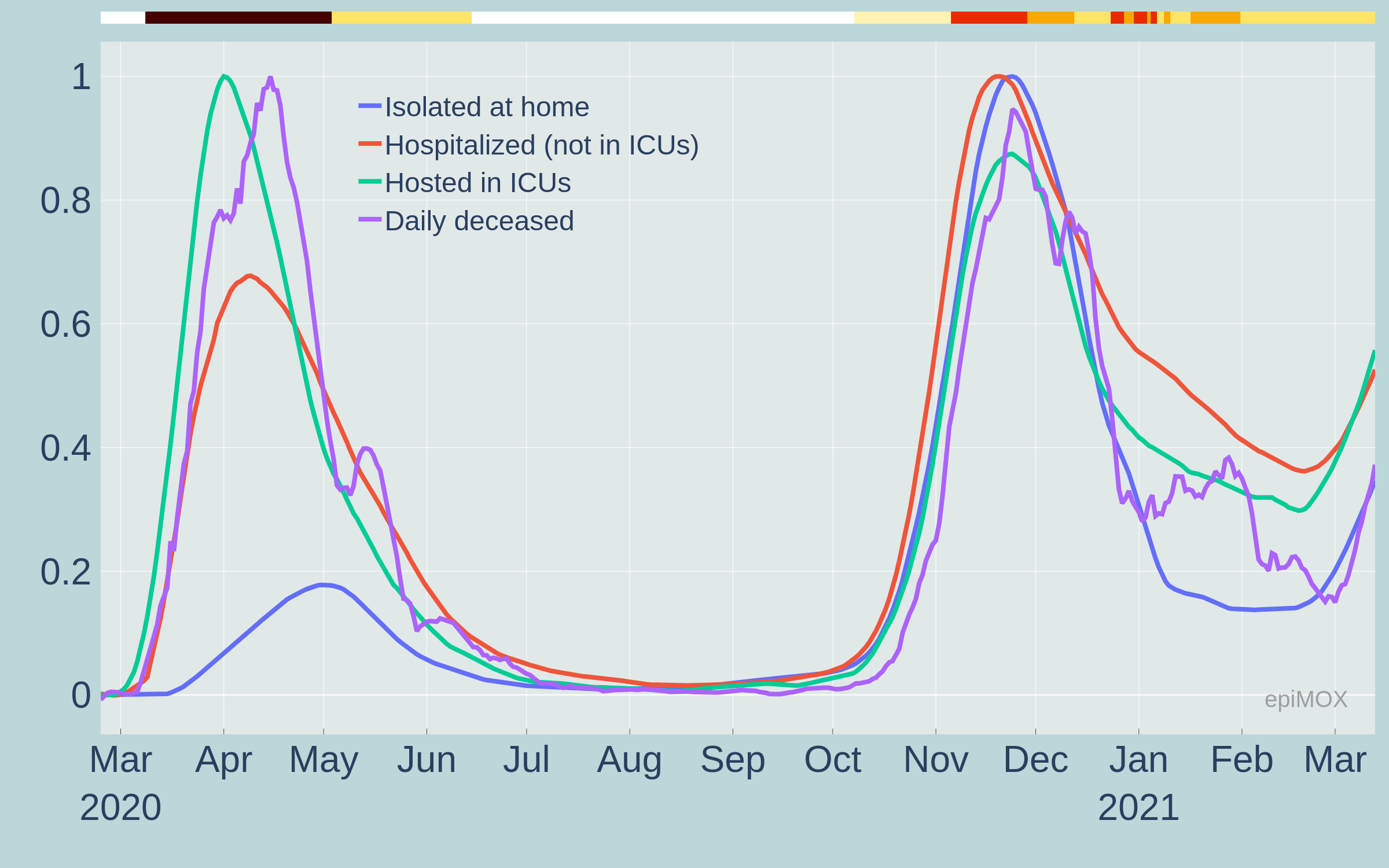}
         \caption{Piemonte}
     \end{subfigure}
     \begin{subfigure}[b]{0.46\textwidth}
         \centering
         \includegraphics[width=\textwidth]{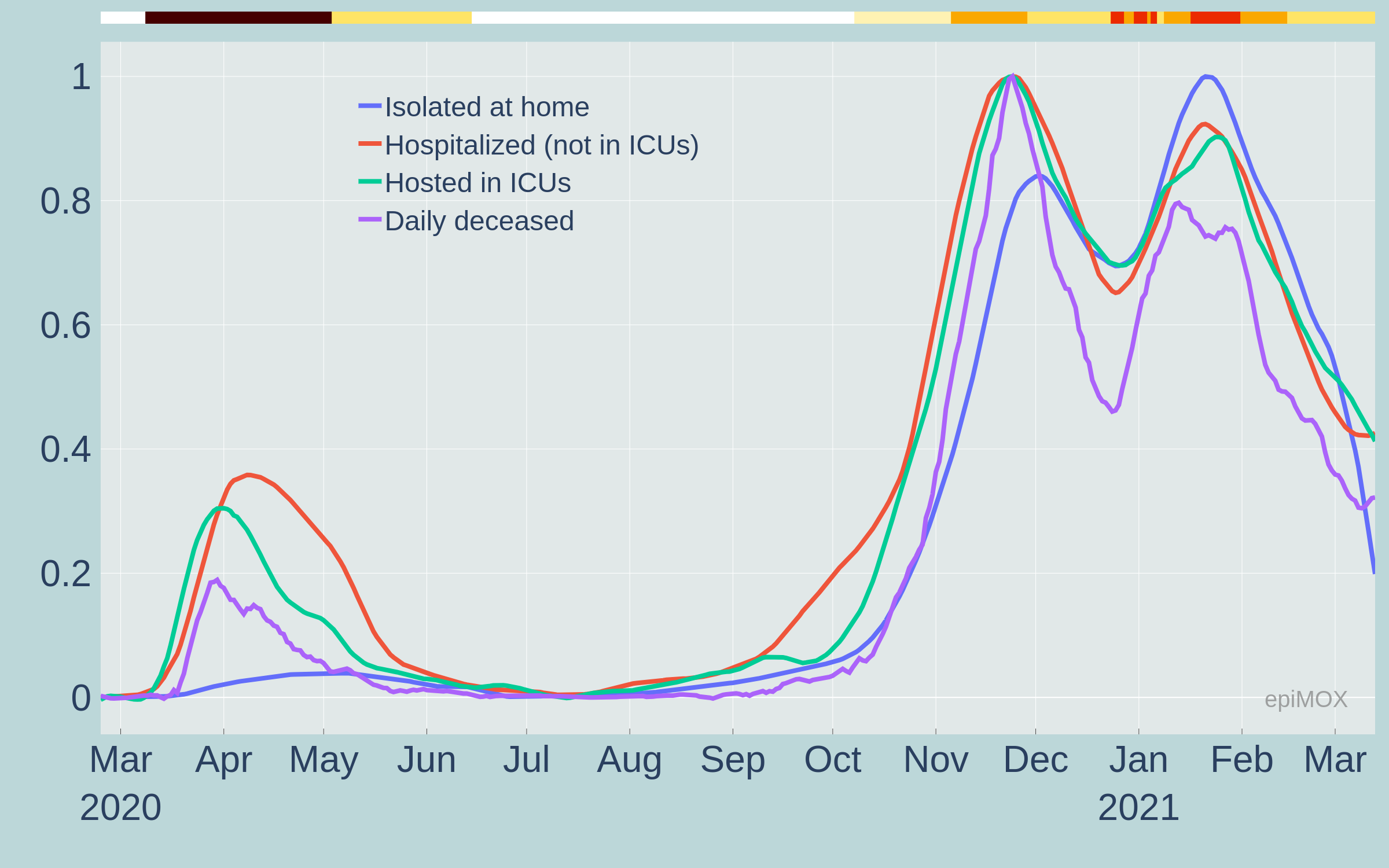}
         \caption{Veneto}
     \end{subfigure}
     \hspace{8mm}
     \begin{subfigure}[b]{0.46\textwidth}
         \centering
     \includegraphics[width=\textwidth]{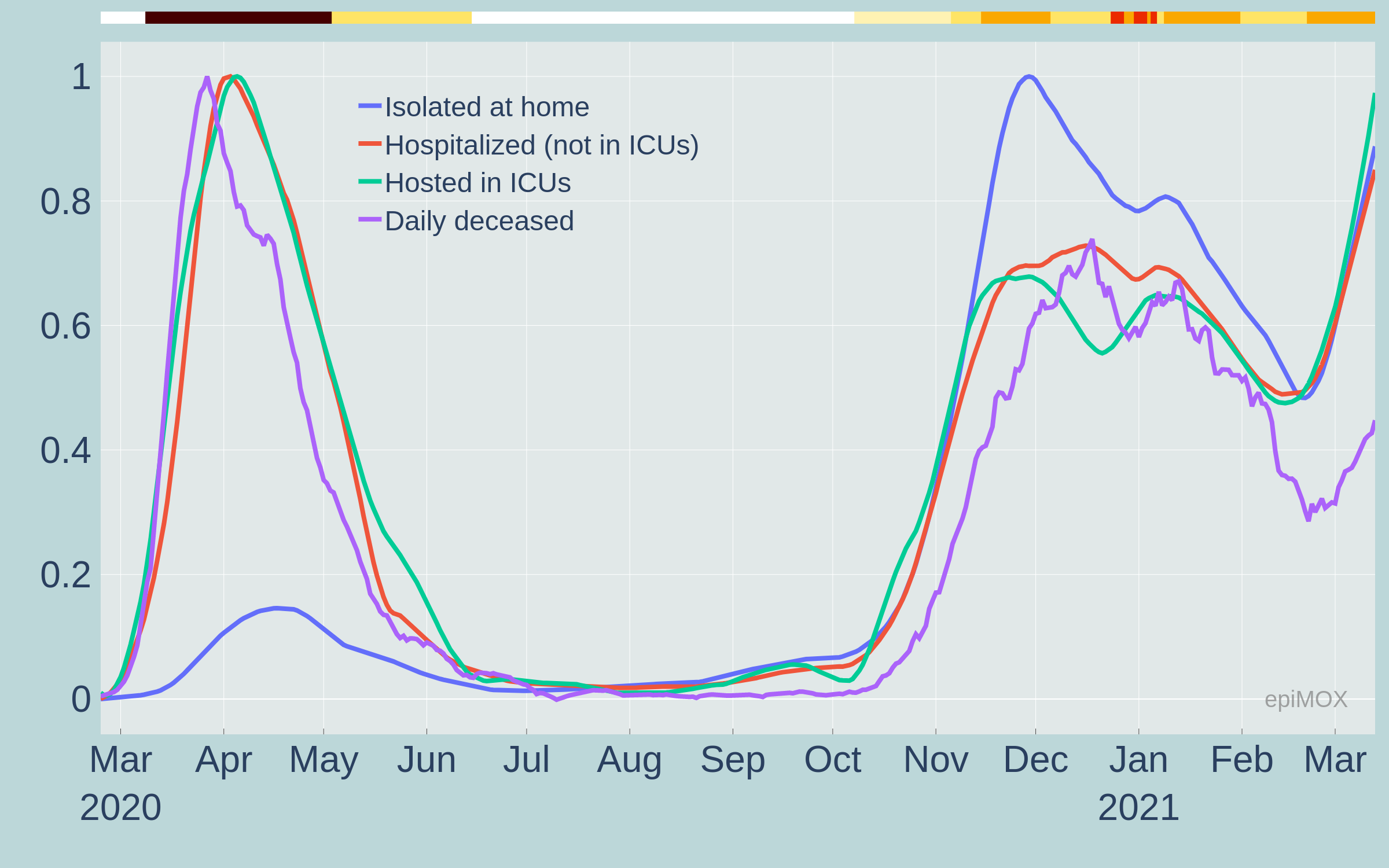}
         \caption{Emilia-Romagna}
     \end{subfigure}
         \begin{subfigure}[b]{0.46\textwidth}
         \centering
         \includegraphics[width=\textwidth]{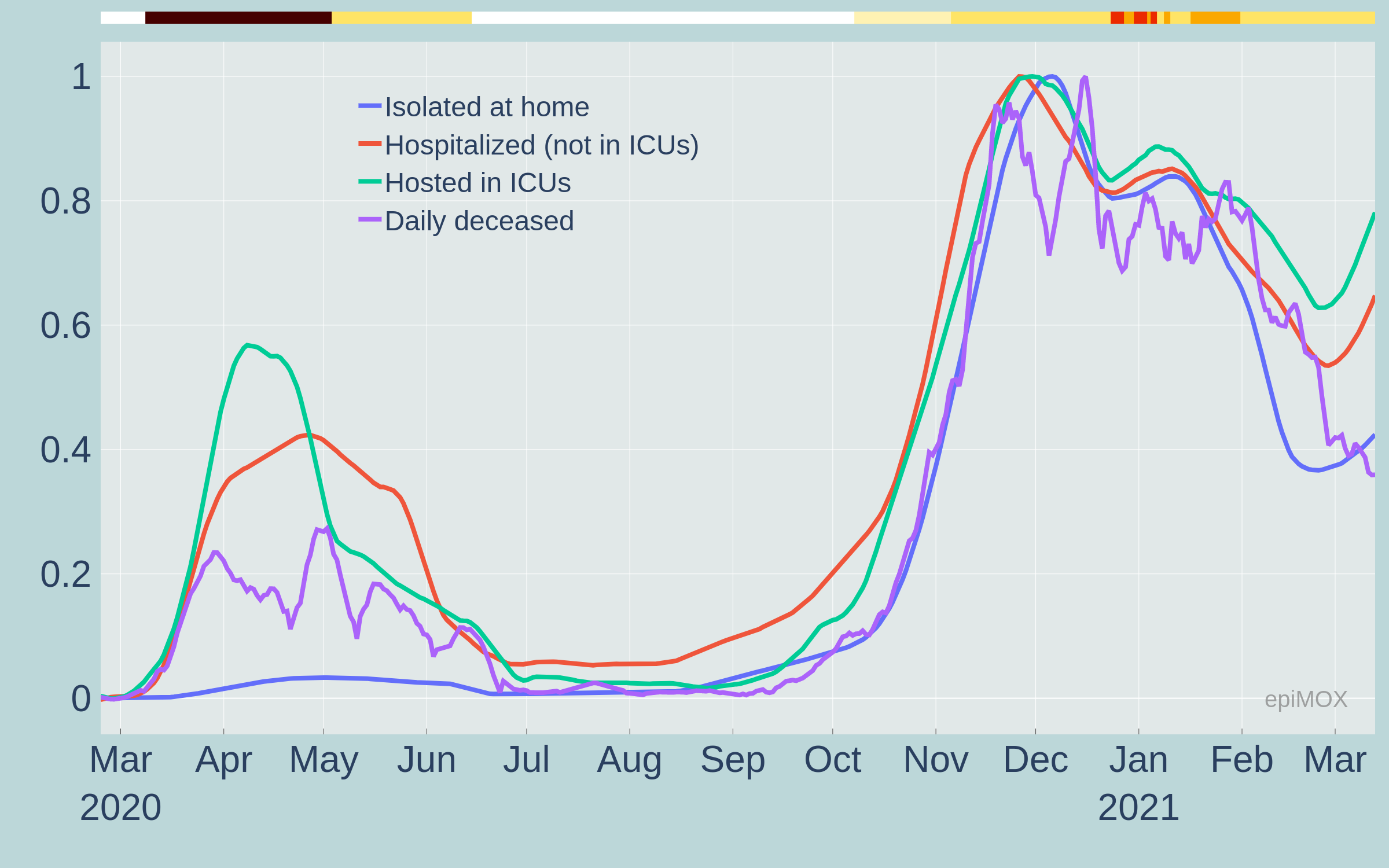}
         \caption{Lazio}
     \end{subfigure}
     \hspace{8mm}
     \begin{subfigure}[b]{0.46\textwidth}
         \centering
         \includegraphics[width=\textwidth]{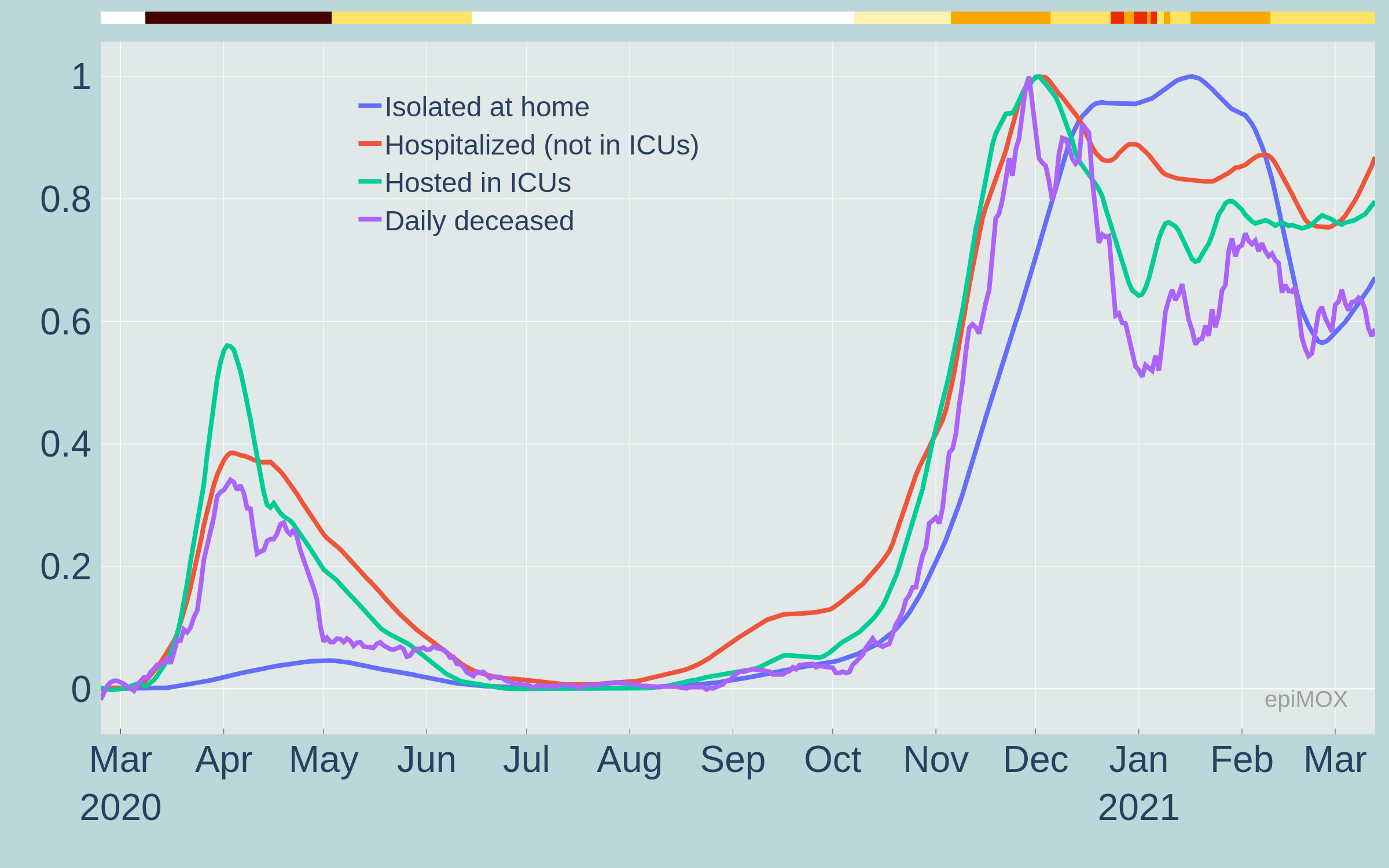}
         \caption{Puglia}
     \end{subfigure}
     \begin{subfigure}[b]{0.46\textwidth}
         \centering
         \includegraphics[width=\textwidth]{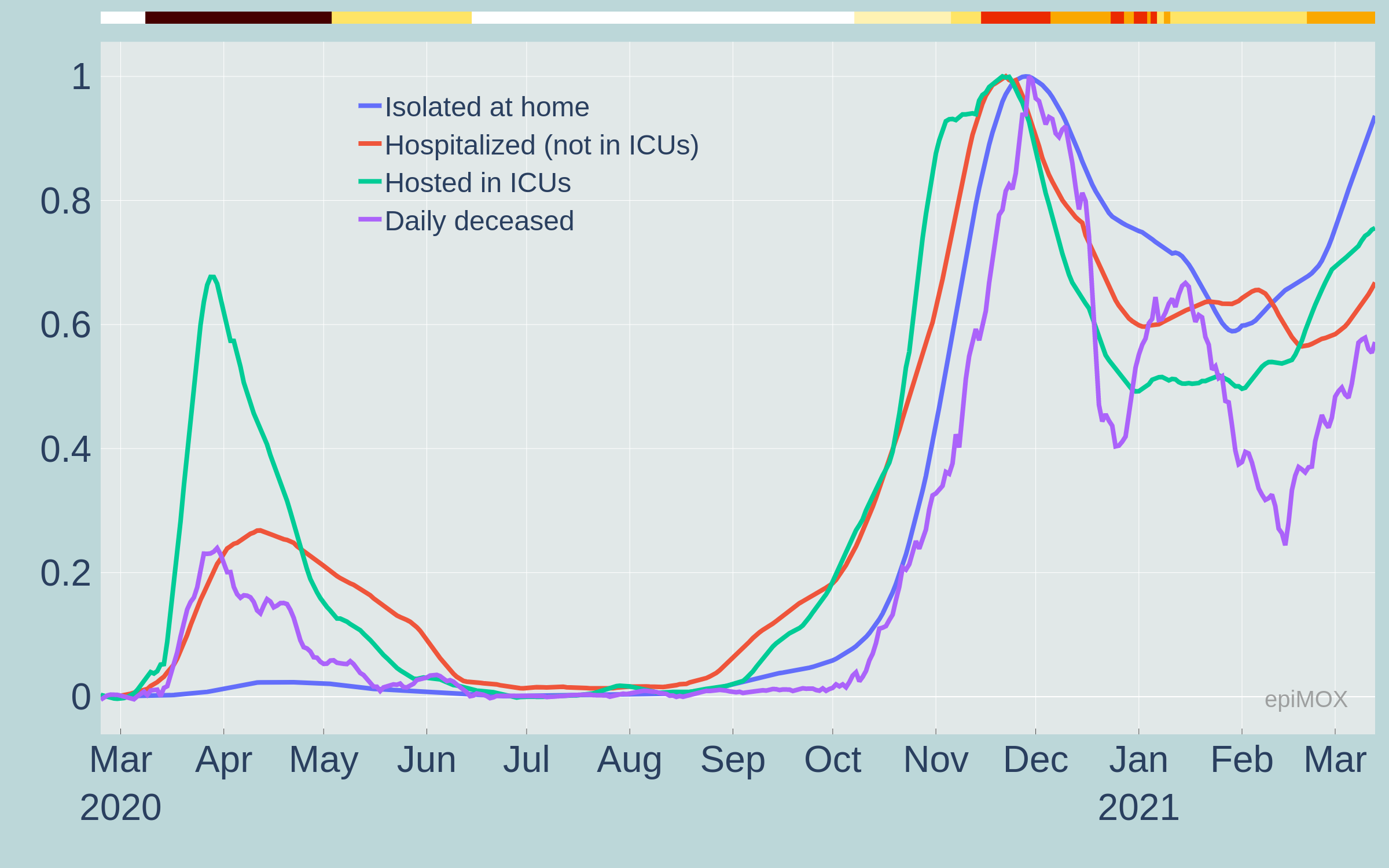}
         \caption{Campania}
     \end{subfigure}
     \hspace{8mm}
     \begin{subfigure}[b]{0.46\textwidth}
         \centering
     \includegraphics[width=\textwidth]{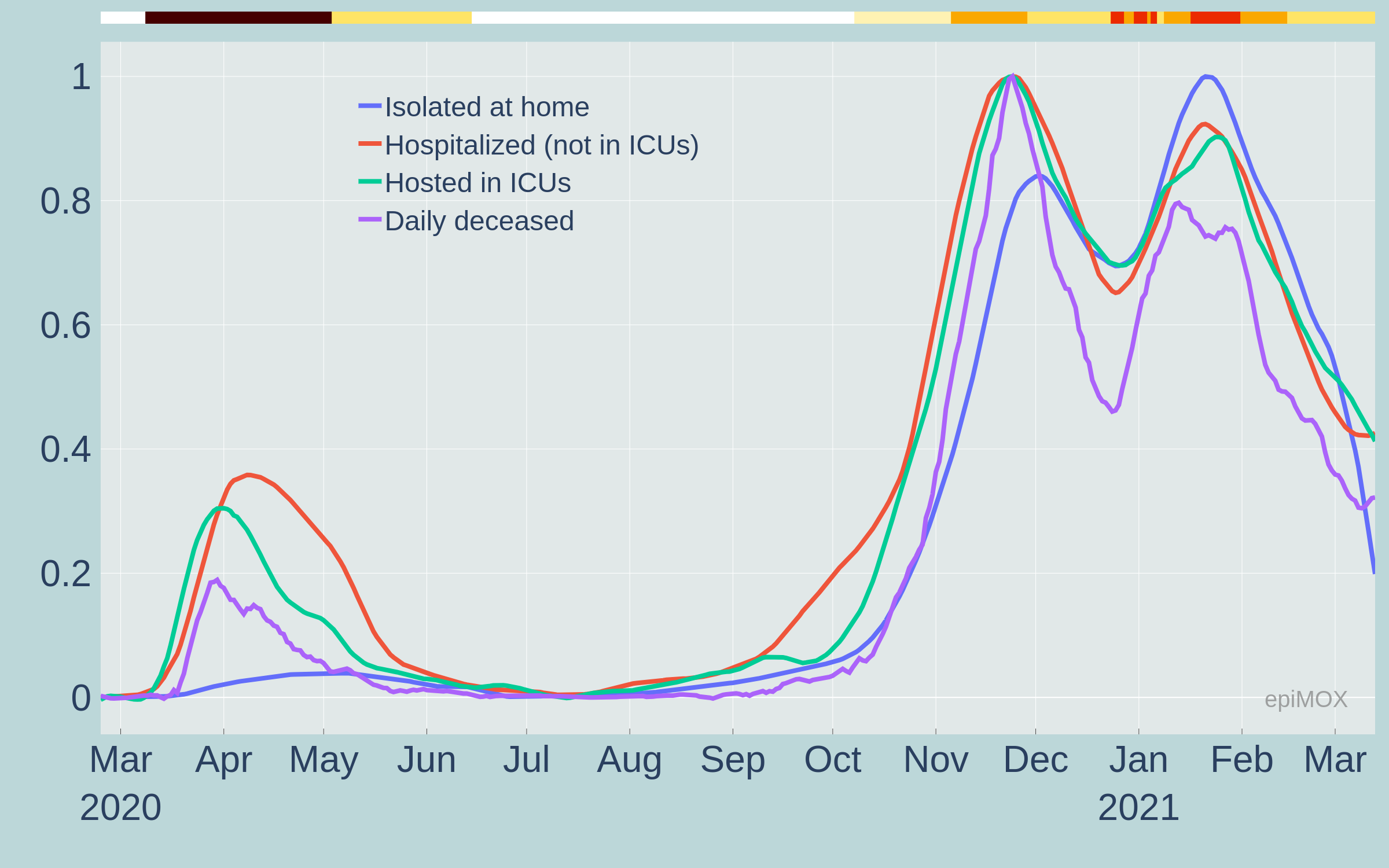}
         \caption{Sicilia}
     \end{subfigure}
\caption{Normalized time series of different compartments during the whole epidemic history in $8$ Italian regions}
\label{fig:regions}
\end{figure}

The same analyses that have been proposed at the national Italian level can be carried out at the regional scale, that is for each of the $19$ Italian Regions and the $2$ autonomous provinces of Trento and Bolzano. In Figure~\ref{fig:regions}, the normalized time series of \textit{\REV{daily positive}} cases, \textit{hospitalized}, \textit{hosted in ICUs} and \textit{\REV{daily deceased}} are presented for $8$ Italian regions, among which Lombardia, Emilia Romagna, Piemonte, Veneto that were severely hit by the first epidemic wave, and  Lazio, Puglia, Campania, Sicilia that are far more evidently affected by the latter epidemic wave than by the former. 

\begin{figure}[h]
\centering
\begin{subfigure}[b]{\textwidth}
         \centering
         \includegraphics[width=\textwidth]{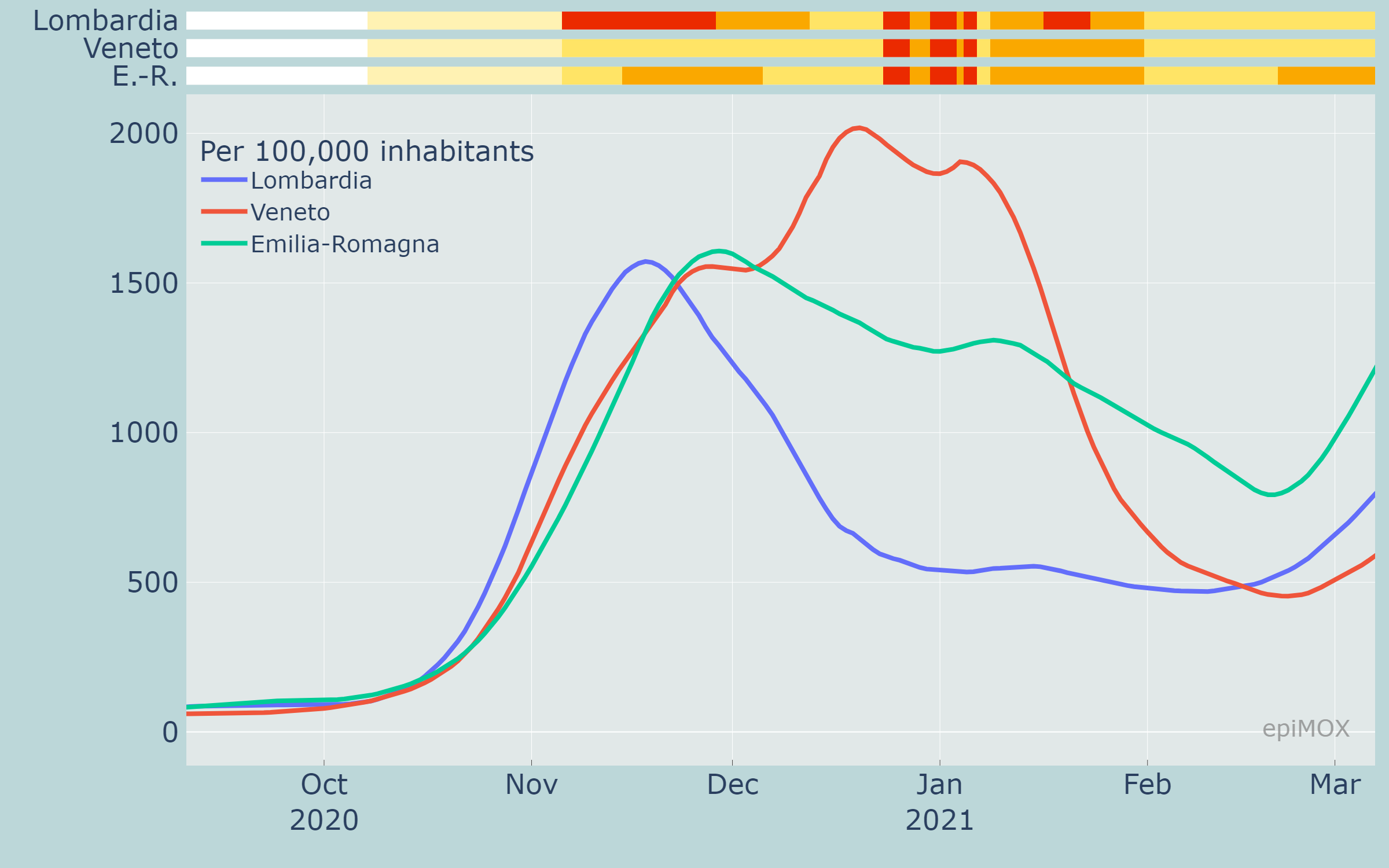}
         \caption{Positives}
     \end{subfigure}
     \hspace{8mm}
     \begin{subfigure}[b]{\textwidth}
         \centering
         \includegraphics[width=\textwidth]{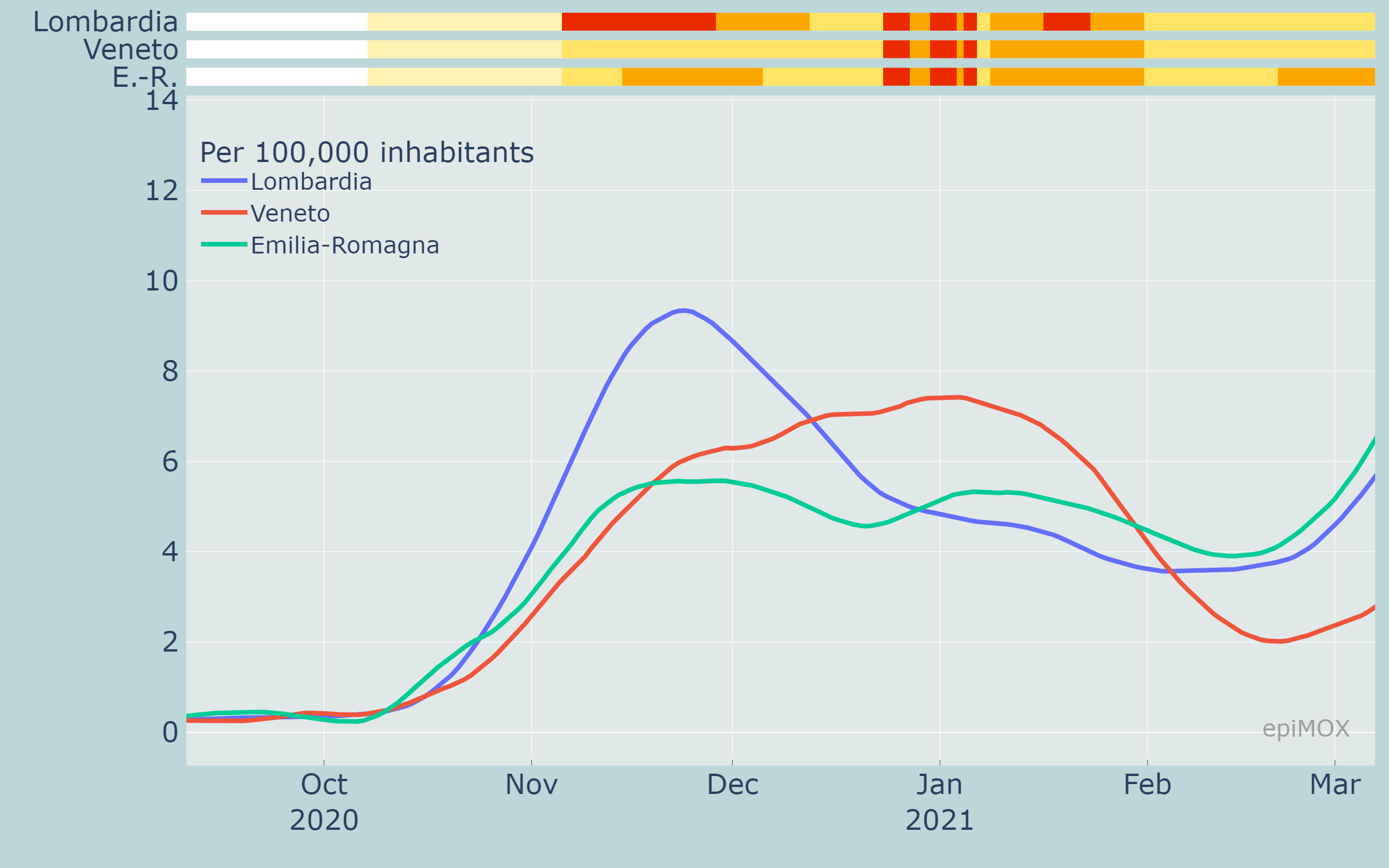}
         \caption{ICUs}
     \end{subfigure}
\caption{Comparison of \textit{\REV{positive}} cases and \textit{hosted in ICUs} per 100,000 inhabitants in Emilia Romagna, Lombardia and Veneto during the second wave in relation with the measures adopted by each region}
\label{fig:nord_comp}
\end{figure}

\begin{figure}[h]
\centering
\begin{subfigure}[b]{\textwidth}
         \centering
         \includegraphics[width=\textwidth]{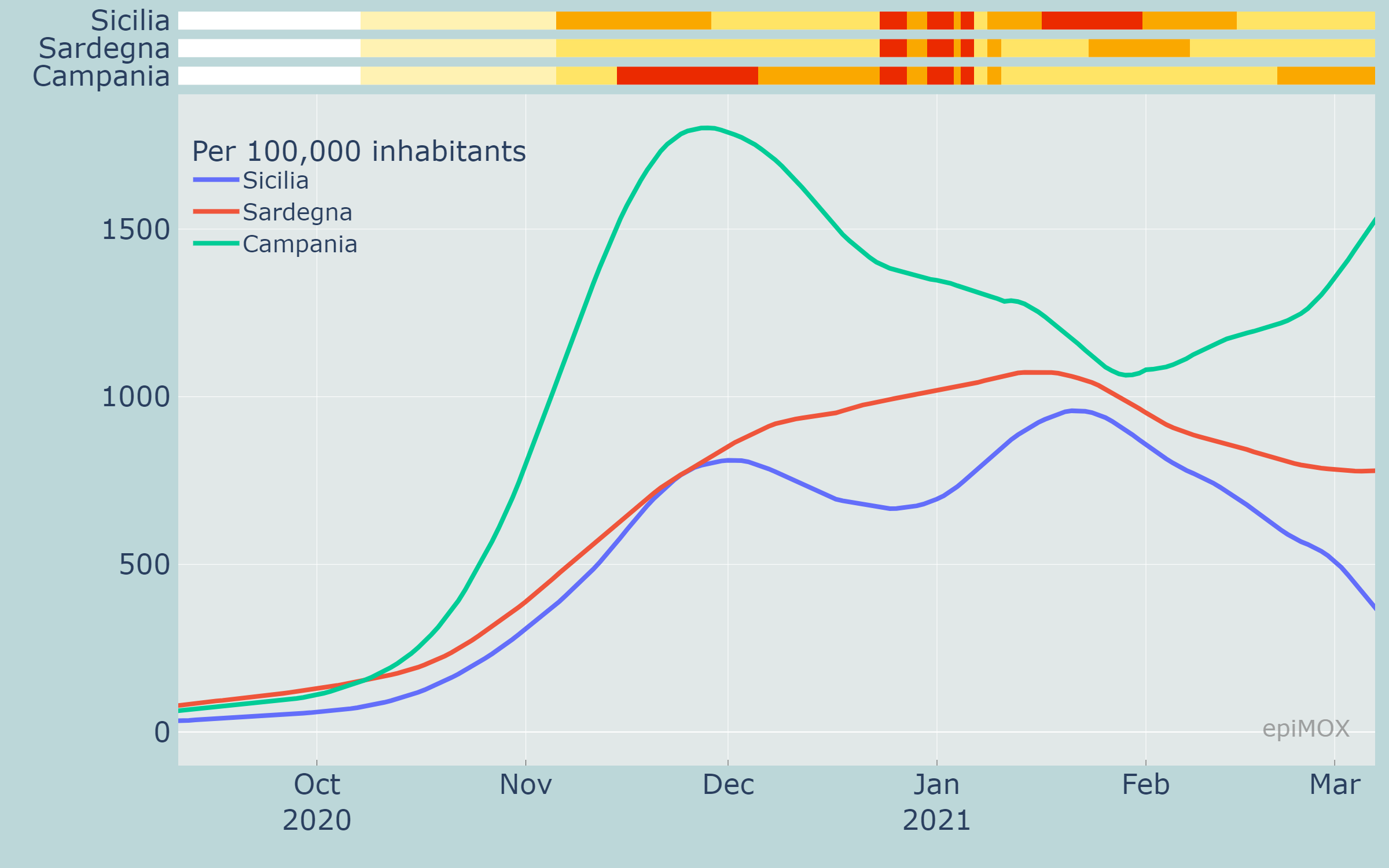}
         \caption{Positives}
     \end{subfigure}
     \hspace{8mm}
     \begin{subfigure}[b]{\textwidth}
         \centering
         \includegraphics[width=\textwidth]{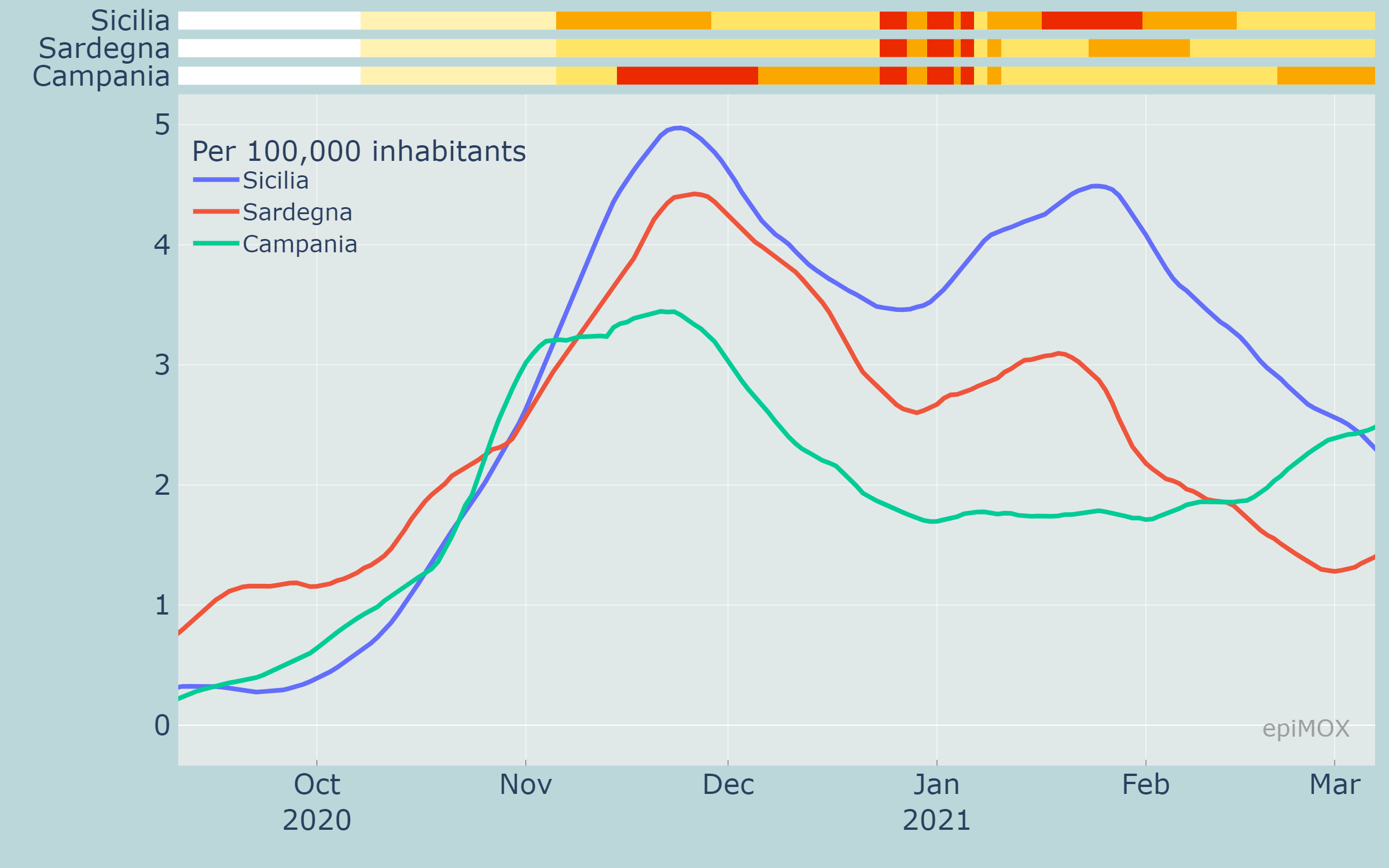}
         \caption{ICUs}
     \end{subfigure}
\caption{Comparison of \textit{\REV{positive}} cases and \textit{hosted in ICUs} per 100,000 inhabitants in Campania, Sardegna and Sicilia during the second wave in relation with the measures adopted by each region}
\label{fig:sud_comp}
\end{figure}

\AQ{In Figure \ref{fig:nord_comp} we display the evolution of the pandemic in \REV{the eight largest} Italian regions since the \LD{early stages of the second wave} (September 2020). We compare two compartments: \REV{\textit{positives}, and \textit{hosted in ICUs}}, in three regions in northern Italy (Lombardia, Veneto, Emilia-Romagna) that were severely affected by the first wave. The same comparison is carried out in Figure \ref{fig:sud_comp} concerning three southern Italian regions (Campania, Sicilia and Sardegna, the latter two being insular) that were very mildly affected by the first wave last Spring 2020. We recall that, in all figures, the upper color bars indicate the non-uniform NPIs. If we focus on the number of \textit{Positives} per 100,000 inhabitants (Figures \ref{fig:nord_comp}(a) and \ref{fig:sud_comp}(a)) that were diagnosed as positive to the virus, we notice that the red regime yields a pronounced decrease of the curves. Both the orange and yellow regimes serve to mitigate the outbreak, without however significantly damp it. This behavior is less evident on the \textit{hosted in ICUs} compartment (Figures \ref{fig:nord_comp}(b) and \ref{fig:sud_comp}(b)). A possible explanation is that this latter compartment is slightly more "indifferent" to the stringency of the NPIs that are being implemented. The \textit{hosted in ICUs} individuals are at a very large extent elder people who were less directly exposed to outdoor contacts as either they were \LD{residents of healthcare facilities} or they spent a substantial part of their time at home.}

\clearpage

\begin{figure}
    \centering
    \includegraphics[width=\textwidth]{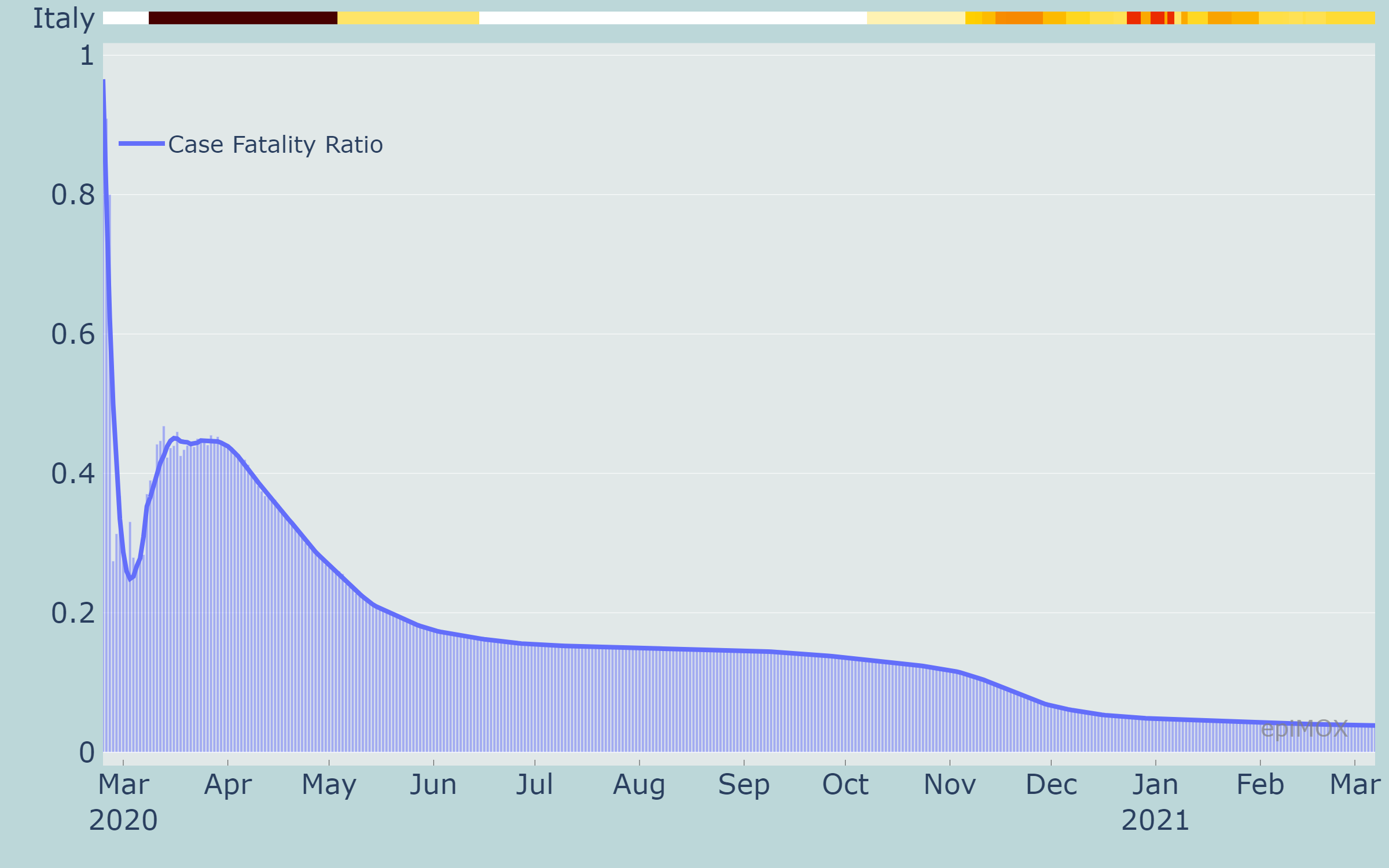}
    \caption{Case Fatality Ratio (CFR) in Italy since the beginning of the epidemic}
    \label{fig:cfr}
\end{figure}

\begin{figure}
    \centering
    \includegraphics[width=\textwidth]{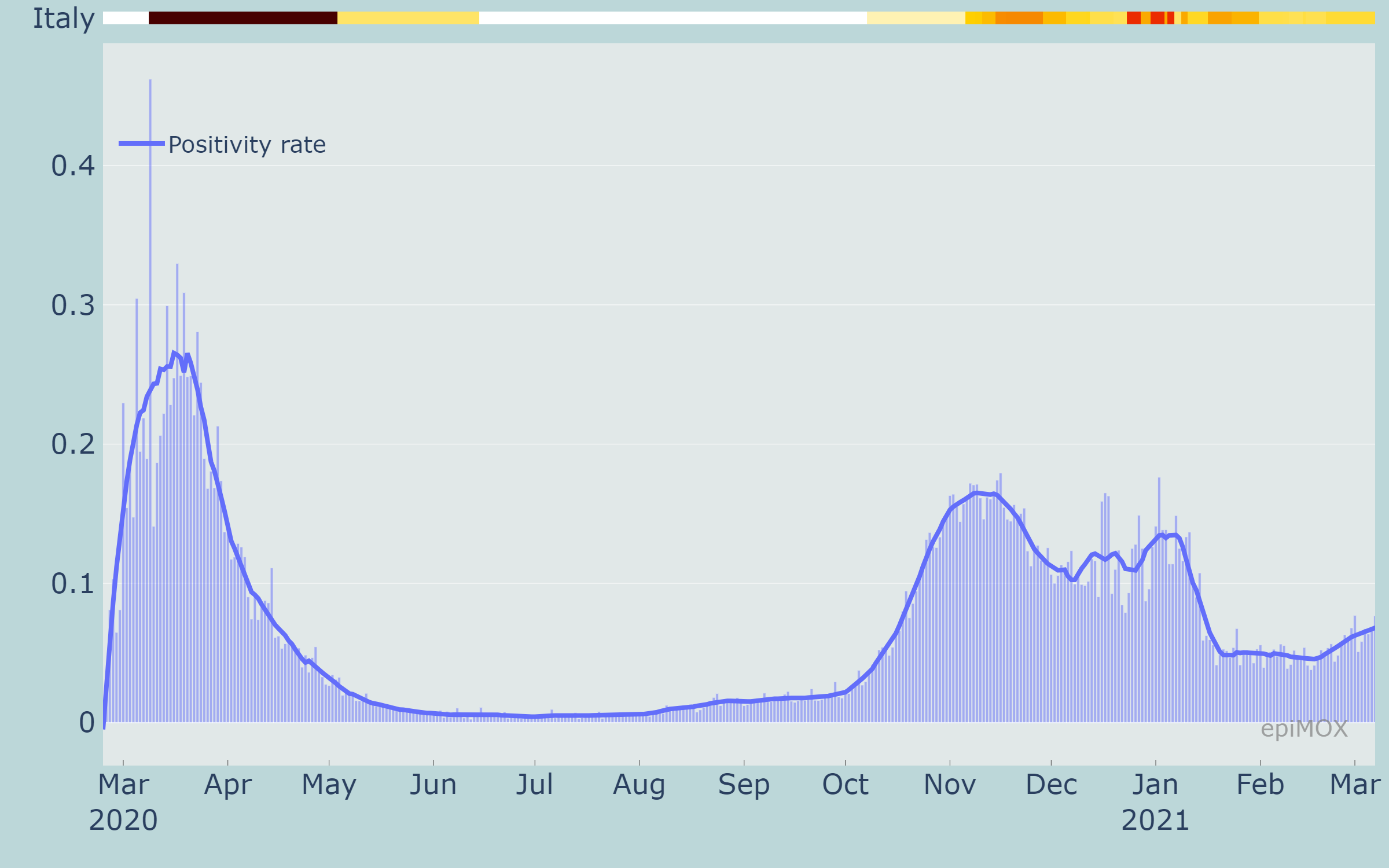}
    \caption{Positivity rate (PR) in Italy since the beginning of the epidemic}
    \label{fig:positivity}
\end{figure}

\begin{figure}
    \centering
    \includegraphics[width=\textwidth]{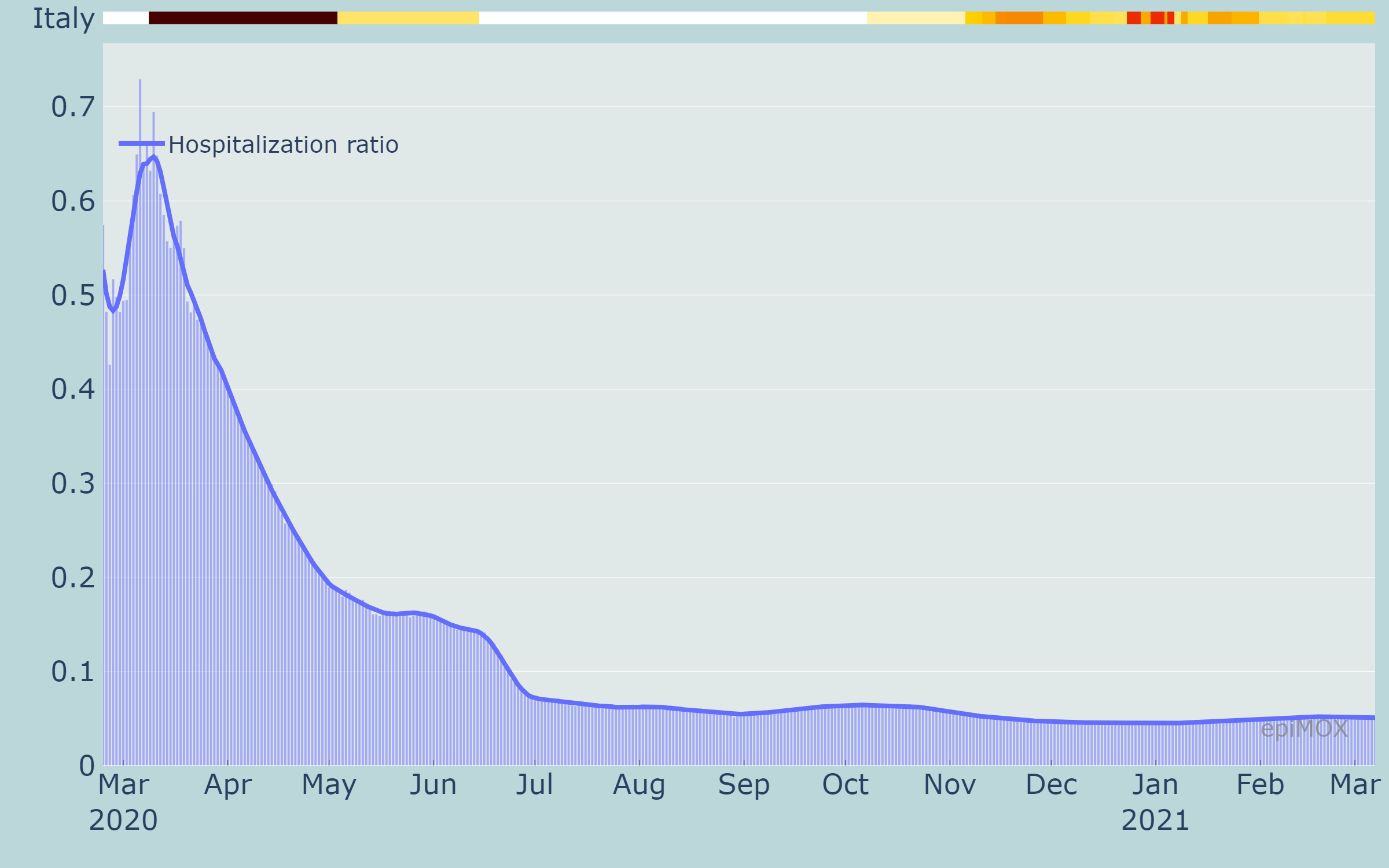}
    \caption{Hospitalization Case Ratio (HCR) in Italy since the beginning of the epidemic}
    \label{fig:hcr}
\end{figure}

\begin{figure}
    \centering
    \includegraphics[width=\textwidth]{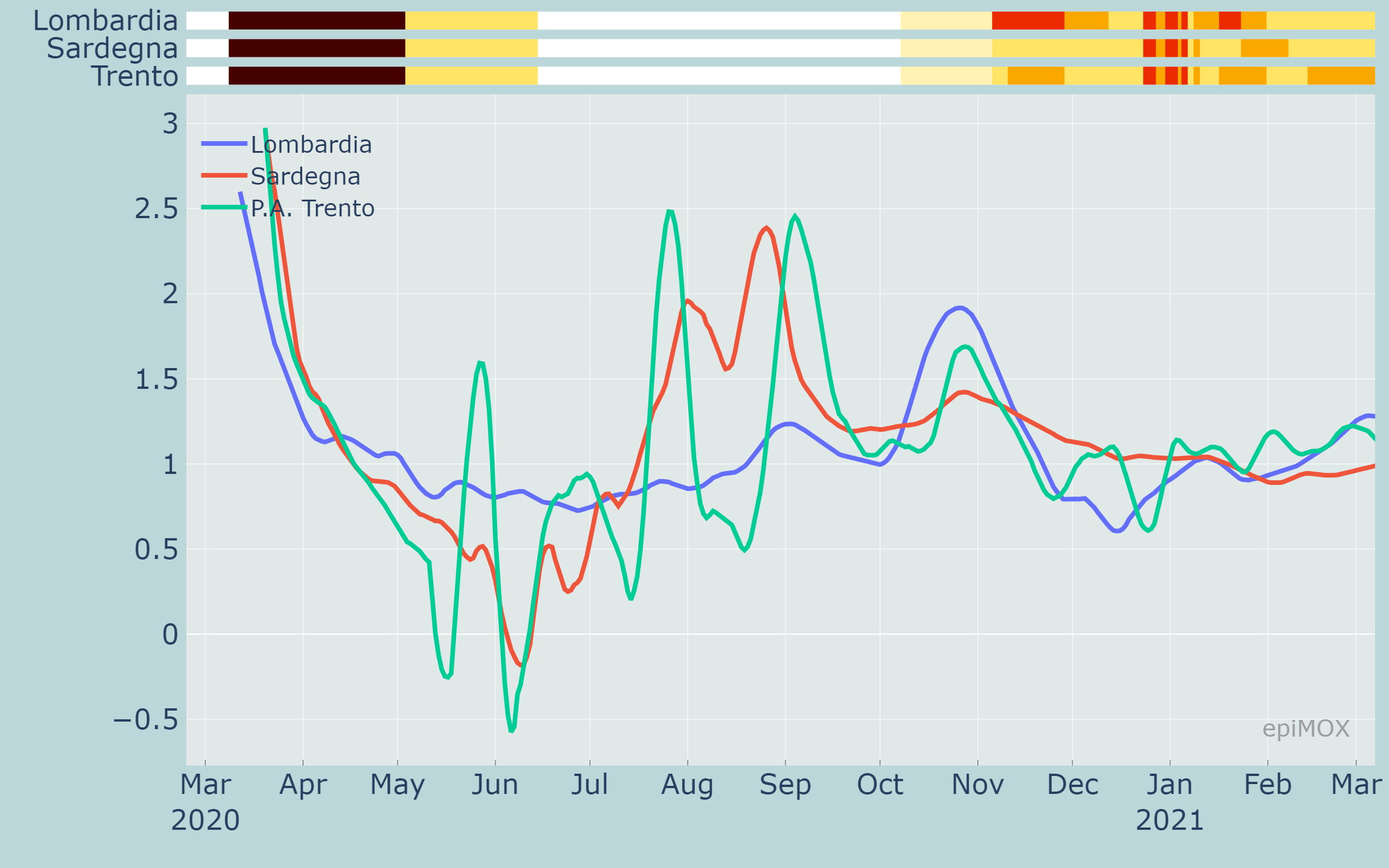}
    \caption{Evolution of R* in three different regions since the beginning of the epidemic}
    \label{fig:rt_reg}
\end{figure}

\AQ{
\section{Analysis of some critical indicators} \label{sec:indicators}

}
\AQ{The {\it epiMOX} dashboard \REV{can display} several critical \REV{indicators for monitoring} the progress of the pandemic, namely:
\begin{itemize}
    \item The daily evolution of R*, an estimate of \REV{the} reproduction number Rt proposed in \LD{\cite{Battiston}}.
\item The daily evolution of the \LD{fatality} rate, estimated through the CFR (Case Fatality Ratio) as the ratio between the number of cumulative deaths at a given date and the number of resolved cases at the same date

$$\text{CFR} = \frac{\text{Deceased}}{\text{Healed + Deceased}}.$$
\item The daily evolution of the positivity rate ($PR$) defined as $$\text{PR} = \frac{\text{Daily positives}}{\text{Daily swabs performed}}.$$
\item The daily evolution of the hospitalized case ratio \REV{(HCR)}, defined as 
$$\REV{\text{HCR} = \frac{\text{Total hospitalized}}{\text{Positives}}=\frac{\text{Hospitalized + Hosted in ICUs}}{\text{Positives}}.}$$
\item The daily evolution of the 14-day notification rate, that is the number of new positive cases emerged in the previous 14 days per 100,000 population, an indicator used by the European Union to monitor the evolution of the epidemic in the different European regions. 
\end{itemize}

\REV{From} Figures \ref{fig:cfr}, \ref{fig:positivity} and \ref{fig:hcr}, we can \REV{notice} that CFR, PR and HCR were far much higher in the first wave than in the second one because of a significant underestimation of the value of the denominators. The $PR$ indicator dramatically dropped around mid January 2021 because of the \LD{different counting strategy for cases} that was adopted \LD{by authorities}. Indeed, since then the denominator accounts for both molecular and genetic swabs. Figure \ref{fig:rt_reg} displays the value of the (estimated) reproduction number R* in three different Italian Regions. We can notice that a significant increase of R* is reported over the summer period in two touristic regions like the Sardegna island and the Trento province in the Dolomites due to a substantial increase of the population and a very likely relaxation of the restrictions (\LD{limited} use of facial masks and \LD{lacking of the social distancing}).}

\section{Empowering \textit{epiMOX} with the SUIHTER epidemiological model} \label{sec:suihter}
\REV{Epidemic forecasting of COVID-19 is difficult because of the intrinsic variability and uncertainty of the pandemic. Incomplete, uncertain, or inaccurate data, in terms of both initial conditions and time series of the different compartments, represent a serious limitation. The partial knowledge of the behavior of the specific infecting agent and the dynamic evolution of environmental and social conditions is a further source of uncertainty.}

Since the seminal work in \cite{kermack}, where the first SIR compartmental model based on a system of nonlinear ordinary differential equations was presented, a large variety of models have been proposed -- see, for instance, \cite{bertuzzo,brauer,Gatto,sidarthe,Hethcote,martcheva} -- \REV{each one attempting} to cope with specific aspects of the problem. If in the original SIR model the epidemic evolution is described by the number of individuals belonging to the susceptible (S), infected (I) and removed (R) compartments, several models consider increasing the number of compartments to include, for instance: 
\begin{itemize}
    \item the exposed individuals (those who have already been exposed to the infecting agent but are not yet infectious);
    \item possible splitting of the infected compartment into different classes according to the actual level of severity;
    \item a distinction of the removed compartment in recovered and dead. 
\end{itemize}

\REV{These models, although typically derived under many simplifying assumptions, enable forecasting analyses that go beyond mere data extrapolation. Moreover, simulations regarding the future dynamics may allow the investigation of different scenarios corresponding to different NPIs.} 
One of the most critical aspects in the development of complex compartmental models is indeed related to their \textit{calibration} based on available data. On the one hand, data related to the different compartments may not be available (or they may not even be collected); on the other hand, the resulting data assimilation problem may suffer from limited identifiability of the parameters, as discussed, \REV{e.g., in the  paper \cite{PTT20} for} SIR-like models of COVID-19. 
A new compartmental model named SUIHTER has been recently \AQ{introduced in \cite{SUIHTER20}} with the goal of facing the first of these two issues, i.e. defining a model best suited for the data actually available. In the context of the COVID-19 epidemic, the time series that are daily collected and made available (which have already been described discussing the dashboard for the data analysis) \REV{led} us to consider for the new model the following compartments: 
\begin{itemize}
\item $S$: \textit{susceptible} (uninfected) individuals;
\item $U$: \textit{undetected} (both asymptomatic and symptomatic) infected individuals;
\item $I$: \textit{isolated} (quarantined) individuals;
\item $H$: \textit{hospitalized} individuals;
\item $T$: \textit{threatened} (acutely symptomatic infected, \REV{hosted in ICUs}) individuals;
\item $E$: \textit{extinct} individuals;
\item $R$: \textit{recovered} individuals,
\end{itemize}

In this paper, we use the following specific version of the SUIHTER model:
\begin{equation}\label{eq:suihter}
\begin{array}{l}
\dot S(t) = - S(t)\frac{\beta_U U(t) + \beta_I I(t)}{N}, \\[3mm]
\dot U(t) =   S(t)\frac{\beta_U U(t) + \beta_I I(t)}{N} - (\delta + \rho_U)U(t), \\[3mm]
\dot I(t) = \delta U(t) - (\rho_I + \omega_I +\gamma_I)I(t),\\[3mm]
\dot H(t) = \omega_I I (t) - (\rho_H + \omega_H) H(t), \\[3mm]
\dot T(t) = \omega_H H (t)  - (\rho_T + \gamma_T)T(t), \\[3mm]
\dot E(t) = \gamma_I I(t) +\gamma_T T(t), \\[3mm]
\dot R(t) =  \rho_U U(t) + \rho_I I(t) + \rho_H H(t) + \rho_T T(t), 
\end{array}
\end{equation}
where $N=S+U+I+H+T+E+R$ denotes the total population (assumed constant).
\REV{The model is characterized by the following $11$ parameters, which are chosen as time dependent piecewise constant functions:
\begin{itemize}
\item $\beta_U$, $\beta_I$ denote the transmission rates due to contacts between a susceptible subject and an undetected infected, a quarantined, or a hospitalized subject, respectively;
\item $\omega_I$ denotes the rate at which $I$-individuals develop clinically relevant symptoms, while $\omega_H$
denotes the rate at which $H$-individuals develop life-threatening symptoms;
\item $\delta$ denotes the probability rate of detection, relative to undetected infected individuals;
\item $\rho_U$, $\rho_I$, $\rho_H$ and $\rho_T$ denote the rate of recovery for four classes ($U$, $I$, $H$ and $T$, respectively) of infected subjects;
\item $\gamma_I$ and $\gamma_T$ denote the mortality rates for the individuals isolated at home, and hosted in ICUs, respectively.
\end{itemize}
}

\REV{Model calibration through data fitting is essential to reproduce the past history of the epidemic and to perform short-term forecasts by inferring the epidemiological characteristics of COVID-19. 
Here we use reported isolated, hospitalized, threatened and extinct cases data to estimate the parameters of the { SUIHTER} model. In particular, we perform the calibration in two steps. At first, we find a set of parameter values by using an (ordinary) least squares (LS) estimator. Then, we perform a Bayesian calibration using the delayed rejection adaptive Metropolis (DRAM) algorithm \cite{DRAM} implemented in the Python library \texttt{pymcmcstat} \cite{Miles2019}, starting from a prior distribution of the parameters centered about the LS estimate.}

\REV{System \eqref{eq:suihter} is solved in the time interval $[t_I,t_F]$ which is subdivided into several sub-intervals, named \textit{phases}. Typically, phases change when the NPIs are modified by the Italian Authorities. 
When $n_{ph}$ phases are considered,  the calibration leads to the optimization {of $n_p= 11 n_{ph}$ parameters in total. Namely,  for each phase of the epidemic, we have the $11$ parameters given by $[\beta_U, \beta_I, \omega_I, \omega_H, \delta, \rho_U, \rho_I, \rho_H, \rho_T, \gamma_I, \gamma_T]$. 

Unfortunately, so many parameters make the calibration process difficult. In what follows, we calibrate our model under the following simplifying assumptions:}
\begin{itemize}
    \item $\beta_I$ is set to be a fraction of $\beta_U$, namely $\beta_I = \alpha \beta_U$ with $\alpha$ constant in all the time frame; 
    \item $\delta$, $\rho_U$, $\rho_I$, $\rho_H$, $\rho_T$, $\gamma_I$, {are constant on $[t_I,t_F]$.}
\end{itemize}
With these restrictions, the total number of parameters to be calibrated is reduced to {$4n_{ph}+7$}.

More details on the initialization and numerical solution of system \eqref{eq:suihter} are reported in \cite{SUIHTER20}.
}

\subsection{\NP{Results'  assessment}}\label{sec:validation}

\begin{figure}
\begin{tabular}{cc}
Hospitalized & Hosted in ICUs \\
\includegraphics[width=0.49\textwidth]{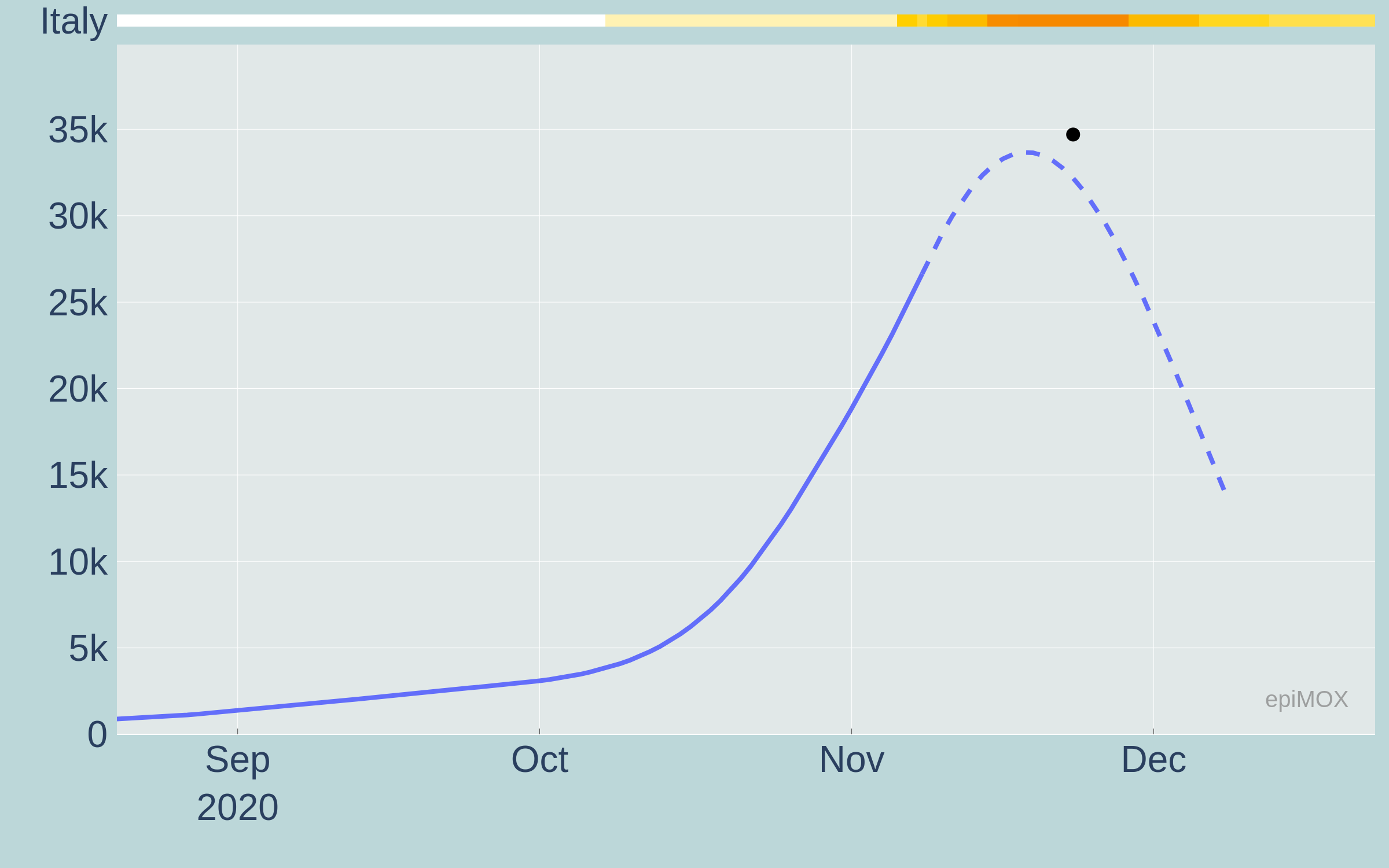} &
\includegraphics[width=0.49\textwidth]{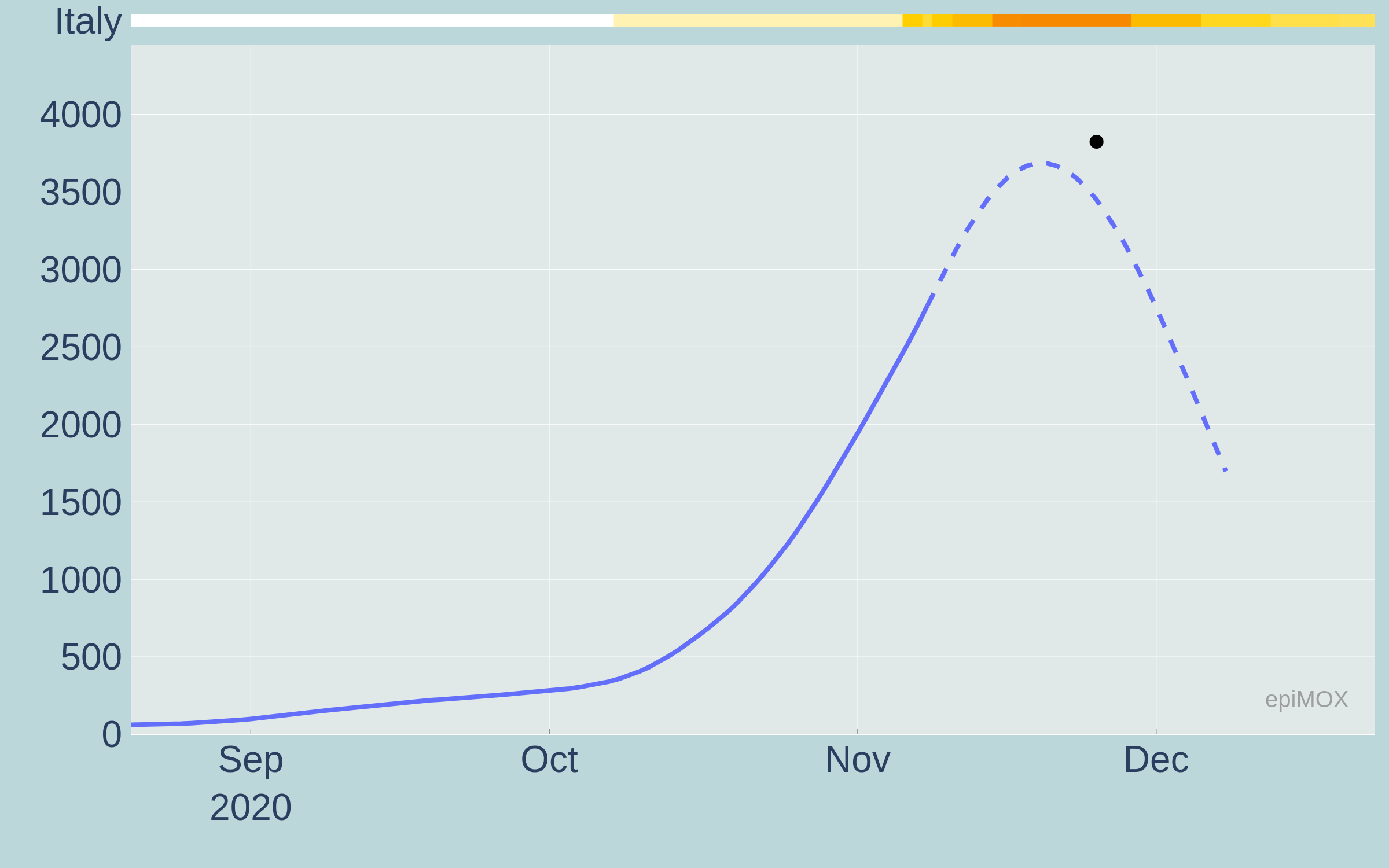} \\
\scriptsize Data until November 9, 2020 & \scriptsize Data until November 9, 2020 \\
\includegraphics[width=0.49\textwidth]{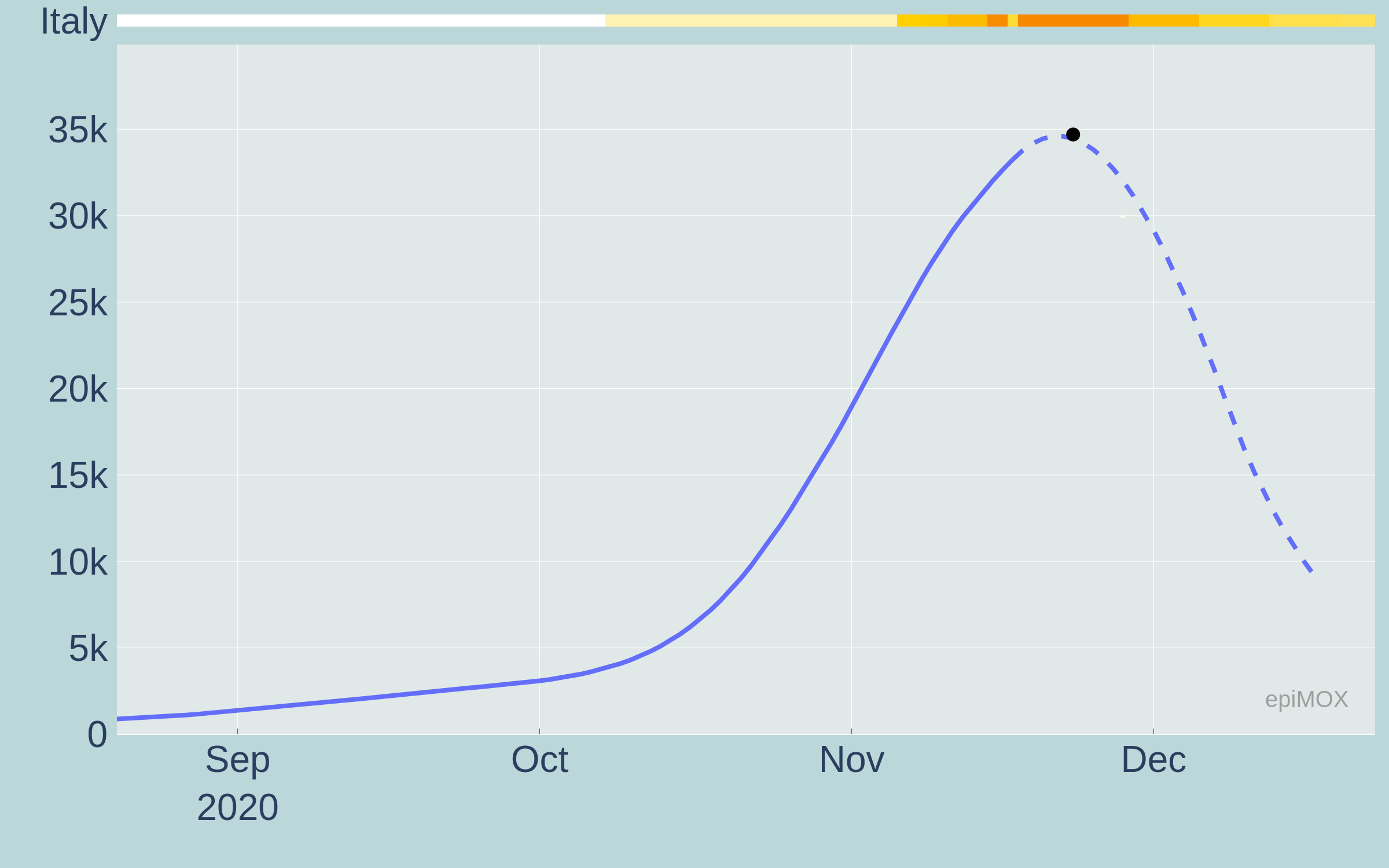} &
\includegraphics[width=0.49\textwidth]{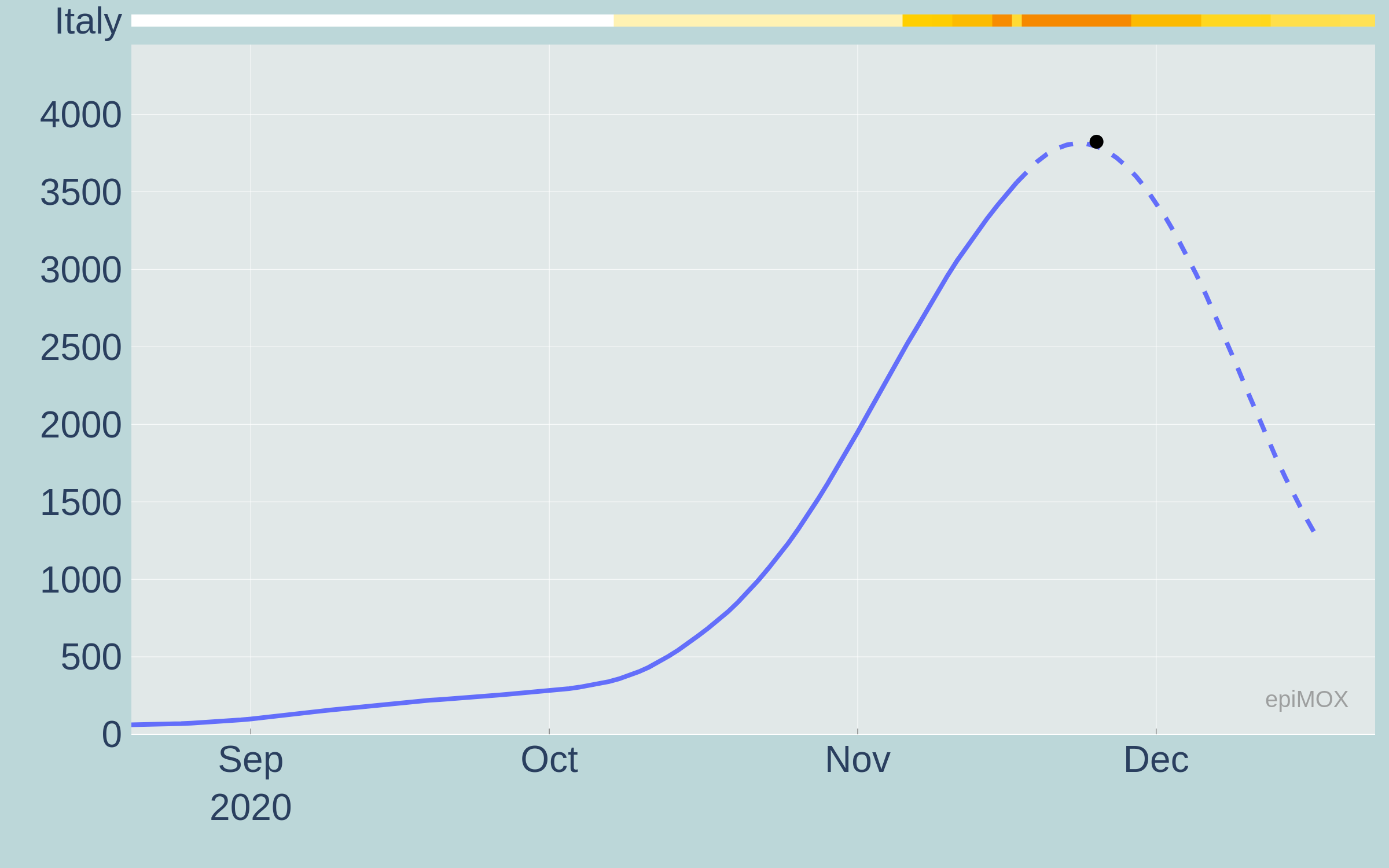} \\
\scriptsize Data until November 18, 2020 & \scriptsize Data until November 18, 2020 \\
\includegraphics[width=0.49\textwidth]{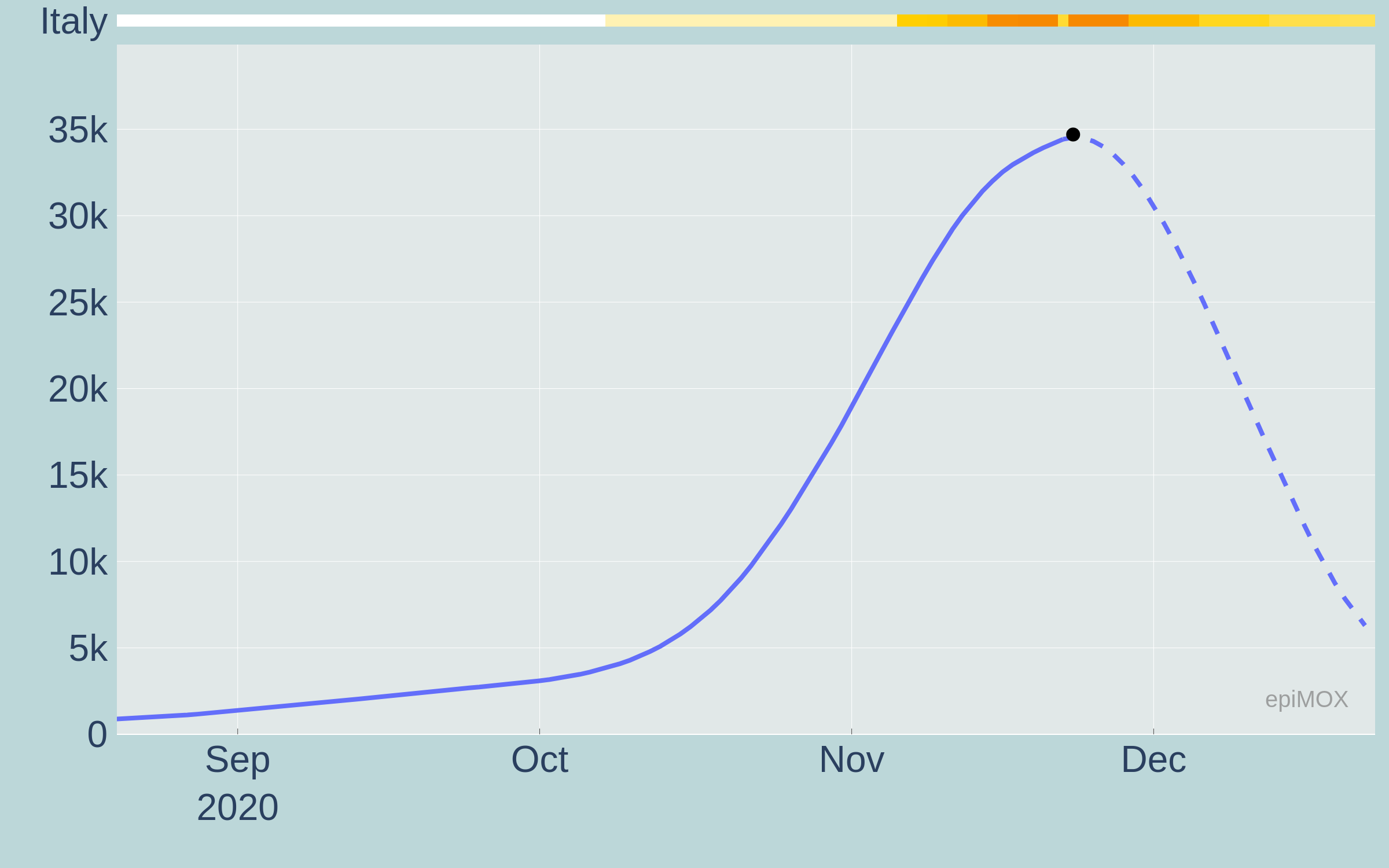} &
\includegraphics[width=0.49\textwidth]{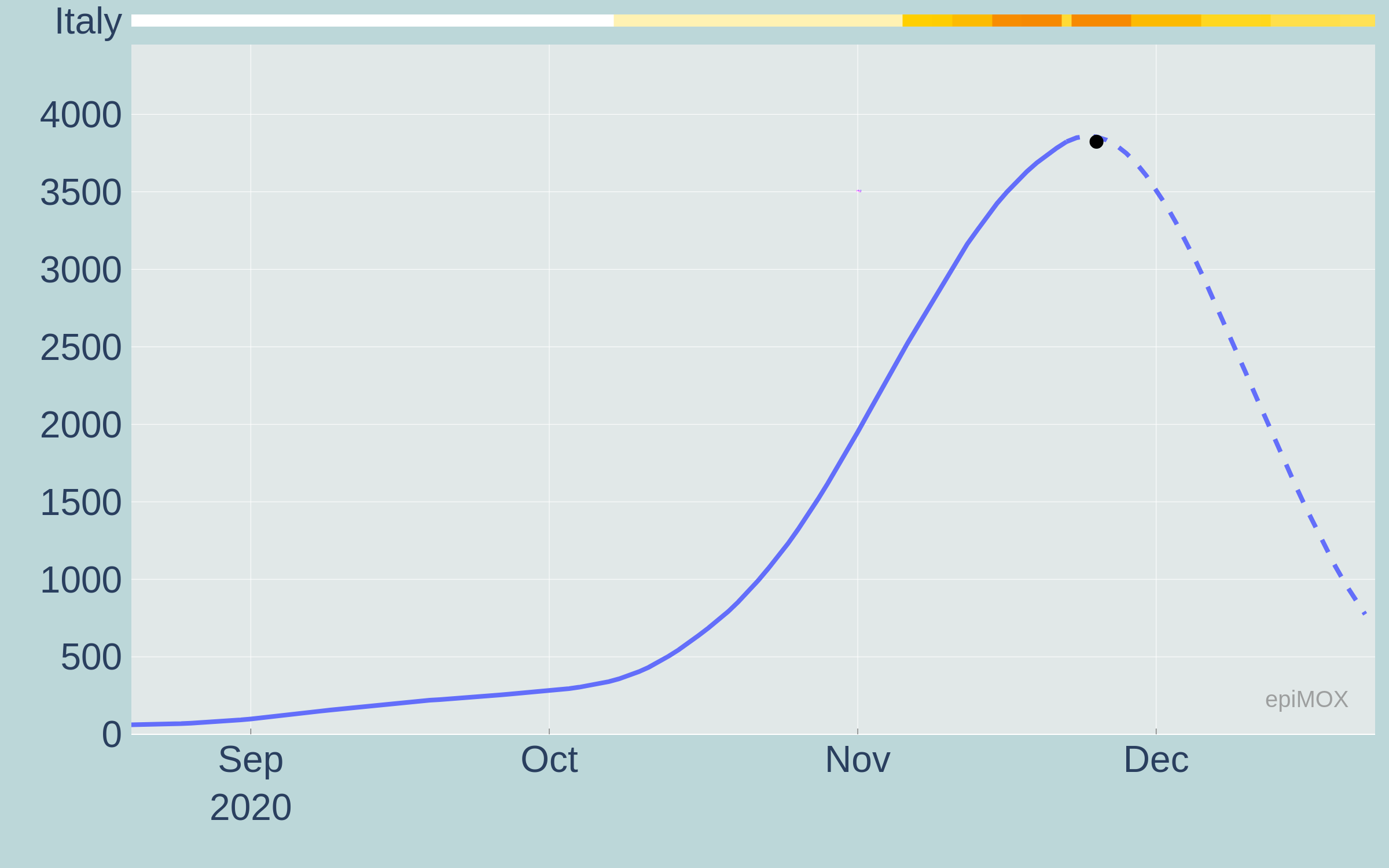} \\
\scriptsize Data until November 23, 2020 & \scriptsize Data until November 23, 2020 \\
\end{tabular}

\caption{Prediction of the peak of the \textit{Hospitalized} (left) and \textit{Hosted in ICUs} (right) compartments during the second epidemic wave estimated by the SUIHTER model at different dates prior the peak. \LD{The black dot indicates the peak actually occurred for the corresponding compartment (day and numbers)}.}
\label{fig:peaks}
\end{figure}

\AQ{Scope of this section is to provide an assessment of the forecasting capability of the SUIHTER model. More specifically, we will analyze suitability of the epidemiological model to predict the date and the height of the peaks of the curves representing the hospitalized and hosted in ICUs compartments, focusing on the second wave at the national scale. For each compartment, by proceeding retrospectively, in Figure \ref{fig:peaks} we display the predictions made by the SUIHTER model at different dates prior to the date at which the corresponding curve has attained its \LD{peak value}. By comparing the predictions with the real data (see Table \ref{tab:peaks}) we can draw the conclusion that the SUIHTER model can provide accurate predictions within a week (with errors lower than 1\% on the peak value and 1 day on the peak date), and reasonably  good predictions two weeks in advance.}



\begin{table}[]
    \centering
    \begin{tabular}{|c|c|c|c|c|}
    \hline
    &  \multicolumn{2}{c|}{Hospitalized}    
    &  \multicolumn{2}{c|}{Hosted in ICUs} \\ \hline  
    Day of forecast & 
    Peak value & Peak date &
    Peak value & Peak date \\
    \hline
    2020-11-09  & 
     33\,176 & 2020-11-19 &
      3\,618 & 2020-11-19 \\ 
    2020-11-18  &
     34\,639 & 2020-11-22 & 
      3\,989 & 2020-11-25 \\ 
    2020-11-23  & 
     34\,527 & 2020-11-23 & 
      3\,859 & 2020-11-24 \\
    \hline
    Actual peaks & 
     34\,697 & 2020-11-23 & 
      3\,823 & 2020-11-25 \\
    \hline
    \end{tabular}
    \caption{Forecast of peak values and days of \textit{Hospitalized} and \textit{Hosted in ICUs} compartments for the second epidemic wave in Italy}
    \label{tab:peaks}
\end{table}

\subsection{\NP{What-if scenarios}}\label{sec:whatif}
\begin{figure}
    \centering
    \begin{subfigure}[b]{0.49\textwidth}
        \centering
        \includegraphics[width=\textwidth]{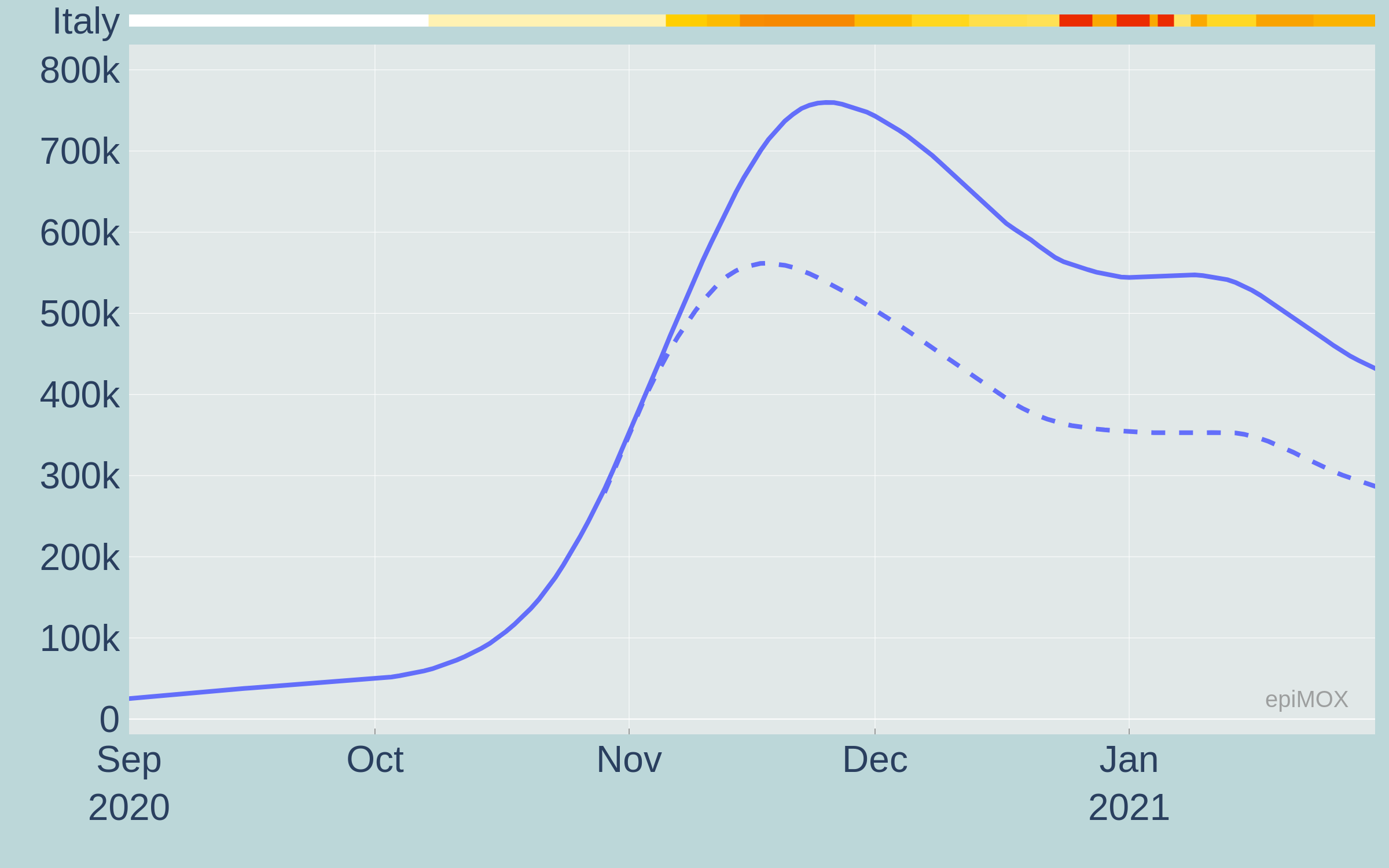}
        \caption{Isolated}
    \end{subfigure}
    \begin{subfigure}[b]{0.49\textwidth}
        \centering
        \includegraphics[width=\textwidth]{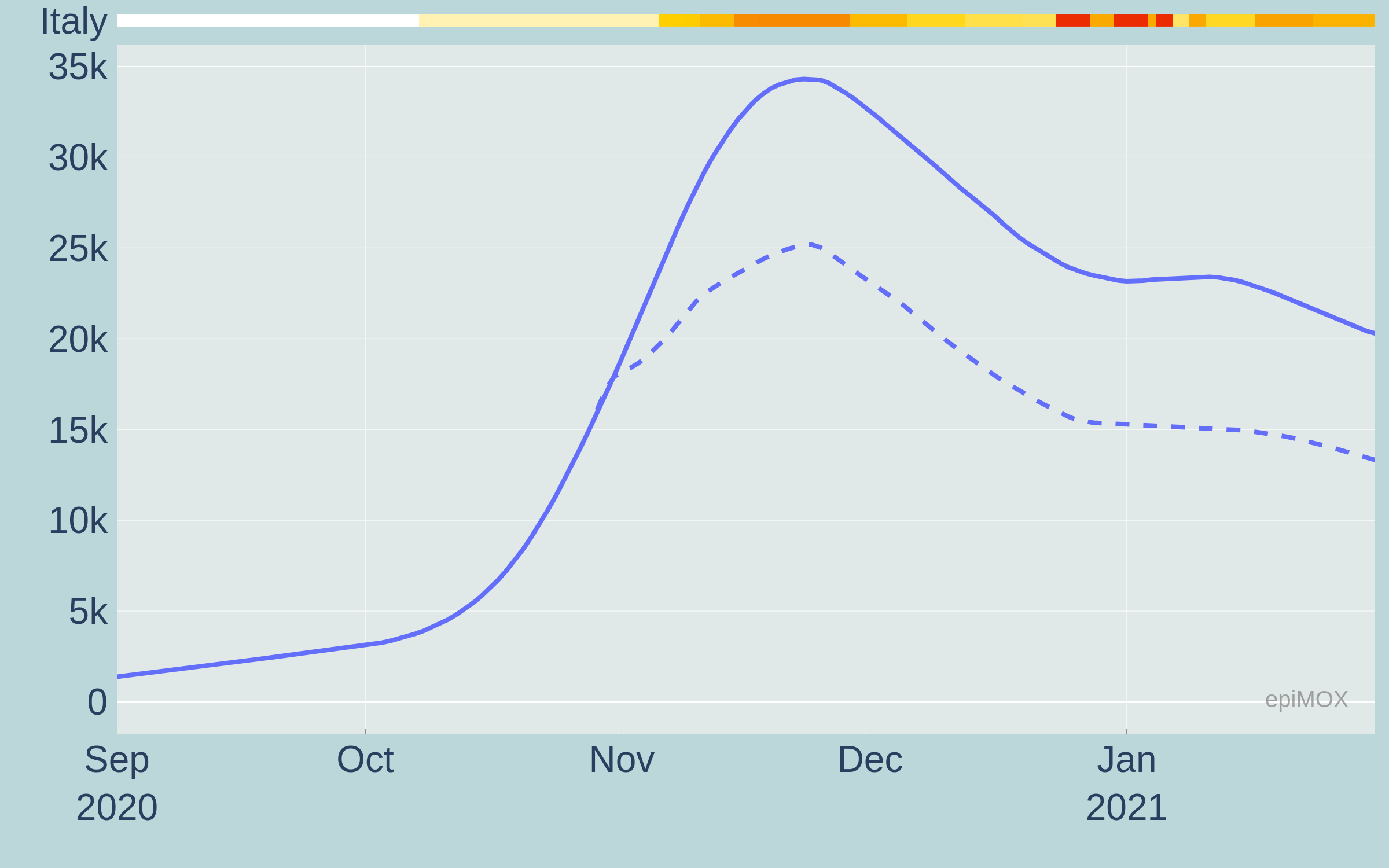}
        \caption{Hospitalized}
    \end{subfigure}
    \begin{subfigure}[b]{0.49\textwidth}
        \centering
        \includegraphics[width=\textwidth]{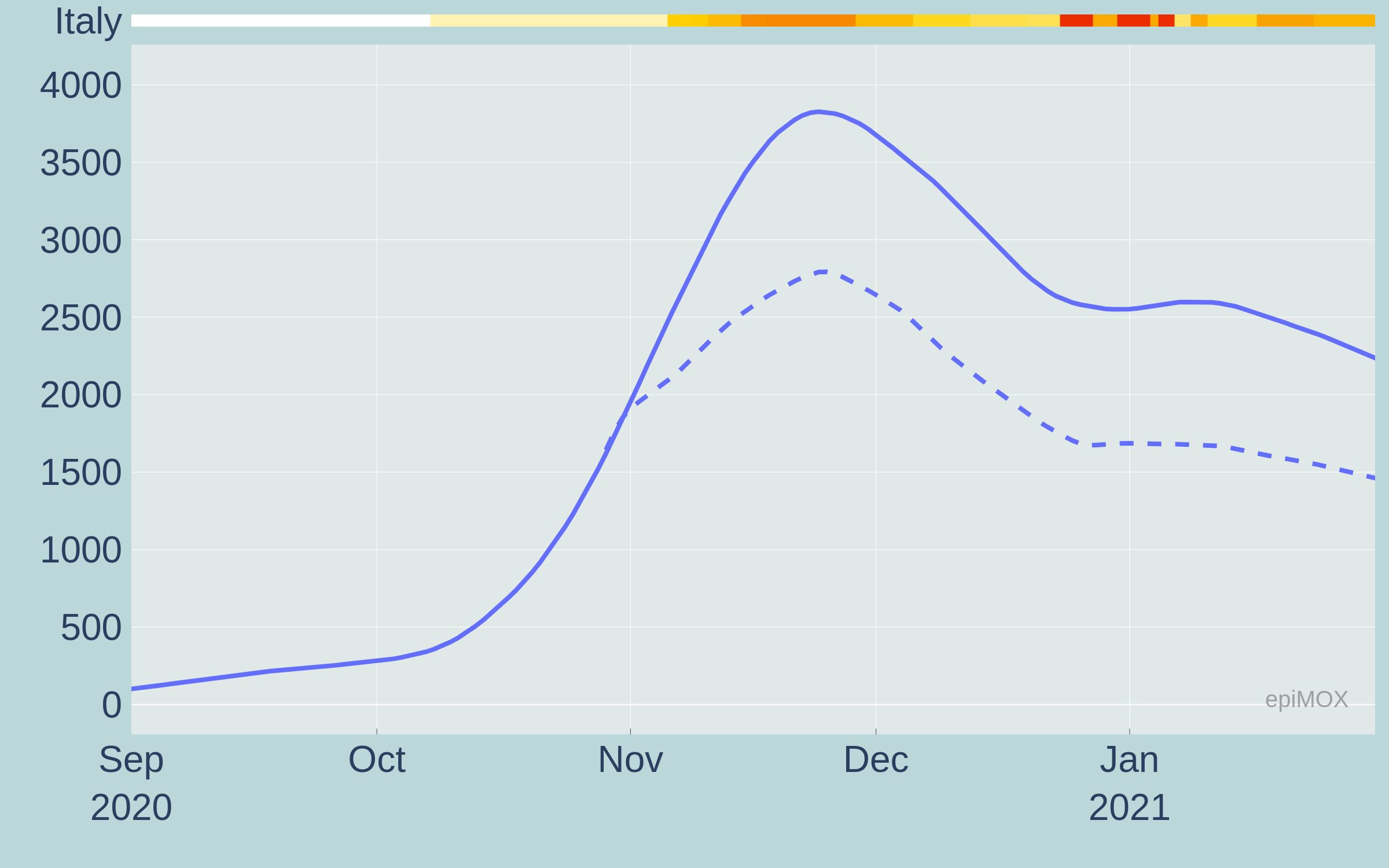}
        \caption{Hosted in ICUs}
    \end{subfigure}    
    \begin{subfigure}[b]{0.49\textwidth}
        \centering
        \includegraphics[width=\textwidth]{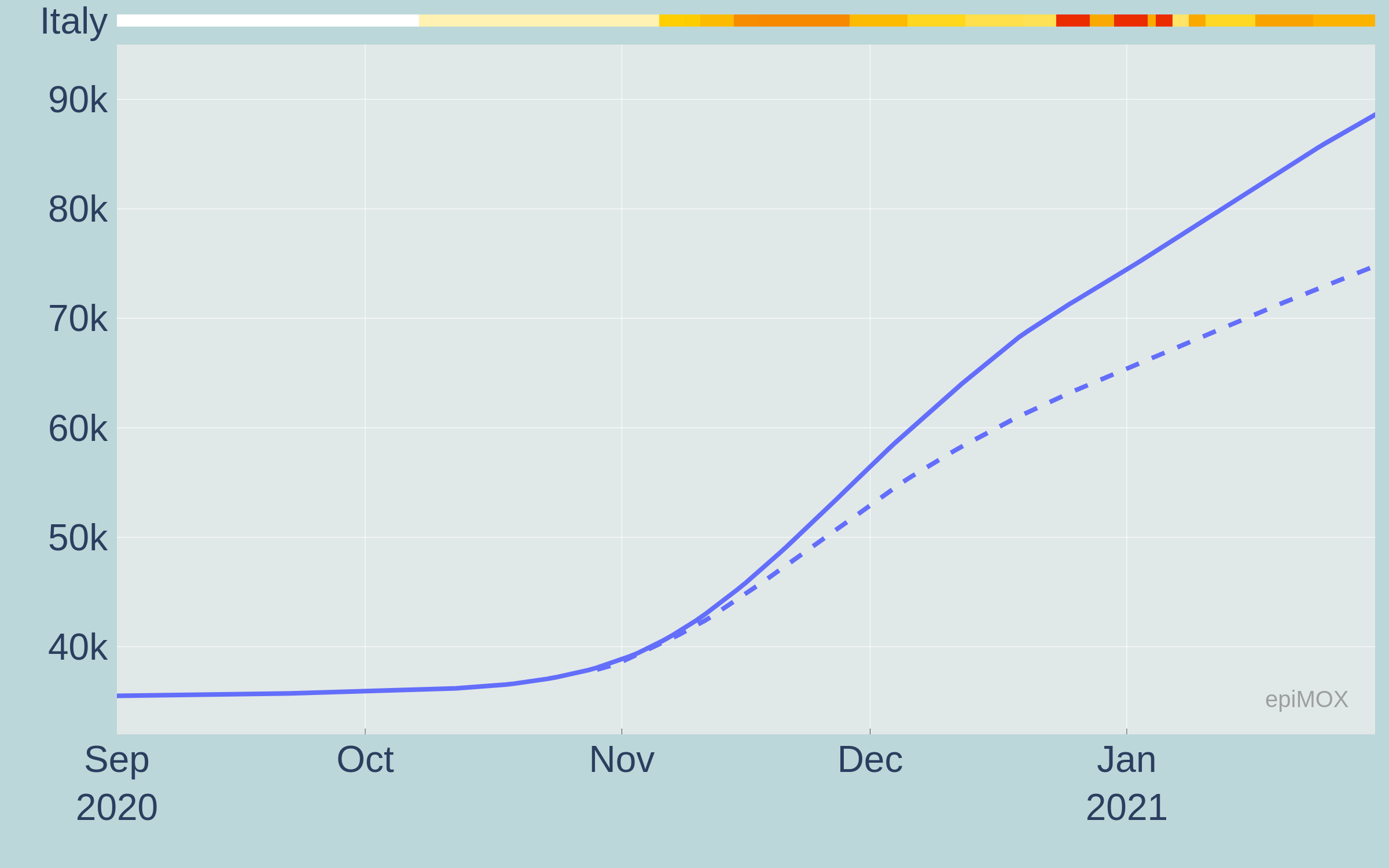}
        \caption{Cumulative Deaths}
    \end{subfigure}    
    \caption{Comparison between the observed time evolution (solid line) of four compartments and a simulated scenario (dashed line) in which the November restrictive measures had been anticipated by 10 days}
    \label{fig:whatif}
\end{figure}

In this section we aim at highlighting another feature of the dashboard, the possibility of carrying out \textit{what-if} scenarios. An example is illustrated in Figure \ref{fig:whatif}. \REV{Here we simulate (by means of the SUIHTER model) what would have happened in the case the Italian Government had anticipated by ten days the restriction measures introduced on November 4, 2020}\footnote{DPCM November 4, 2020, \url{https://www.gazzettaufficiale.it/eli/gu/2020/11/04/275/so/41/sg/pdf}}. The difference is \REV{indeed} quite striking: the peak value for the \textit{Isolated} during the second epidemic wave would have been 561\,714 instead of 759\,993, for the \textit{Hospitalized} 25\,169 instead of 34\,697, for the \textit{Hosted in ICUs} 2\,793 instead of 3\,823. \REV{More importantly, according to our estimates, a remarkable number of human lives would have likely been saved. Obviously, once again we point out that these conclusions should be taken with a grain of salt, because of the accuracy limitations of the epidemiological model. These impressive figures prove that the most effective strategy in the early phase of an outbreak is to enforce  very strict confinement measures as early as possible. This strategy, even if maintained for a short period, is indeed far more effective than adopting more permissive measures (or even the very same measures) for a much longer duration at a \textit{later} stage.}

\subsection{\NP{Impact of the UK variant}}\label{sec:variants}
\begin{figure}
    \centering
    \begin{subfigure}[b]{0.49\textwidth}
        \centering
        \includegraphics[width=\textwidth]{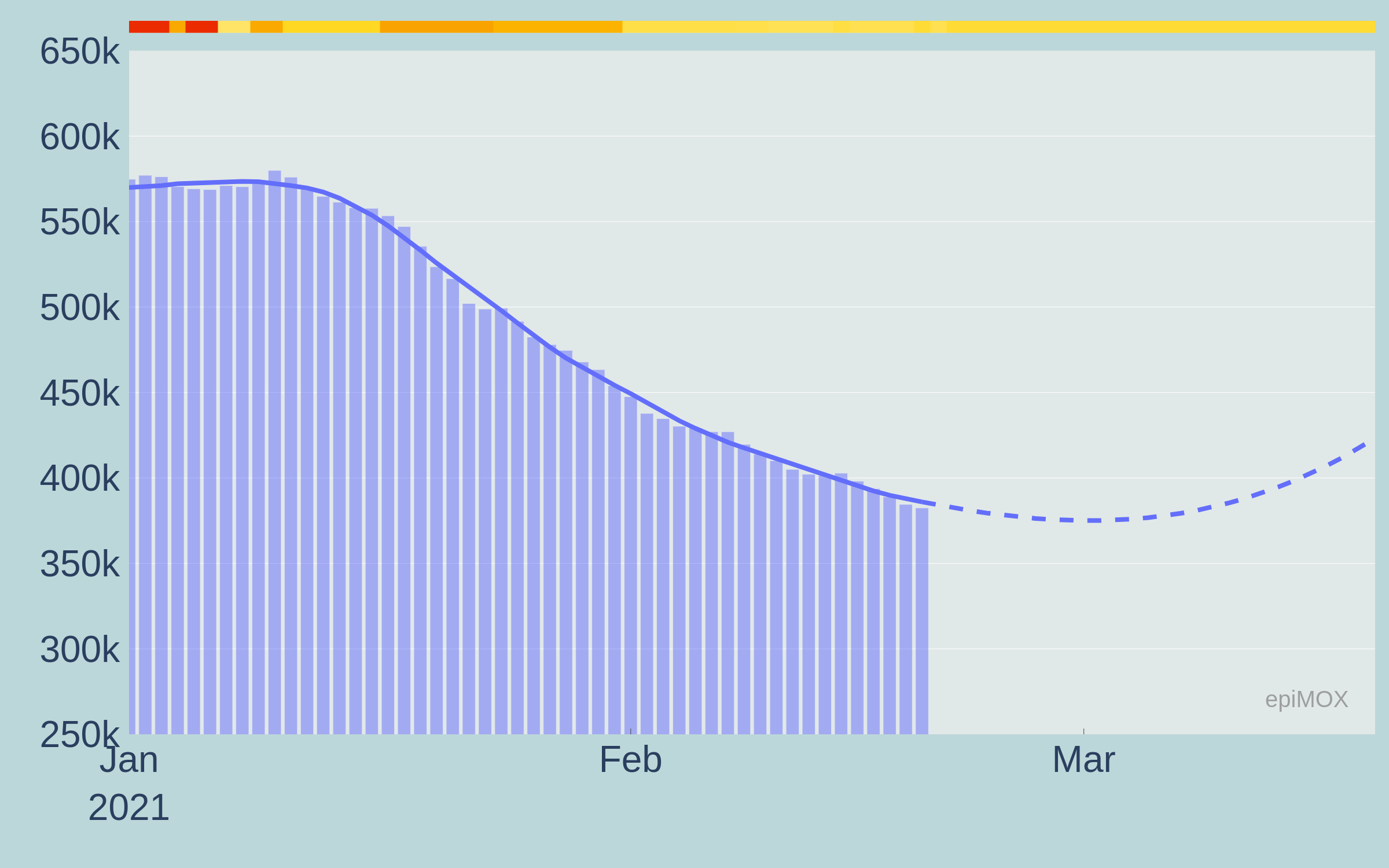}
        \caption{With UK variant}
    \end{subfigure}
    \begin{subfigure}[b]{0.49\textwidth}
        \centering
        \includegraphics[width=\textwidth]{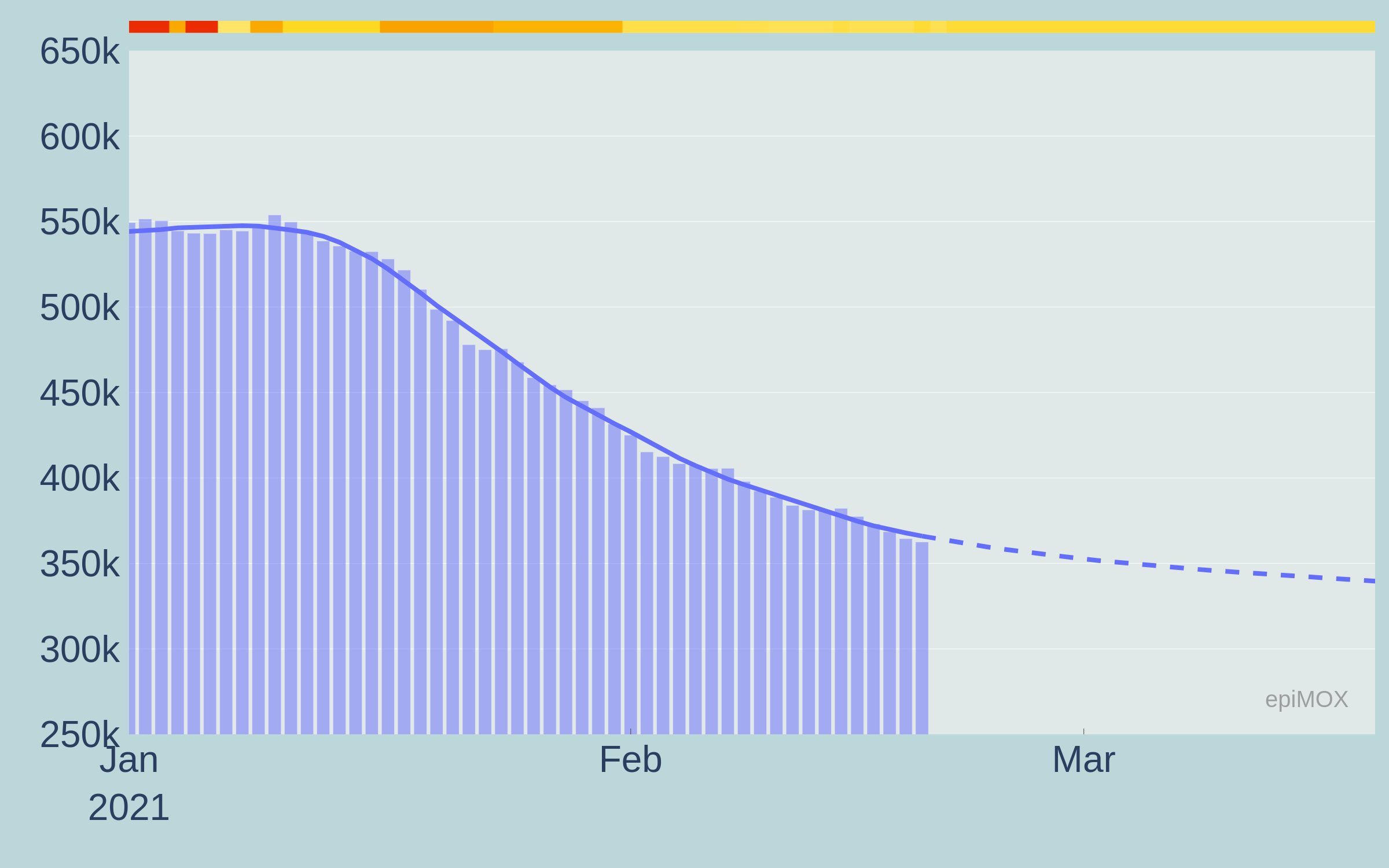}
        \caption{Without UK variant}
    \end{subfigure}
    \caption{30-day forecast of the \textit{Isolated} compartment with (left) and without (right) the effect of the UK virus variant}
    \label{fig:variant_isolated}
\end{figure}

\begin{figure}
    \centering
    \begin{subfigure}[b]{0.49\textwidth}
        \centering
        \includegraphics[width=\textwidth]{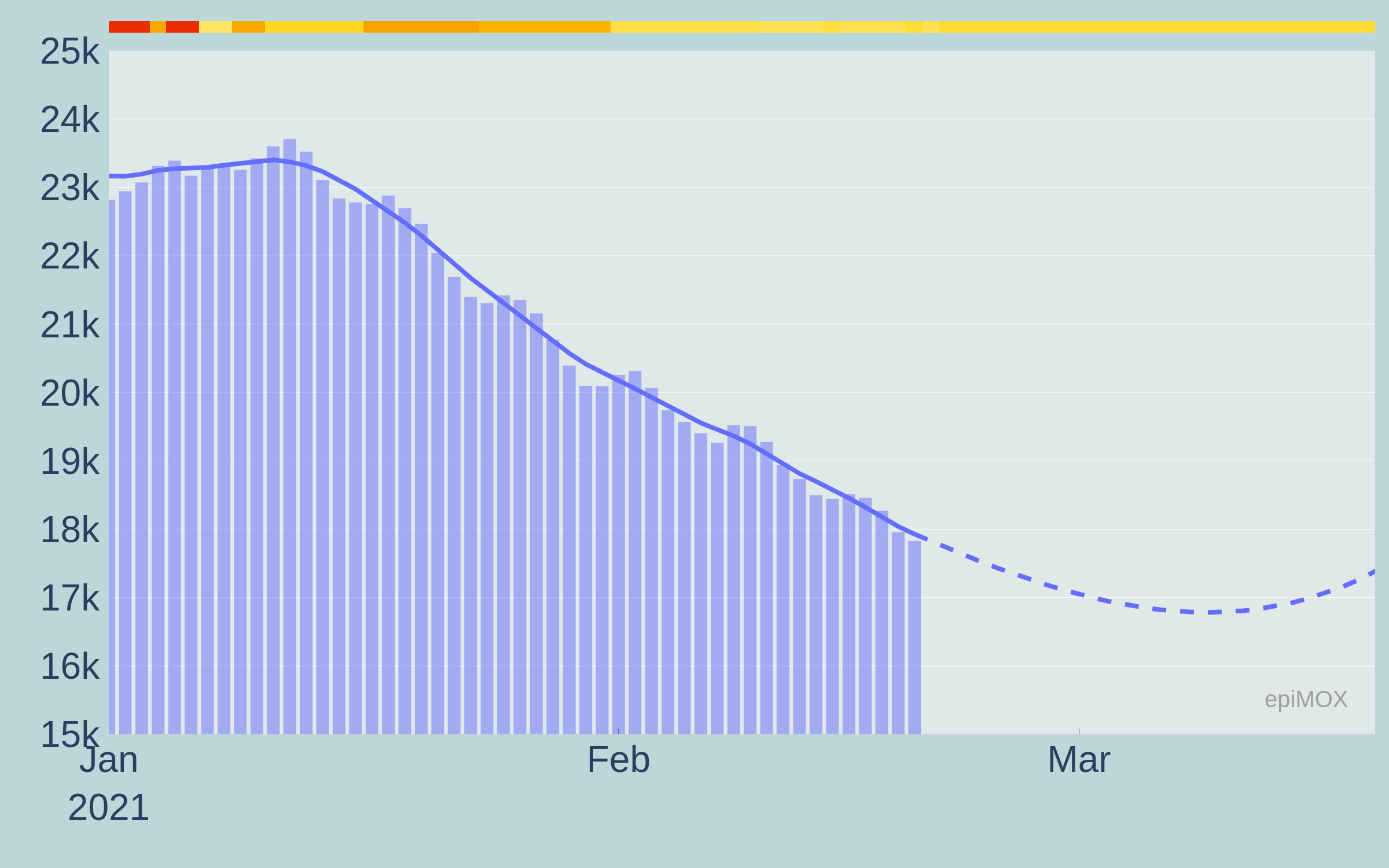}
        \caption{With UK variant}
    \end{subfigure}
    \begin{subfigure}[b]{0.49\textwidth}
        \centering
        \includegraphics[width=\textwidth]{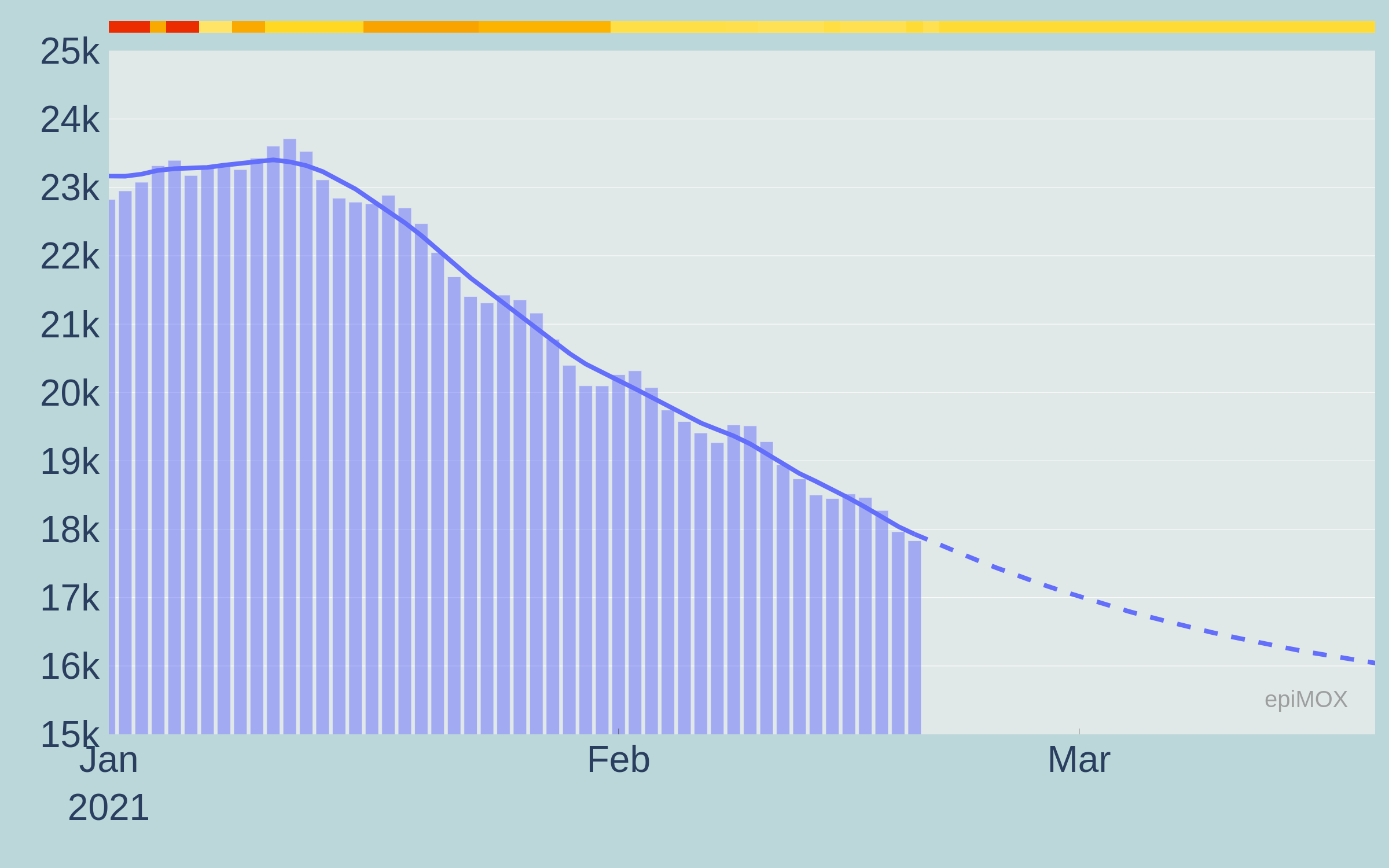}
        \caption{Without UK variant}
    \end{subfigure}
    \caption{30-day forecast of the \textit{Hospitalized} compartment with (left) and without (right) the effect of the UK virus variant}
    \label{fig:variant_hospitalized}
\end{figure}

\begin{figure}
    \centering
    \begin{subfigure}[b]{0.49\textwidth}
        \centering
        \includegraphics[width=\textwidth]{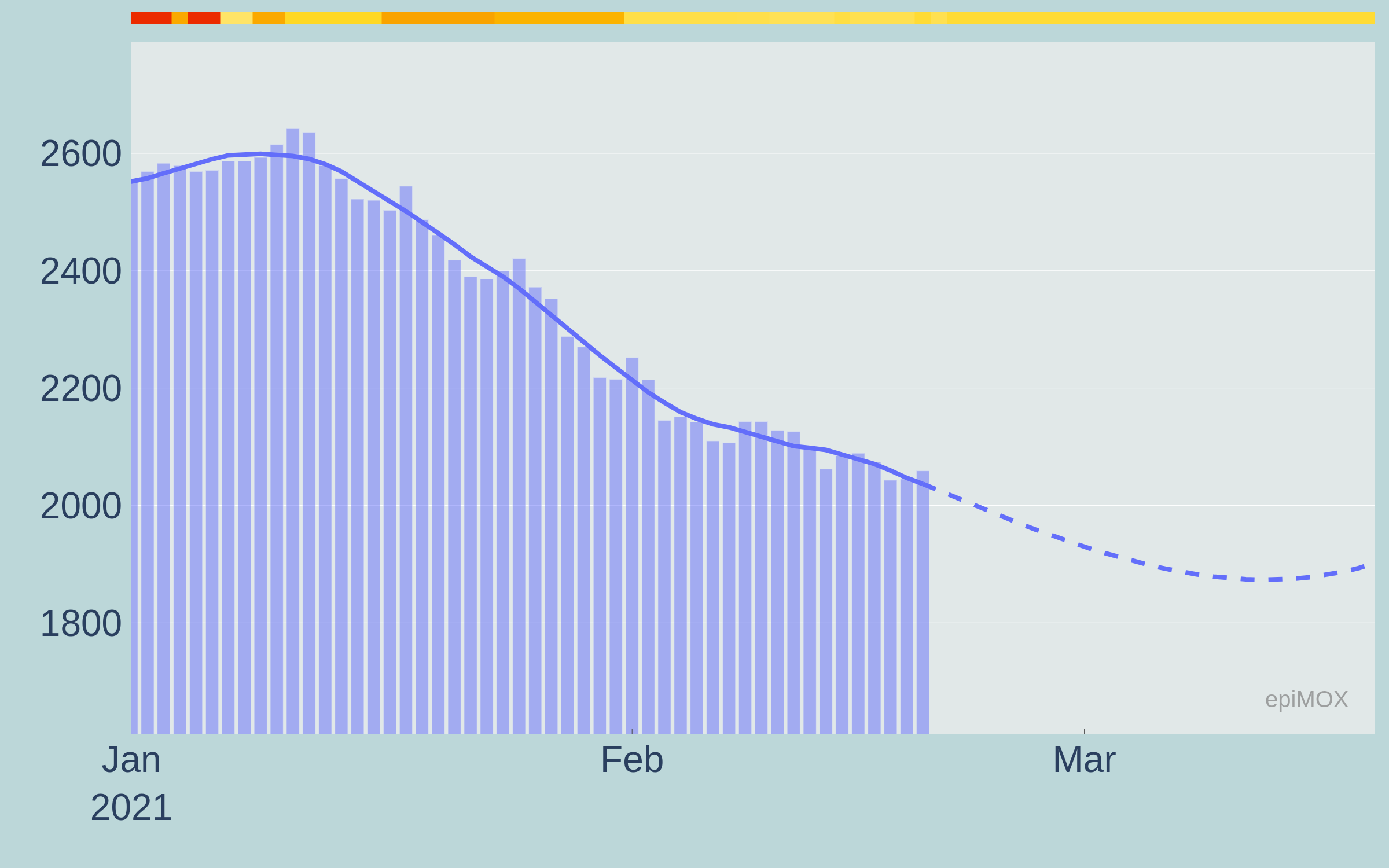}
        \caption{With UK variant}
    \end{subfigure}
    \begin{subfigure}[b]{0.49\textwidth}
        \centering
        \includegraphics[width=\textwidth]{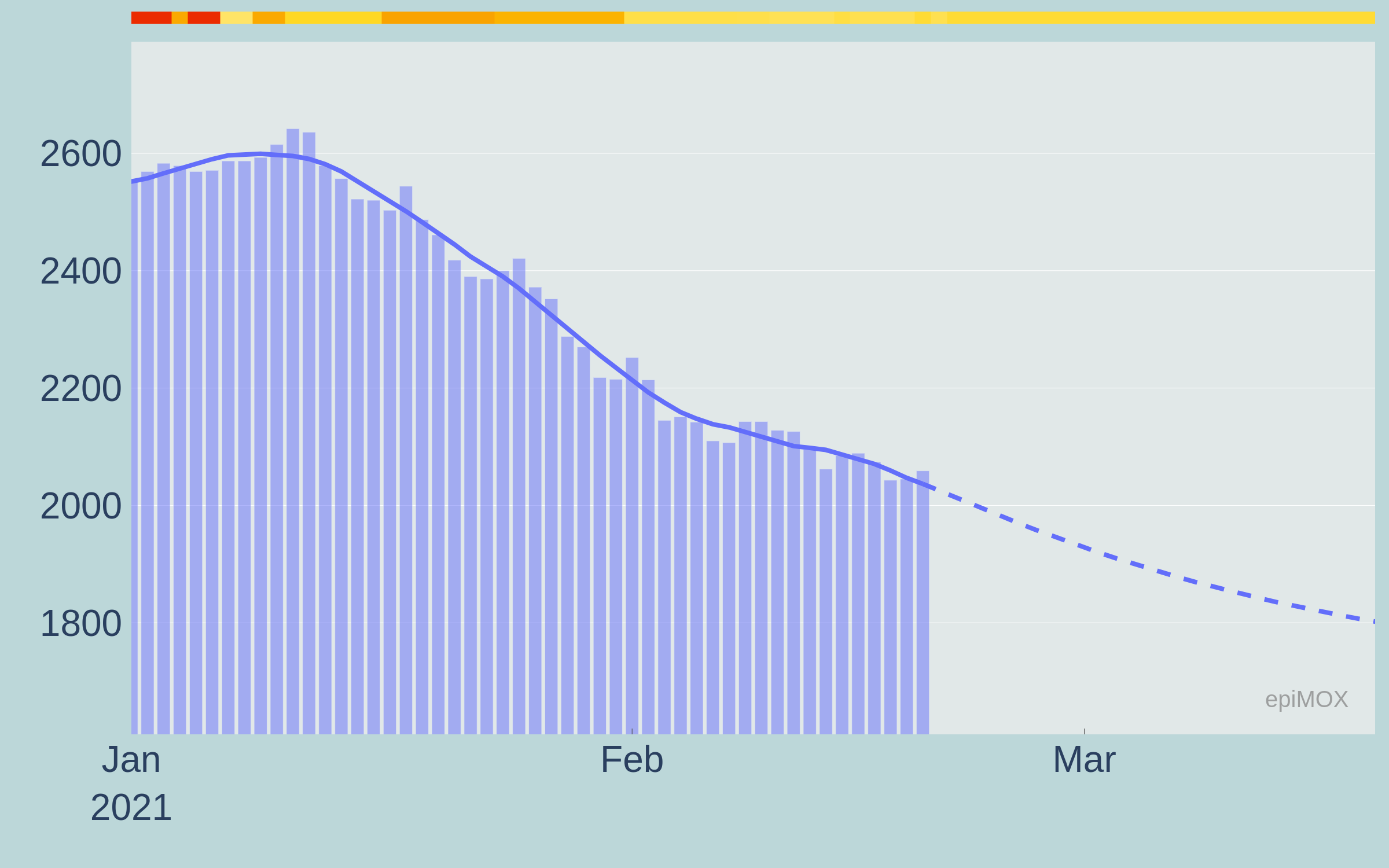}
        \caption{Without UK variant}
    \end{subfigure}
    \caption{30-day forecast of the \textit{Hosted in ICUs} compartment with (left) and without (right) the effect of the UK virus variant}
    \label{fig:variant_icus}
\end{figure}

\AQ{The appearance of different virus variants (UK, Brazilian, South African) at the end of 2020 has introduced an additional level of complexity on the managing of the epidemic as well as on the effort of the scientific community in modeling its dynamics (see e.g. \cite{giordano2021vaccination}).
\REV{A recent study in \cite{D1SC01203G} indicates that ``SARS-CoV-2 is at an advanced stage of evolution for human infection''. Other than questioning the vulnerability of available vaccines and current therapeutics to virus variants, the mutations can lead to different characterizations of the epidemic, in particular of the transmission rates  (see \cite{UKvariantSeverity}). In this respect, in the present paper, we limit our study to the effect of virus mutations on the transmission rates that can be readily incorporated in existing compartmental epidemiological models.}

\REV{In particular, in this paper, we adapt the SUIHTER model to this new challenge} by splitting each of the compartments that spread the infection (in particular the \textit{Undetected} and the \textit{Isolated} compartments) into two sub-compartments, one carrying the baseline virus variant and the other the UK virus variant (VOC
202012/01, lineage B.1.1.7). Recent studies (see e.g. \cite{UKvariantISS}) indicate that the UK variant reached a prevalence of 17.8\% in Italy at the beginning of February 2021 and. Due to its higher transmission rate (recently estimated in Italy as 37\% higher than the one of the baseline virus \cite{UKvariantSeverityItaly}), it is going to become the prevalent virus variant \LD{in March 2021}.

The updated SUIHTER model accounting for the UK virus variant is governed by the following system of \LD{ODEs} (see Figure~\ref{fig:suihter_var})
\begin{equation}\label{eq:suihterUKvariant}
\begin{array}{l}
\dot S(t) = - S(t)\, \frac{\beta_U^b \, U^b(t) +\beta_U^v \, U^v(t) + \beta_I^b \, I^b(t)+\beta_I^v \, I^v(t)}{N}, \\[3mm]
\dot U^b(t) =   S(t)\, \frac{\beta_U^b \, U^b(t) + \beta_I^b \, I^b(t)}{N} - (\delta + \rho_U)\, U^b(t), \\[3mm]
\dot U^v(t) =   S(t)\, \frac{\beta_U^v \, U^v(t) + \beta_I^v \, I^v(t)}{N} - (\delta + \rho_U)\, U^v(t), \\[3mm]
\dot I^b(t) = \delta \, U^b(t) - (\rho_I + \omega_I +\gamma_I)\, I^b(t),\\[3mm]
\dot I^v(t) = \delta \, U^v(t) - (\rho_I + \omega_I +\gamma_I)\, I^v(t),\\[3mm]
\dot H(t) = \omega_I \, (I^b (t)+I^v(t)) - (\rho_H + \omega_H) \, H(t), \\[3mm]
\dot T(t) = \omega_H \, H (t)  - (\rho_T + \gamma_T)\, T(t), \\[3mm]
\dot E(t) = \gamma_I \, (I^b (t)+I^v(t)) +\gamma_T \, T(t), \\[3mm]
\dot R(t) =  \rho_U \, (U^b (t)+U^v(t)) + \rho_I \, (I^b (t)+I^v(t)) + \rho_H \, H(t) + \rho_T \, T(t), 
\end{array}
\end{equation}
\LD{endowed with suitable initial conditions},
where the indices $b$ and $v$ respectively denote the base and UK variant splitting of the \textit{Undetected} and \textit{Isolated} compartments and a 37\% higher transmission rate for the UK variant is considered, namely
$$\beta_U^v=1.37 \beta_U^b, \quad \quad \quad \beta_I^v=1.37 \beta_I^b.$$

\begin{figure}
    \centering
    \includegraphics[width=\textwidth]{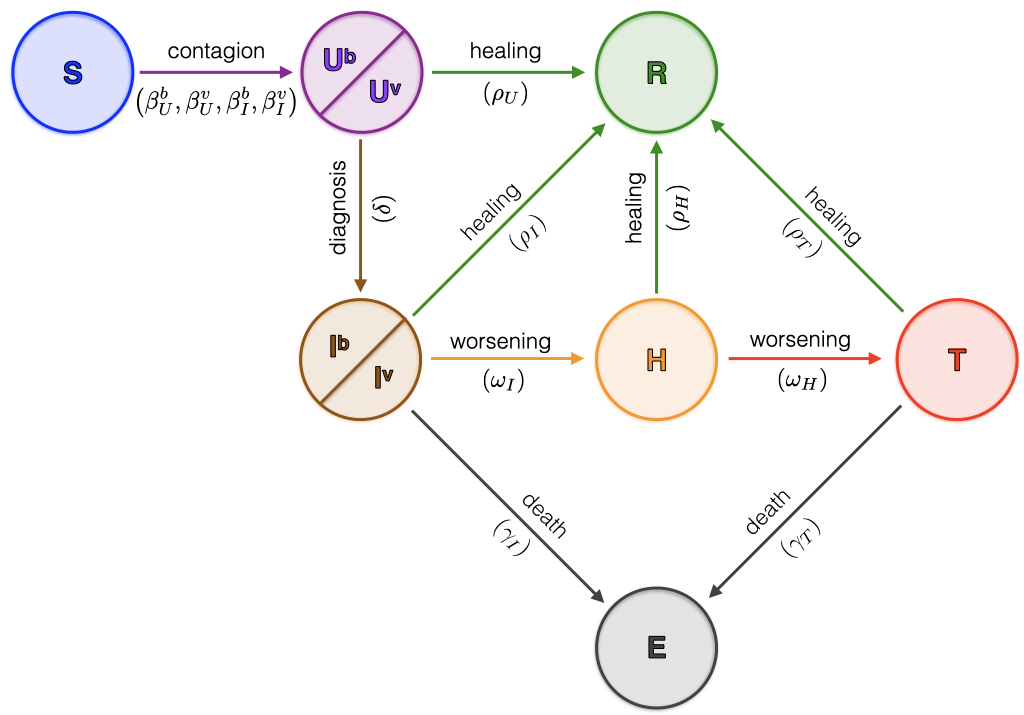}
    \caption{Sketch of the compartmental epidemiological model SUIHTER used to account for the UK virus variant}
    \label{fig:suihter_var}
\end{figure}

A comparison between the 30-day forecast obtained using the original SUIHTER model and the updated model accounting for the effect of the UK virus variant starting on February 20, 2021 is presented in Figures \ref{fig:variant_isolated}, \ref{fig:variant_hospitalized} and \ref{fig:variant_icus} for the \textit{Isolated}, \textit{Hospitalized} and \textit{Hosted in ICUs} compartments, respectively.

The most remarkable differences are noticed on the \textit{Isolated}, the  \textit{Hospitalized}, and, at a lesser extent, the \textit{Hosted in ICUs} compartments. This analysis does not account for the vaccination campaign. Moreover, by the time we write this paper, the national spreading of the UK variant is still highly non-uniform, with a pronounced prevalence in a few local environments. The numerical simulation at the national scale provides a homogenization effect which is perhaps rather inaccurate.
We highlight that the model accounting for the variant was able to predict the outbreak of the third epidemic wave observed in Italy in late February 2021.
}


\newpage

\subsection{\NP{30-day forecast with different scenarios}}\label{sec:forecast}
Modeling the epidemic evolution by a mathematical epidemiological model such as SUIHTER allows not only to analyse backward \textit{what-if} scenario investigating the effectiveness of past political choices (as discussed in Section \ref{sec:whatif}), but also to forecast the evolution of the epidemic in the near future.

As previously mentioned, the outbreak of the third epidemic wave has been observed in Italy since the second half of February 2021. At the time of writing, this outbreak is in its exponential growth phase and its evolution will strongly depend on the \LD{NPIs} that will be imposed in the coming weeks.

We have considered three different scenarios with different level of severity in the \LD{NPIs}:
\begin{itemize}
\item the first scenario considers the restriction currently imposed (on March 7, 2021) in which some limitations are adopted at national level (use of mask, limitation of mobility among regions, limitation on recreational activities, \dots), while additional stricter limitations (confinement within municipality limits, distance learning for school and universities, \dots) are adopted with increasing level of severity in \textit{yellow}, \textit{orange} and \textit{red} regions) \footnote{A detailed summary of the evolution of the \LD{NPIs introduced} at the regional level has been collected    on this webpage \url{https://it.wikipedia.org/wiki/Gestione_della_pandemia_di_COVID-19_in_Italia}};
\item the second scenario considers that all the region become \textit{yellow} starting on March 8, 2021;
\item the second scenario considers that all the region become \textit{red} starting on March 8, 2021.
\end{itemize}

\begin{figure}
    \centering
    \includegraphics[width=\textwidth]{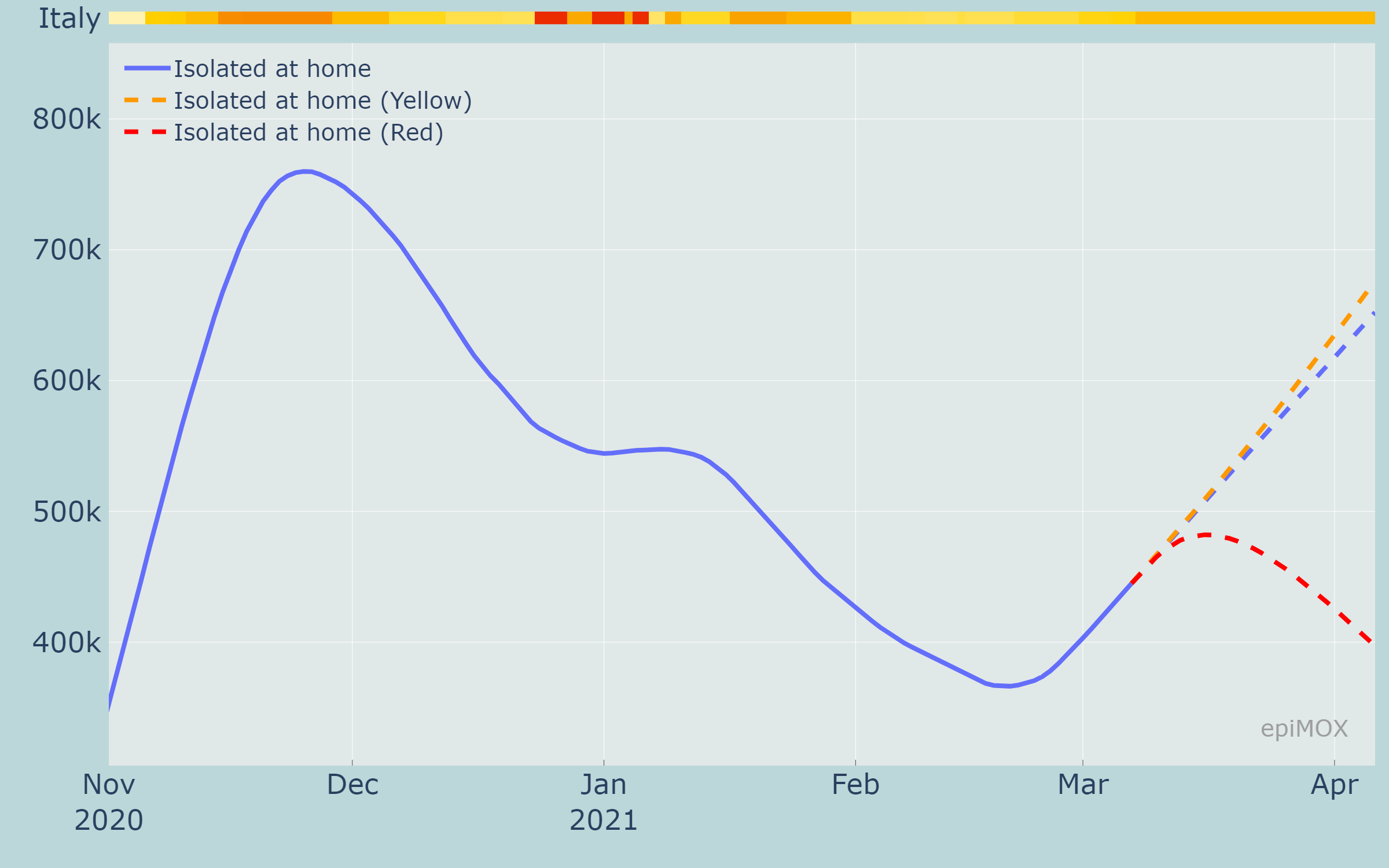}
    \caption{30-day forecast of the \textit{Isolated} compartment with current restrictions (at March 7, 2021) and with 3 alternative restriction scenarios}
    \label{fig:forecast_isolated}
\end{figure}

\begin{figure}
    \centering
    \includegraphics[width=\textwidth]{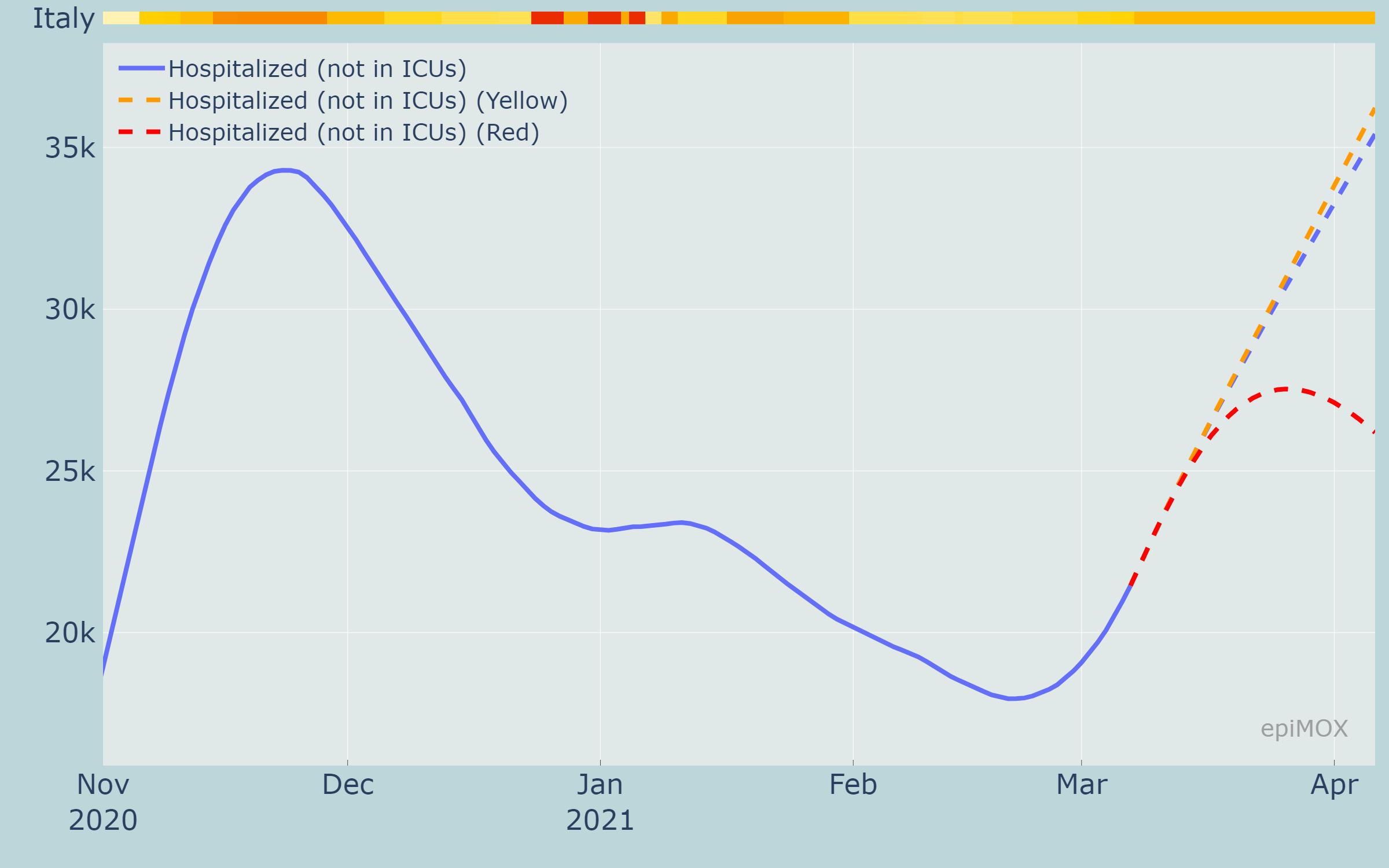}
    \caption{30-day forecast of the \textit{Hospitalized} compartment with current restrictions (at March 7, 2021) and with 3 alternative restriction scenarios}
    \label{fig:forecast_hospitalized}
\end{figure}

\begin{figure}
    \centering
    \includegraphics[width=\textwidth]{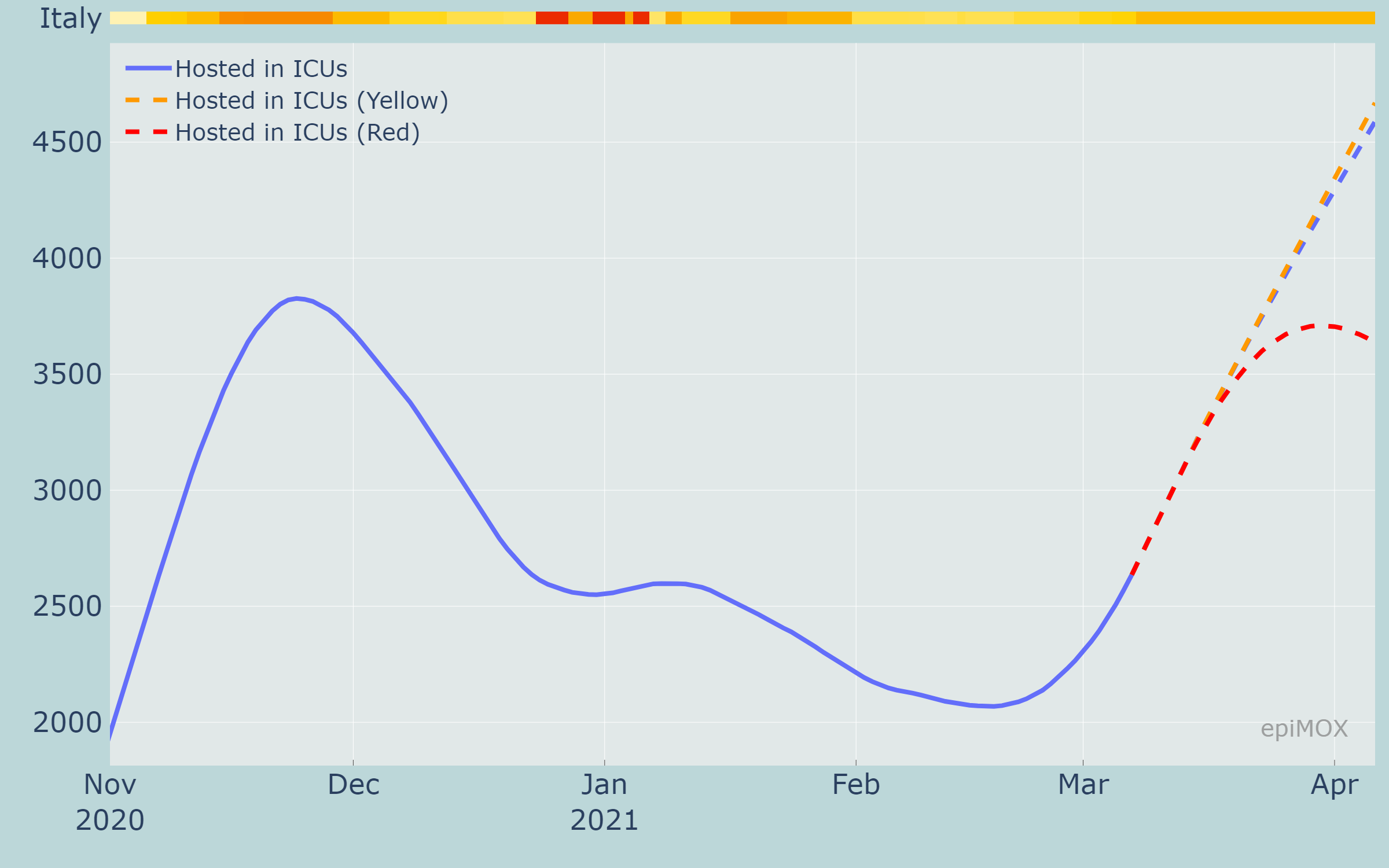}
    \caption{30-day forecast of the \textit{Hosted in ICUs} compartment with current restrictions (at March 7, 2021) and with 3 alternative restriction scenarios}
    \label{fig:forecast_icus}
\end{figure}

A comparison \LD{among} the time evolution of the \textit{Isolated}, \textit{Hospitalized} and \textit{Hosted in ICUs} compartments for the three scenarios are displayed in Figures
\ref {fig:forecast_isolated}, \ref {fig:forecast_hospitalized} and \ref {fig:forecast_icus}, respectively.
As expected, when more severe restrictions are imposed the growths of all the infecting compartments are slowed down. The results clearly highlight that the ongoing outbreak that has been activated by the increased transmission rate of the new virus variants could be controlled only should stricter restriction measures be imposed as soon as possible. Without a prompt intervention the risk of a third epidemic wave that will end up to be worse than the previous two is extremely high.

The model results displayed on the dashboard are obtained real-time running the model any time the dashboard is interrogated, thus allowing to switch between different forecast or what-if scenarios. 

As \REV{mentioned in Section \ref{sec:suihter} and} discussed in details in \cite{SUIHTER20}, the SUIHTER model can be calibrated using a Monte-Carlo Markov Chain (MCMC) approach for a robust estimate of the parameters.
For each parameter considered in the calibration, the MCMC calibration starts from a prior, in our case a uniform distribution around an initial guess computed by minimizing with a least-square minimization procedure the difference between data and model results, and provides its posterior probability density function. 

Based on the posterior estimate of the model calibration it is possible to supply confidence interval for the time evolution of the different compartments.

\begin{figure}
    \centering
    \includegraphics[width=\textwidth]{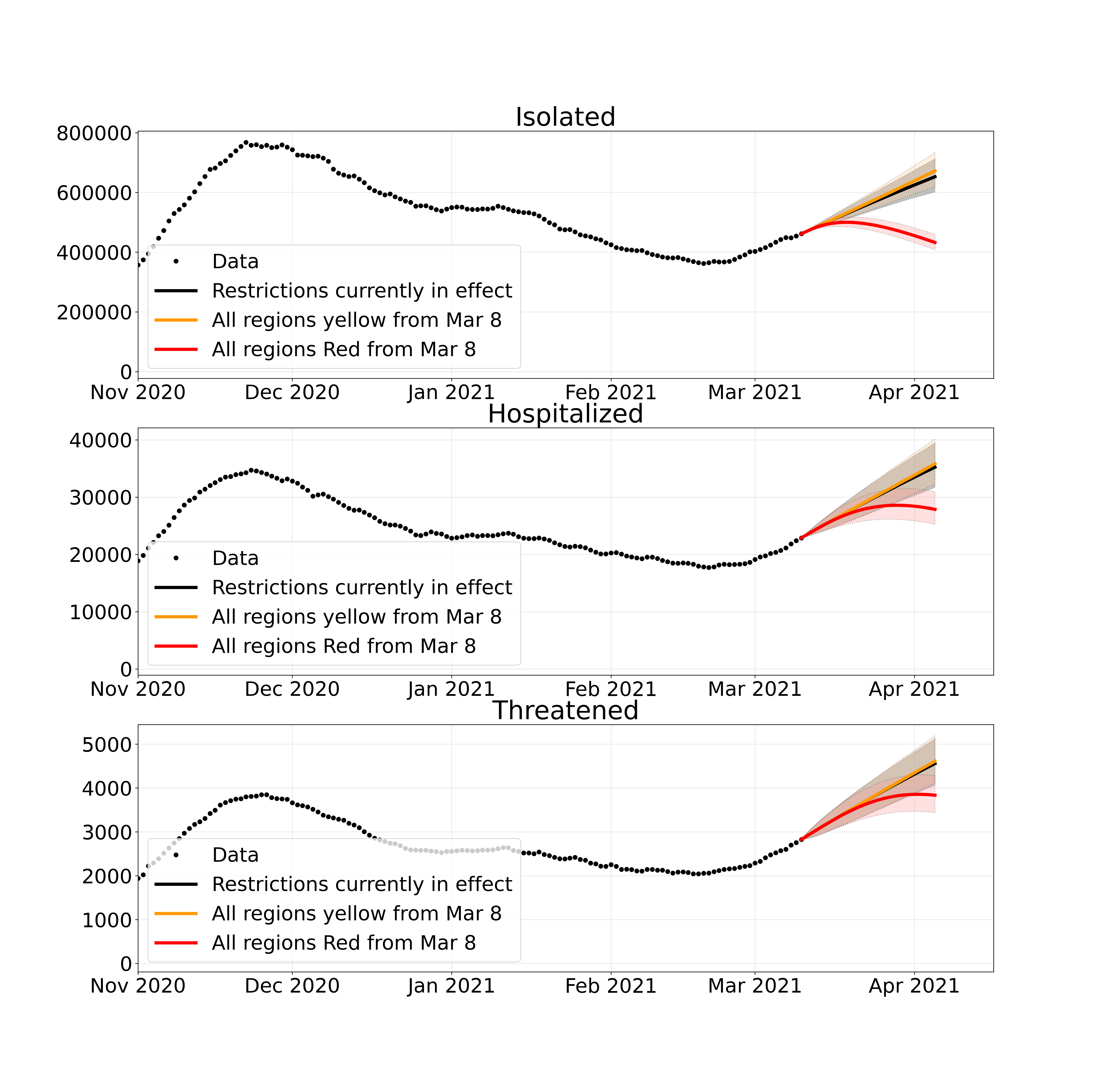}

    \caption{30-day forecast of four compartments with current restrictions (at March 7, 2021) and with 2 alternative restriction scenarios and 95\% confidence interval}
    \label{fig:forecast_mcmc}
\end{figure}

The solid curves displayed in Figure \ref{fig:forecast_mcmc} represent the time evolution of the mean value of the different compartments in the three considered scenarios, while the shaded region surrounding each curve represents the 95\% confidence interval associated with the probabilistic characterization of the model parameters. 

\newpage

\section{Conclusions}\label{sec:conclusions}
\LD{In this paper, we presented the {\it epiMOX} dashboard for the analysis of the COVID-19 epidemic that hit northern Italy in early Spring 2020 and that is still severely affecting the entire Country in March 2021.} Data were reported for several compartments (\textit{total and \REV{daily positive}} cases, \textit{isolated at home}, \textit{hospitalized}, \textit{hosted in ICUs}, \textit{cumulative deaths}, \textit{recovered}) that allow to provide a synthetic yet informative description of the epidemic evolution. Several analyses \LD{were carried out by means of the dashboard}. \LD{A careful comparison between the first two epidemic waves (the one of Spring 2020, and the one that took off in Fall 2020) is made. Finally, we have reported results, mainly concerning predictions, obtained by using the novel epidemiological differential mathematical model SUIHTER}. \REV{This allowed us to provide} short term forecasts on the epidemic trends, as well to examine some what-if scenarios related to the possible implementations of different NPIs than those that had been actually enforced.

\bibliographystyle{abbrv}
\bibliography{paperEpiMOXdashboard}

\begin{thebibliography}{10}

\bibitem{UKvariantSeverity}
Public {H}ealth {E}ngland ({PHE}), {NERVTAG} note on {B}.1.1.7 severity.

\bibitem{Battiston}
{R. Battiston} webpage.
\newblock \url{https://www.robertobattiston.it/}.
\newblock Accessed: 2021-03-12.

\bibitem{UKvariantISS}
Istituto {S}uperiore di {S}anit\`a, {P}revalenza della variante {VOC}
  202012/01, lineage {B}.1.1.7 in {I}talia. {S}tudio di {P}revalenza 4-5
  {F}ebbraio 2021, 2021.

\bibitem{UKvariantSeverityItaly}
Istituto {S}uperiore di {S}anit\`a, {Q}ual \`e la trasmissibilit\`a della
  ``variante inglese" in {I}talia?, 2021.

\bibitem{bertuzzo}
E.~Bertuzzo, L.~Mari, D.~Pasetto, S.~Miccoli, R.~Casagrandi, M.~Gatto, and
  A.~Rinaldo.
\newblock The geography of {COVID}-19 spread in {I}taly and implications for
  the relaxation of confinement measures.
\newblock {\em medRxiv}, 2020.

\bibitem{brauer}
F.~Brauer, C.~Castillo-Chavez, and Z.~Feng.
\newblock {\em Mathematical models in epidemiology}.
\newblock Springer, 2019.

\bibitem{D1SC01203G}
J.~Chen, K.~Gao, R.~Wang, and G.-W. Wei.
\newblock Prediction and mitigation of mutation threats to covid-19 vaccines
  and antibody therapies.
\newblock {\em Chemical Science}, pages~--, 2021.

\bibitem{Gatto}
M.~Gatto, E.~Bertuzzo, L.~Mari, S.~Miccoli, L.~Carraro, R.~Casagrandi, and
  A.~Rinaldo.
\newblock Spread and dynamics of the {COVID}-19 epidemic in {Italy}: Effects of
  emergency containment measures.
\newblock {\em Proceedings of the National Academy of Sciences},
  117(19):10484--10491, 2020.

\bibitem{sidarthe}
G.~Giordano, F.~Blanchini, R.~Bruno, P.~Colaneri, A.~Di~Filippo, A.~Di~Matteo,
  and M.~Colaneri.
\newblock Modelling the {COVID}-19 epidemic and implementation of
  population-wide interventions in {Italy}.
\newblock {\em Nature Medicine}, pages 1--6, 2020.

\bibitem{giordano2021vaccination}
G.~Giordano, M.~Colaneri, A.~D. Filippo, F.~Blanchini, P.~Bolzern, G.~D.
  Nicolao, P.~Sacchi, R.~Bruno, and P.~Colaneri.
\newblock Vaccination and {SARS-CoV-2} variants: how much containment is still
  needed? a quantitative assessment, 2021.

\bibitem{DRAM}
H.~Haario, M.~Laine, A.~Mira, and E.~Saksman.
\newblock {DRAM}: Efficient adaptive {MCMC}.
\newblock {\em Statistics and Computing}, 16:339–354, 2006.

\bibitem{Hethcote}
H.~W. Hethcote.
\newblock The mathematics of infectious diseases.
\newblock {\em SIAM Rev.}, 42(4):599--653, 2000.

\bibitem{kermack}
W.~O. Kermack and A.~G. McKendrick.
\newblock A contribution to the mathematical theory of epidemics.
\newblock {\em Proceedings of the royal society of london. Series A, Containing
  papers of a mathematical and physical character}, 115(772):700--721, 1927.

\bibitem{martcheva}
M.~Martcheva.
\newblock {\em An introduction to mathematical epidemiology}, volume~61.
\newblock Springer, 2015.

\bibitem{Miles2019}
P.~R. Miles.
\newblock pymcmcstat: A {P}ython package for {B}ayesian inference using
  {D}elayed {R}ejection {A}daptive {M}etropolis.
\newblock {\em Journal of Open Source Software}, 4(38):1417, 2019.

\bibitem{SUIHTER20}
N.~Parolini, L.~Dede', P.~F. Antonietti, G.~Ardenghi, A.~Manzoni, E.~Miglio,
  A.~Pugliese, M.~Verani, and A.~Quarteroni.
\newblock {SUIHTER}: A new mathematical model for {COVID-19}. {A}pplication to
  the analysis of the second epidemic outbreak in {I}taly, 2021.

\bibitem{PTT20}
C.~Piazzola, L.~Tamellini, and R.~Tempone.
\newblock A note on tools for prediction under uncertainty and identifiability
  of {SIR}-like dynamical systems for epidemiology.
\newblock 2020.

\bibitem{QDP20Springer}
A.~Quarteroni, L.~Dede', and N.~Parolini.
\newblock {\em Data Analysis and Predictive Mathematical Modeling for COVID-19
  Epidemic Studies}, pages 1--7.
\newblock Springer International Publishing, Cham, 2020.

\bibitem{SG64}
A.~Savitzky and M.~J.~E. Golay.
\newblock {Smoothing and Differentiation of Data by Simplified Least Squares
  Procedures.}
\newblock {\em Analytical Chemistry}, 36(8):1627--1639, 1964.

\end{thebibliography}

\end{document}